\documentclass [prb,reprint,notitlepage,superscriptaddress,floatfix] {revtex4-2}
\usepackage{amsmath}
\usepackage{graphicx}
\usepackage{lmodern}
\usepackage{amsmath}

\usepackage{color}
\usepackage{amssymb}
\usepackage{bm}
\usepackage{braket}
\usepackage{mathtools}
\newcommand{\beq} {\begin{equation}}
\newcommand{\eeq} {\end{equation}}
\newcommand{\bea} {\begin{eqnarray}}
\newcommand{\eea} {\end{eqnarray}}
\newcommand{\be} {\begin{equation}}
\newcommand{\ee} {\end{equation}}

\newcommand{\mr}{moir\'{e} }

\usepackage[dvipsnames]{xcolor}
\usepackage[colorlinks=true]{hyperref}
\hypersetup{
    colorlinks=true,
    linkcolor=blue,
    citecolor=blue,
    urlcolor=blue
} 

\begin{document}

\title {Pair density wave and loop current promoted by van Hove singularities in \mr systems }
\author{Zhengzhi Wu}
\author{Yi-Ming Wu}
\affiliation{Institute for Advanced Study, Tsinghua University, Beijing 100084, China}
\author{Fengcheng Wu}
\affiliation{School of Physics and Technology, Wuhan University, Wuhan 430206, China}
\affiliation{Wuhan Institute of Quantum Technology, Wuhan 430206, China}
\date{\today}

\begin{abstract}
We theoretically show that in the presence of conventional or higher order van Hove singularities(VHS), the bare finite momentum pairing, also known as the pair density wave (PDW), susceptibility can be promoted to the same order of the most divergent bare BCS susceptibility through a valley-contrasting flux 3$\phi$ in each triangular plaquette at $\phi=\frac{\pi}{3}$ and $\phi=\frac{\pi}{6}$ in \mr systems. This makes the PDW order a possible leading instability for an electronic system with repulsive interactions. We confirm that it indeed wins over all other instabilities and becomes the ground state under certain conditions through the renormalization group calculation and a flux insertion argument. Moreover, we also find that a topological nontrivial loop current order becomes the leading instability if the Fermi surface with conventional VHS is perfectly nested at $\phi=\frac{\pi}{3}$. Similar to the Haldane model, this loop current state has the quantum anomalous Hall effect. If we dope this loop current state or introduce a finite next-nearest neighbour hopping $t^{\prime}$, the chiral $d$-wave PDW becomes the dominant instability. Experimentally, the flux can be effectively tuned by an out-of-plane electric field in \mr systems based on graphene and transition metal dichalcogenides.
\end{abstract}
\maketitle

\section{Introduction} 
\label{sec:introduction}
Although superconductivity from the condensation of zero center-of-mass momentum Cooper pairs is commonly observed in many superconducting materials, that with finite momentum Cooper pairs, also known as pair density wave (PDW), stays rare in nature\cite{Agterberg}. The PDW can be thought as a superconducting state with periodic spatial modulations in the order parameter, which vanishes on average. Its rareness can be ascribed to the fact that, for a conventional Fermi liquid with time-reversal and inversion symmetry, the pairing susceptibility $\chi_{\text{sc}}(\bm{q},T)$ diverges only at $\bm{q}=0$ in low energy limit. The first proposal for the finite $\bm{q}$ pairing is the Fulde-Ferrell-Larkin-Ovchinnikov (FFLO) state\cite{FF,LO}, which is predicted to exist in a clean superconductor in the presence of a high magnetic field if the orbital pairing breaking effect is negligible, i.e. without the creation of Abrikosov vortcies,  and the superconducting state persists up to the Pauli limit. So stringent are these conditions that very few materials can realize this FFLO state. Nonetheless, some experimental evidence for its existence have been reported in organic superconductors\cite{cryst8070285}, heavy fermion compounds\cite{Matsuda2007}, and iron-based superconductors\cite{Gurevich,FeSC}. 

In underdoped cuprates, the PDW has been proposed as a competitor to $d$-wave uniform superconductivity\cite{Berg2007,Wang2015a,Wang2015b,Wang2018}. Unlike the FFLO state,  this PDW has zero coupling to the external magnetic field, and arises from strong electronic correlation. As a result, it coexists or neighbors with other charge or spin orders in the phase diagram\cite{Tranquada1995,Fujita2004,Zimmermann2011}. The complicated interplay among PDW and other electronic orders makes it hard to identify which order is primary. Because of this complicated interplay, another scenario is also proposed, where the PDW is argued to be the mother order, while other orders are descendants from it\cite{Berg_2009,PALee2014}. Indeed, through partial melting of a PDW, there may be other vestigial orders such as charge density wave or nematicity, appearing through a cascade of finite $T$ transitions\cite{Agterberg2008,Nie2014,Fradkin2015}. Probably the most nontrivial result from this scenario is the development of the highly exotic charge-$4e$ or even charge-$6e$ superconductors\cite{Berg2009,Agterberg2011}, of which the experimental signatures have not been reported until very recently\cite{ge2022}. With these many unsettled yet interesting puzzles, it is highly valuable to find a platform which can realize the PDW order as a unique ground state, and the interplay between different orders at finite temperature can be investigated in depth. 

Here in this paper, we argue that a certain class of two dimensional moir\'e band structures with van Hove singularities (VHS)\cite{Li2010,Brihuega2012,Wu2021} and enhanced PDW susceptibilities can serve as the promising platform for hosting the PDW in the ground state. The bare PDW susceptibilities are enhanced to the same order of the bare BCS susceptibilities in these systems. Conventionally, a van Hove singularity occurs in a two dimension system when the Fermi level is tuned to the energy dispersion saddle point. Formally, the condition can be expressed as $\nabla_{\bm{k}}\epsilon(\bm{k})=0$ and $\det D<0$ where $D$ is the $2\times2$ Hessian matrix defined by $D_{ij}=\frac{1}{2}\partial_{k_i}\partial_{k_j} \epsilon(\bm{k})$. An example which satisfies this constraint is $\epsilon(\bm{k})=k_x^2-k_y^2$. Close to this VHS, the electron density of states diverges logarithmically, and more often than not, the VHS is associated with Fermi surface nesting. These facts indicate that there has to be a competition among different electronic orders, in both particle-hole channel and particle-particle channel. In an unbiased analysis all these orders must be treated on equal footing. There have been plenty of discussions of competing orders near conventional van Hove filling in the literature \cite{Furukawa,Nandkishore2012,PhysRevX.8.041041,PhysRevB.100.085136,PhysRevB.102.085103,PhysRevB.104.195134}. However, even with nested Fermi surface, it is various particle-hole density wave orders, such as charge density wave and spin density wave orders, that generally win over uniform superconducting orders in the previous studies \cite{Furukawa,PhysRevX.8.041041,PhysRevB.100.085136}. This is due to the fact that a nested Fermi surface often enhances the bare particle-hole susceptibility $\Pi_{ph}(\bm{Q},T)$ at finite momentum such that it diverges at the same order as the bare particle-particle susceptibility $\Pi_{pp}(0,T)$, which scales as $\ln^2(W/T)$ with $W$ being the bandwidth. Nevertheless, it is very rare that the bare $\Pi_{pp}(\bm{Q},T)$ can be enhanced to the same order of $\Pi_{pp}(0,T)$, which is a necessary condition for the PDW order to be the leading instability in the weak coupling regime. In this work, we show that the bare PDW susceptibility can be enhanced to the same order of the BCS susceptibility by tuning the effective valley-dependent flux in the moir\'e system, which can be controlled, for example, by an applied out-of-plane displacement field. This system can favour the stable PDW order as the ground state with repulsive interactions.

 
Besides the conventional van Hove singularity (CVHS), the moir\'e system also allows for feasible realizations of the higher order van Hove singularity (HOVHS)\cite{Yuan2019,Wufengcheng2020,Guerci}. Using the notations above, a HOVHS can be formally defined as the case when $\det D=0$. This can be satisfied when one of the eigenvalues of $D$ vanishes and the other stays nonzero, or when both of these two eigenvalues vanish. Examples of these two types are $\epsilon(\bm{k})=k_y^2-k_x^4$ and $\epsilon(\bm{k})=k_x(k_x^2-\sqrt{3}k_y^2)$, respectively. Following the terminology in Ref \onlinecite{Yuan2019} we name the first case as type-I and the second as type-II. It is obvious that the type-I HOVHS is parity even and the type-II is parity odd. Like in the CVHS, the electron density of states near a HOVHS also diverges. But instead of a logarithmic way, it diverges in a power-law manner in the case of HOVHS. This behavior strongly promotes the competition among different orders, as well as their transition temperature $T_c$. There is a crucial difference between type-I and type-II HOVHS: the bare PDW susceptibility diverges only in the latter, due to the simple fact that the type-II dispersion is odd in momentum. We therefore mainly focus on type-II when discussing the HOVHS case. The type-II HOVHS also promotes the bare PDW susceptibility to diverge at the same order of the BCS susceptibility.

As discussed above, different orders seriously compete with each other near a VHS due to the divergent density of states, and it is likely that under certain circumstances the PDW stands out, wining over all other orders. We will show that this indeed can occur. To inspect this competition in an unbiased way, we thereby employ the parquet renormalization group (pRG) analysis. This was first introduced in the discussion of messon scattering\cite{osti_4338008}, and was later successfully applied to discuss competing orders in interacting one-dimensional electron gas\cite{oneD1979}, cuprates\cite{Furukawa,Dzyaloshinskii1997}, iron-based superconductors\cite{Chubukov2008}, graphene\cite{Nandkishore2012}, and more recently, moir\'e band structures with van Hove fermiology\cite{PhysRevB.100.085136,PhysRevB.102.085103,PhysRevB.104.195134}. It involves in identifying the leading divergent free susceptibilities, and inserting them as building blocks to renormalize different interactions, which helps to identify the leading instability in low energy limit. 

The model we use is inspired by the moir\'e band structure of the twisted bilayer transition metal dichalcogenides (TMD)\cite{Wufc2018,Wufc2019,Zhang2020,Shabani2021,Weston2020,Devakul2021,ZhangYang2021,doi:10.1073/pnas.2021826118,Tran_2020,Vitale_2021,Bi2021,Scherer2021}, the ABC stacked trilayer graphene on h-BN\cite{Schrade2019,PhysRevB.99.205150,PhysRevB.101.035122}, twisted double bilayer graphene\cite{Wufengcheng2020,PhysRevB.99.075127,PhysRevB.99.235417,PhysRevB.99.235406,PhysRevX.9.031021,Lee2019,Haddadi2020}, and also twisted bilayer graphene \cite{PhysRevX.8.041041,PhysRevX.8.031087,PhysRevB.98.045103,PhysRevX.8.031088,Dmitry2022}. In all these systems, the small twist angle or the small lattice constant mismatch results in a large scale triangular moir\'e pattern. The \mr superlattices profoundly changes the low energy band structure, and in some cases lead to nearly flat band, where the interaction plays an important role. Therefore, these system have been suggested to simulate the Hubbard physics\cite{Wufc2018}. Apart from the twist angle, a valley-contrasting flux is another important tuning parameter in these systems. This flux changes the non-interacting band structure dramatically. For example, it changes the location of the van Hove singularities and the Fermi surface nesting in both the particle-particle and particle-hole channels, and thus controls the bare susceptibilities of various orders. This flux can be modeled by endowing the nearest neighbor hopping $t$ with a complex phase factor $\phi$ and modifies $t$ into $t e^{i\phi}$. Experimentally, the valley-contrasting flux can be effectively tuned by an out-of-plane electric field \cite{Wang2020,Pan2020,Ghiotto2021,Zhang2021}. 

\begin{figure}
    \includegraphics[width=8.5cm]{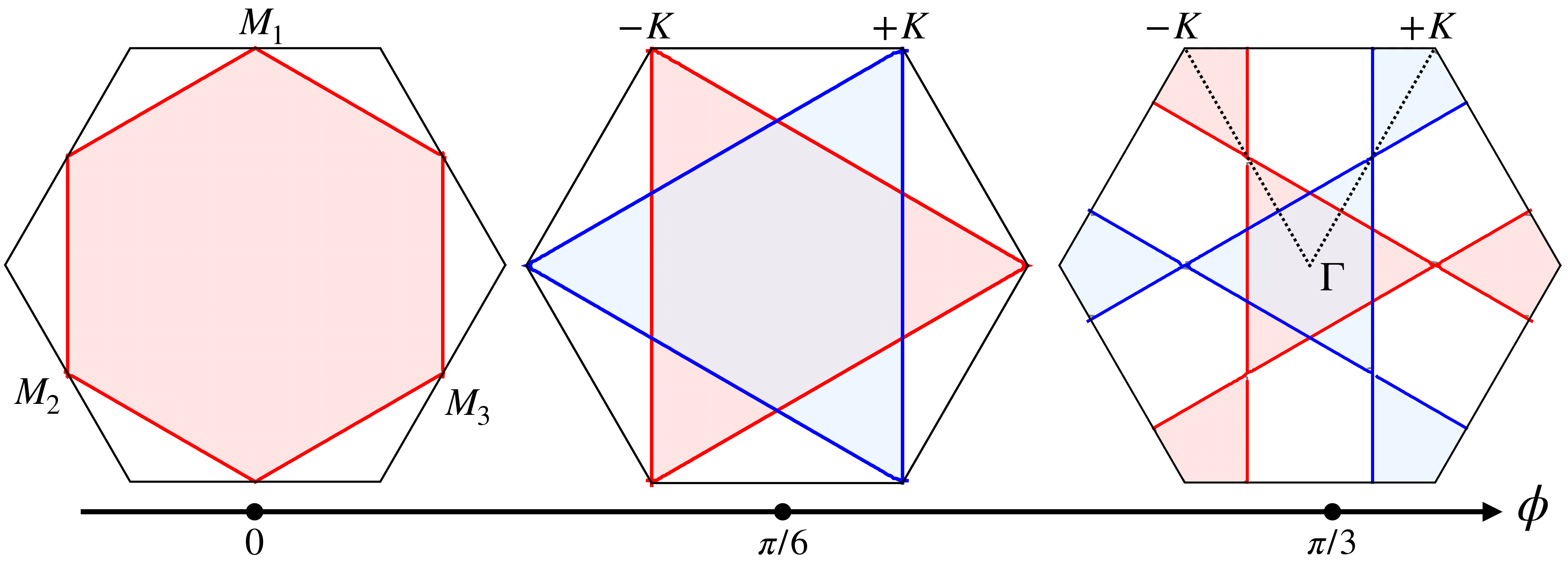}
    \caption{Perfectly nested low energy Fermi surfaces of moir\'e systems for different valleys (red and blue lines) in the presence of a valley-contrasting flux. In the absence of the flux, van Hove singularity appears when the system is doped to the $\bm{M}$ point. As $\phi$ increases,  there can be two higher order van Hove singularities located at the Brillouin zone corners, or six conventional van Hove singularities at the middle point between $\Gamma$ (Brillouin zone center) and $\pm \bm{K}$.}\label{fig:intro}
\end{figure}

In Fig.\ref{fig:intro} we show how the Fermi surfaces with only nearest neighbor hopping for different valleys evolve with different $\phi$. For the realistic system, the next nearest neighbor hopping $t'$ should also been considered, as it reduces the Fermi surface nesting and suppresses the particle-hole channel density waves. We will discuss this point in RG calculations. In the perfect nesting model with $\phi=0$, electronic bands from different valleys can hardly be distinguished. In the ideal case, the valley components, together with the spin components, form four fermion flavors, giving rise to an (emergent) $\text{SU(4)}$ symmetry in the non-interacting Hamiltonian. A nonzero $\phi$ breaks this symmetry and also spoils some degeneracy which is protected by the $\text{SU(4)}$ symmetry. We show that there is a symmetry transformation regarding $\phi$, which turns the valley-polarized pairing at $\phi=0$ to a PDW order at $\phi=\pi/3$, for which there are six CVHS in the moir\'e Brillouin zone at van Hove doping. The case with $\phi=\pi/6$ is of special interest because there are two type-II HOVHS points located at the moir\'e Brillouin zone corners $\pm\bm{K}$.

Based on the non-interacting band structure, we further introduce different initial interactions as the input of our RG analysis. Our results are obtained for the conventional and higher order VHS separately. In the conventional case with $\phi=\pi/3$ and $\text{SU(4)}$ preserving interactions, we find that PDW loses to a topological loop current order in the perfect nesting limit, i.e. with a vanishing $t'$ and the filling fraction approaching the van Hove filling. This loop current order results from an imaginary $d$-wave particle-hole condensate, and is similar to the Haldane model for quantum anomalous Hall effect \cite{PhysRevLett.61.2015}. This loop current has been previously discussed in the context of a $\phi\to0$ model with the fermion flavor equal to 4. We show that indeed in our model, the loop current order at $\phi=\pi/3$ is related to that at $\phi=0$ via a symmetry transformation. When the system is tuned away from the perfect nesting, we find that the ground state favors a chiral $d$-wave PDW instead. If we include $\text{SU}(4)$ breaking interactions, such as Hund coupling, the chiral-$d$ wave or $s$-wave PDW exists in a wide parameter space as a stable phase even in the perfect nesting limit. Again, via symmetry transformation, this PDW is related to the chiral $d$-wave valley-polarized uniform superconductivity at $\phi=0$. As a result, PDW phase can also exist for a generic filling $\nu=n$ ($\nu$ is the number of electrons per site) at $\phi=\frac{\pi}{3}$ (although not necessarily the chiral $d$-wave PDW), as long as the same interactions favour the valley polarized uniform superconductivity at $\phi=0$ and filling $\nu=4-n$. For the higher order VHS, our results suggest that PDW can be the leading instability once a valley splitting field is introduced. The resulting PDW phase has a relatively high transition temperature and is thus more promising to be observed experimentally. 

Our paper is organized as following. In Sec.\ref{sec:model} we introduce the lattice model inspired by the moir\'e band structure of twisted bilayer TMD, which is also relevant to the twisited double bilayer graphene system and ABC stacked trilayer graphene system. In Sec.\ref{sec:class} we discuss the symmetries of our model and symmetry classification of the superconducting orders. In Sec.\ref{sec:CVHS}  we first introduce a flux insertion operation to connect the orders between $\phi=0$ and $\phi=\frac{\pi}{3}$, which generates the ground state at $\phi=\frac{\pi}{3}$ from a corresponding result at $\phi=0$. After that, we perform the concrete RG analysis for the conventional VHS, and the results are consistent with those obtained via the flux insertion operation from previous analysis. In Sec.\ref{sec:higer_order_van_hove_singularities_and_bare_susceptibilities}, we conduct the parquet RG analysis for the higher order van Hove singularities of two patch model and one patch model, and find prevailing PDW order in the latter. Concluding remarks are presented in Sec.\ref{sec:discussion}.

\section{The Lattice Model}
\label{sec:model}
We consider the following triangular lattice model $\hat{H}=\hat{H}_0+\hat{H}_I$, where the single particle Hamiltonian $\hat{H}_0$ is given by
\begin{equation}
\begin{aligned}
    \hat{H}_0=&-t\sum_{i,\hat{a}_j}\sum_{v=\pm,f} e^{iv\phi}\hat{c}^{\dagger}_{v,f}(i) \hat{c}_{v,f}(i+\hat{a}_j)\\
    &-t^{\prime}\sum_{\braket{\braket{ij}}}\sum_{v=\pm,f} \hat{c}^{\dagger}_{v,f}(i) \hat{c}_{v,f}(j)+\text{h.c.},
    \label{lattice}
    \end{aligned}
\end{equation}
where $t$ and $t^{\prime}$ are the nearest-neighbour and next-nearest neighbour hopping amplitudes. $\hat{a}_j=\hat{a}_1,\hat{a}_2,\hat{a}_3$ are the three nearest-neighbour unit vectors on the triangular lattice (with lattice constant $a=1$): $\hat{a}_1=(1,0),\hat{a}_2=(-\frac{1}{2},\frac{\sqrt{3}}{2}),\hat{a}_3=(-\frac{1}{2},-\frac{\sqrt{3}}{2})$. $f=1,2,...,N_f$ is the fermion flavour. This model describes spinless or spinful fermions (for each fixed $v$), where $N_f$ is equal to one or two. A larger $N_f$ can effectively describe multi-orbital physics. The most interesting part is the phase factor $\phi$, which induces a flux $\pm 3\phi$ in each elementary triangle plaquette, and the flux is opposite for different valleys:  $v=\pm$ represents the valley degree of freedom (DOF). This nontrivial flux pattern can be experimentally realized in the twisted homo-bilayer TMD (corresponding to $N_f=1$) \cite{PhysRevB.104.075150}, twisted ABC trilayer graphene/h-BN, and twisted double bilayer graphene(corresponding to $N_f=2$) \cite{PhysRevB.99.205150}.
The role of $\phi$ is to move the location of Van Hove singularities in the Brillouin zone. When $\phi=\frac{\pi}{6}$ and $t^{\prime}=0$, three Van Hove singularities within the same valley will merge into a higher-order Van Hove singularity. 

The interacting Hamiltonian $\hat{H}_I$ can be any symmetry-allowed four fermion interactions, including the Hubbard interaction, Heisenberg interaction, Hund interaction, etc. Explicitly $\hat{H}_I$ with $N_f=2$ can be written as
\begin{equation}
\begin{aligned}
    \hat{H}_I= &\frac{U}{2} \sum_i n_{i}^{2}+J \sum_{\braket{ij},n} c_i^{\dagger}T^nc_ic_j^{\dagger}T^nc_j\\
    &+V_h \sum_{i} (c_i^{\dagger}\vec{S}c_i)^2+K \sum_{i} (c_i^{\dagger}\vec{L}c_i)^2,
    \end{aligned}
    \label{lat_int}
\end{equation}
where $T^n$ are the fifteen generators of the $\text{SU(4)}$ group, and $J$ is the coupling constant of the Heisenberg interaction. The $V_h$ and K are the coupling constants of spin and orbital Hund couplings, and $\vec{S}=\vec{\sigma},\vec{L}=\vec{\tau}$ are the spin-$\frac{1}{2}$ Pauli matrices acting on the spin and valley degrees of freedom respectively. 
 
\section{Classification of orders}
\label{sec:class}
In this section, we classify all the possible particle-hole and particle-particle orders which spontaneously break the global symmetries of Eq. \eqref{lattice}. We will focus on $N_f=2,  \phi=
\frac{\pi}{3}$ or $\phi=
\frac{\pi}{6}$. Besides the $\text{U(1)}_c$ symmetry corresponding to the charge conservation and the lattice translation symmetry, the global symmetries of the lattice model is $\text{SU}(2)_s\times \text{U(1)}_v\times C_{3v}$, where $\text{SU}(2)_s$ is the spin rotation symmetry; $\text{U(1)}_v$ is the $\text{U(1)}$ valley ($v$) rotation symmetry, and $\text{C}_{3v}$ is the point symmetry group of the lattice model. 

We note that if $\phi=0$ and $N_f=2$, the lattice model Eq.\eqref{lattice}  enjoys the $\text{SU(4)}$ symmetry. Therefore, we start with the spontaneous symmetry breaking of $\text{SU(4)}$ symmetry, and then break this symmetry down to $\text{SU}(2)_s\times \text{U(1)}_v$ with a nonzero $\phi$. The fermions $c_f$ serve as the fundamental representation of the $\text{SU(4)}$ internal symmetry. We physically view the four flavours which form the fundamental representation as spin $\frac{1}{2}$ and valley pseudospin $1\over2$ DOF.  The tensor product of two fundamental representations of $\text{SU(4)}$ group satisfies: $\mathbf{4}\otimes \mathbf{4}=\mathbf{6}\oplus\mathbf{10}, \bar{\mathbf{4}}\otimes \mathbf{4}=\mathbf{1}\oplus\mathbf{15}$, where $\mathbf{4}$ is the fundamental representation and $\bar{\mathbf{4}}$ is the complex conjugation of $\mathbf{4}$; $\mathbf{6}$ is the vector representation of $\text{SO(6)}$
and $\mathbf{10}$ is the antisymmetric tensor representation of $\text{SO(6)}$ \cite{ramond2010group}. 
The decomposition of $\mathbf{4}\otimes \mathbf{4}=\mathbf{6}\oplus\mathbf{10}$ means that the superconducting orders which spontaneously break the internal $\text{SU(4)}$ symmetry can only be degenerate between spin-singlet-valley-triplet (ST) and spin-triplet-valley-singlet (TS), corresponding to the vector representation $\mathbf{6}$ \cite{PhysRevB.101.035122,PhysRevLett.121.087001}; or between spin-singlet-valley-singlet (SS) and spin-triplet-valley-triplet (TT), corresponding to the tensor representation $\mathbf{10}$ \cite{PhysRevB.101.035122}. The vector representation $\bm{6}$ is  parity even while the tensor representation $\bm{10}$ is parity odd. These superconducting orders include uniform superconductors and PDW orders,  which may further break the $C_{3v}$ and translational symmetry. Meanwhile, the decomposition $\bar{\mathbf{4}}\otimes \mathbf{4}=\mathbf{1}\oplus\mathbf{15}$ 
constraints the particle-hole orders. If the translation symmetry is broken, the identity representation $\mathbf{1}$ means the charge density wave order (CDW), and the representation $\mathbf{15}$ represents the degenerate spin/valley density wave. If the translation symmetry is intact, the identity representation $\mathbf{1}$ is the chemical potential, and the representation $\mathbf{15}$ represents the degenerate spin/valley magnetism. 

Furthermore, a nonzero $\phi$ explicitly breaks the $\text{SU(4)}_f$ symmetry down to $\text{SU(2)}_s\times \text{U(1)}_v$ symmetry, as mentioned above. The degeneracy between the superconducting orders and particle-hole orders is all broken.
We can use the valley quantum number $L_z=-1,0,1$ and total spin quantum number of the Cooper pair to label the superconducting orders. In other words, we will have spin singlet or triplet pair with $L_z=-1,0,1$. The particle-hole orders are divided into spin orders and charge orders.

\section{Conventional van Hove singularities: Six patch model }
\label{sec:CVHS}
In this section, we focus on the $N_f=2$ and $\phi=\frac{\pi}{3}$ case of Eq. \eqref{lattice}. 
The Fermi surface at van Hove doping is shown in Fig. \ref{fig:intro}. There are six conventional van Hove singularities with log-divergent density of states. Interestingly, the physics for $\phi=0$ and $\phi=\frac{\pi}{3}$ are closely connected with each other. Concretely, there is an invertible local transformation between the systems in Eq.\eqref{lattice} with $\phi=0$ and $\phi=\frac{\pi}{3}$ at van Hove doping, which is proposed in \cite{https://doi.org/10.48550/arxiv.2203.05480} for the single flavour case i.e., $N_f=1$ of Eq.\eqref{lattice}. This local transformation guarantees that there is a one-to-one correspondence between the orders at $\phi=0$ and $\phi=\frac{\pi}{3}$. Below we first discuss this transformation, and then use pRG analysis to identify the leading orders at $\phi=\pi/3$. We find consistency with earlier results on $\phi=0$ case, by virtue of the local transformation. 


\subsection{Flux insertion}\label{sec:flux}
Before we dive into the detailed calculations, it's worth noticing that the physics at $\phi=\frac{\pi}{3}$ is closely connected with that at $\phi=0$. Actually, there exists a local transformation between $\phi=0$ and $\phi=\frac{\pi}{3}$ by inserting $\pi$ flux in each triangular plaquette and preserving the gauge choice of the kinetic energy in Eq.\eqref{lattice} of the lattice model. We label the transformation as G. The transformation G not only maps the Hamiltonian, which includes both the tight binding and the interaction terms, from $\phi=0$ to $\phi=\frac{\pi}{3}$, but also the ground state orders. This implies that if we know the leading order at $\phi=0$, we can immediately arrive at the leading order with $\phi=\frac{\pi}{3}$, which is just the order at $\phi=0$  acted by G. 

\begin{figure}
    \centering
    \includegraphics[width=8cm]{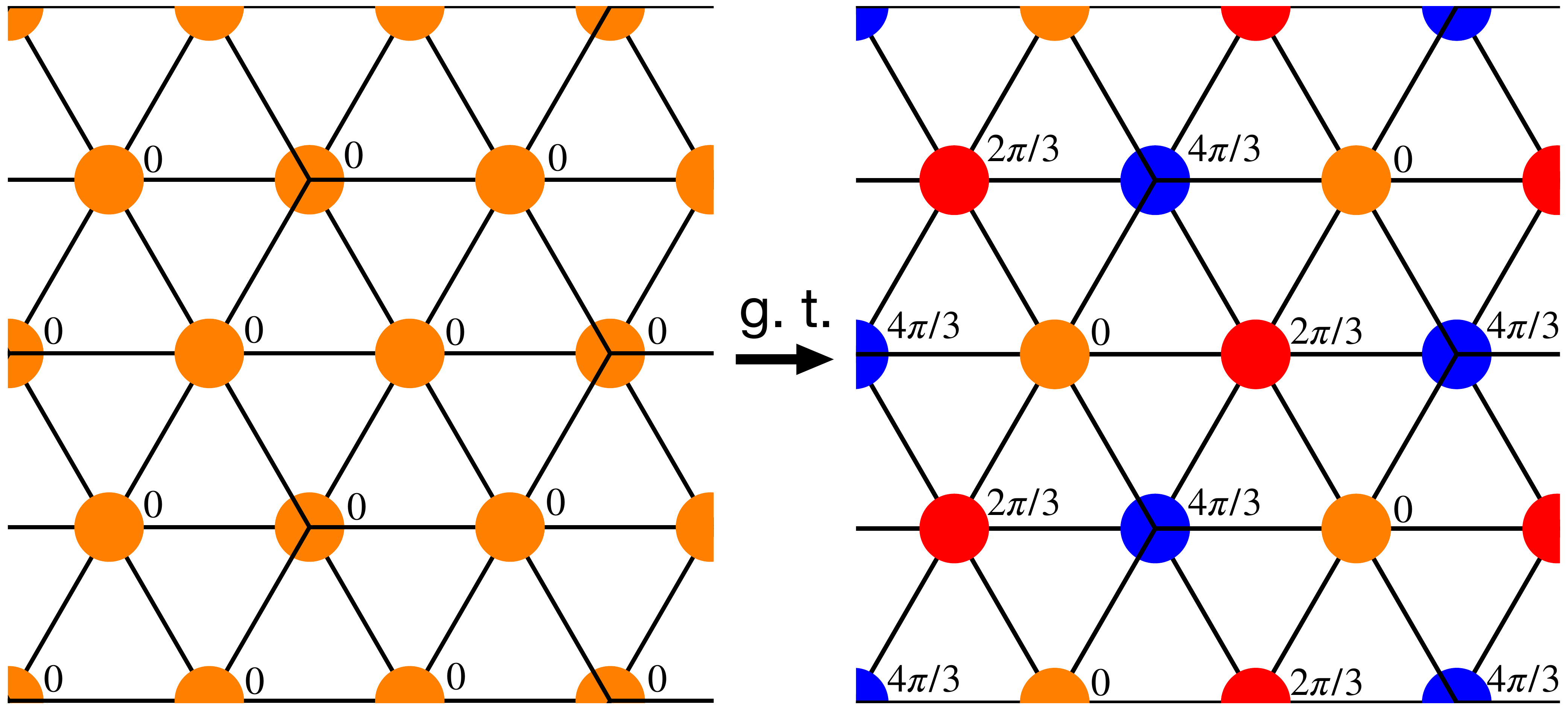}
    \caption{Real space illustration of the local gauge transformation defined in Eq. \eqref{eq:gt}.}
    \label{fig:gt}
\end{figure}


\begin{table*}
  \centering
  \begin{tabular}{lcccc}
    \hline\hline
    Flux &~~~~~\text{Intervalley pairing }~~~~~&~~~~~ \text{Intravelley pairing}~~~~~~~~&\text{Intervalley density wave}~~~~~&~~~~~\text{Intravalley density wave}\\\hline
    $\phi=0$ &$\bm{Q}=0\text{ SC}$ & $\bm{Q}=0\text{ SC}$ &$\bm{Q}=\bm{M}\text{~SDW/CDW}$&$\bm{Q}=\bm{M}\text{~SDW/CDW}$\\
   $\phi=\frac{\pi}{3}$   &$\bm{Q}=0\text{ SC}$ & $\bm{Q}=\pm\bm{K}$\text{~PDW} &$\bm{Q}=\bm{M}\pm\bm{K}\text{~SDW/CDW}$  &$\bm{Q}=\bm{M}\text{~SDW/CDW}$\\ \hline
  \end{tabular}
  	\caption{The correspondence between the orders in both particle-particle and particle-hole channels at $\phi=0$ and $\phi=\frac{\pi}{3}$ under the flux insertion G. The loop current order we discuss in this work is the intravalley CDW order.}
  	
	\label{order}
	\end{table*}

Concretely, the transformation G composes two parts. Starting from $\phi=0$, we first do the particle hole transformation: $c_{v,\sigma}(i)\rightarrow c^{\dagger}_{v,\sigma}(i)$, which inserts $\pi$ flux in each plaquette and the filling is changed from $n$ to $4-n$, where $n$ is the number of fermions per site. Then we perform the local gauge transformation
\begin{equation}
    c_{v,\sigma}(j)\to e^{i v\eta_j}c_{v,\sigma}(j)\label{eq:gt}
\end{equation}
with $\eta_j=\bm{K}\cdot \bm{r}_j$ and $\bm{K}=(\frac{4\pi}{3},0)$, which preserves the gauge choice of the tight binding term in Eq.\eqref{lattice}, such that the nearest-neighbor hopping terms along the $\hat{a}_1$, $\hat{a}_2$, and $\hat{a}_3$ directions have the same phase $\phi$. In Fig.\ref{fig:gt} we illustrate this gauge transformation on the lattice. This flux insertion G does not change the long-range hoppings and interactions we considered here, such as electron density interactions, Heisenberg exchange couplings, Hund couplings etc.

Since G is invertible, the orders at $\phi=0$ and $\phi=\frac{\pi}{3}$ also have a one-to-one correspondence. The results is summarized in the Table~\ref{order}. We first apply G to the superconducting orders. If we start from one valley polarized component of the valley-triplet uniform paring at $\phi=0$: $\Delta(\mathbf{r_i}-\mathbf{r_j})c_{+,\sigma}(i)(i\sigma_y)_{\sigma,\sigma^{\prime}}c_{+,\sigma^{\prime}}(j)$. This is uniform pairing as the pairing amplitude only depends on the relative coordinate $\mathbf{r_i}-\mathbf{r_j}$, which is invariant under the lattice translation symmetry. This pairing amplitude can have any form factors with respect to the relative coordinates, such as the s-wave, d-wave, etc. This order is mapped to the PDW order at $\phi=\frac{\pi}{3}$ under the transformation G: 
\begin{equation}
    \Delta(\mathbf{r_i}-\mathbf{r_j})e^{-i\mathbf{K}\cdot \frac{\mathbf{r_i}+\mathbf{r_j}}{2} }c^{\dagger}_{+,\sigma}(i)(i\sigma_y)_{\sigma,\sigma^{\prime}}c^{\dagger}_{+,\sigma^{\prime}}(j),\label{pdw}
\end{equation}
of which the pairing amplitude gains a phase factor $e^{i\frac{2\pi}{3}}$ under the elementary lattice translation. Similarly, the valley polarized component polarized with the other valley is mapped to the finite momentum pairing with $\mathbf{K}$ in Eq.\eqref{pdw} replaced with $-\mathbf{K}$. However, the intervalley pairing component $\Delta_{+-}(\mathbf{r_i}-\mathbf{r_j})c_{+,\sigma}(i)(i\sigma_y)_{\sigma,\sigma^{\prime}}c_{-,\sigma^{\prime}}(j)$ is mapped to intervalley uniform pairing at $\phi=\frac{\pi}{3}$.

This flux insertion operation G enables us to know the orders at $\phi=\frac{\pi}{3}$ from the results at $\phi=0$. Previous parquet RG calculations \cite{PhysRevB.100.085136}, functional RG (fRG) calculations \cite{PhysRevB.99.195120} and mean field calculations \cite{PhysRevLett.121.087001} on the $\text{SU(4)}$ Hubbard model on the triangular lattice with $\phi=0$ have revealed that the superconducting order near the van Hove doping has the chiral d-wave valley polarized component. The superconducting instability is the leading instability if the system is away from the perfect nesting limit, which is realized by introducing a finite next-nearest neighbour hopping $t^{\prime}$ or a finite doping from the perfect nesting.  Using the flux insertion G above, we can immediately arrive at the conclusion that chiral-d wave PDW order becomes the leading instability at $\phi=\frac{\pi}{3}$ away from the perfect nesting limit. The previous RG calculations with $\phi=0$ are reliable in the weak and moderate coupling regime, which means the chiral d-wave PDW also exists at least in this regime for $\phi=\frac{\pi}{3}$.  

We briefly discuss whether the PDW order is present at other fillings with $\phi=\frac{\pi}{3}$. The flux insertion operation G is applicable to any fillings. As a result, if the valley polarized uniform pairing is favoured at $\phi=0$ and filling $n$, then the PDW order will also be the leading order at $\phi=\frac{\pi}{3}$ and filling 4-$n$, regardless of whether the van Hove singularities are present or not. If the fermion interaction at $\phi=0$ preserves the $\text{SU}(4)$ symmetry, such as the Hubbard or Heisenberg interaction, then the Kohn-Luttinger instability will always become the leading instability with repulsive interactions at generic fillings. Further, recalling the representation decomposition of $\text{SU}(4)$: $\mathbf{4}\otimes \mathbf{4}=\mathbf{6}\oplus\mathbf{10}$, the valley polarized pairing is always degenerate with other pairings in both the representation $\mathbf{6}$  and $\mathbf{10}$. This means that the PDW order is always present at $\phi=\frac{\pi}{3}$ with repulsive $\text{SU}(4)$ preserving interactions, due to the Kohn-Luttinger instability at $\phi=0$. Moreover, from the fRG calculations in \cite{PhysRevB.99.195120}, the valley triplet pairing can be favoured as the unique leading instability in a certain regime of additional $\text{SU}(4)$ breaking Hund couplings, so the PDW order at $\phi=\frac{\pi}{3}$ will also exist in the same coupling parameter regime with additional $\text{SU}(4)$ breaking Hund couplings. To sum up, the PDW order at $\phi=\frac{\pi}{3}$ is not fine tuned to the van Hove doping, but is a stable phase at generic fillings with repulsive interactions.

There is another interesting state, i.e., loop current phase, near van Hove doping reported in previous RG calculations at $\phi=0$ \cite{PhysRevB.99.195120,PhysRevB.100.085136}. If the Fermi surface is nearly perfect nested, the loop current phase will be favoured as the leading instability by the repulsive Hubbard interaction and Heisenberg interaction. The order parameter of the loop current phase is the imaginary CDW order, which spontaneously breaks the time reversal symmetry, and induces a $\frac{\pi}{2}$ or $-\frac{\pi}{2}$ flux in each triangle plaquette. We can also use the flux insertion argument to construct the corresponding order at $\phi=\frac{\pi}{3}$ with repulsive Hubbard and Heisenberg interaction at van Hove doping and a nearly perfect nested Fermi surface. The order parameter of the loop current order at $\phi=0$ is \cite{PhysRevB.99.195120,PhysRevB.100.085136}:
\begin{equation}
    \Delta_{\text{loop current}}=\sum_{\bm{k},a}f_a(\bm{k})\hat{c}^{\dagger}(\bm{k}+\bm{M}_a)c(\bm{k}),\label{eq:loopc}
\end{equation}
where the sum over spin and valley is implicitly assumed henceforth. $\bm{M}_a$ with $ a=1,2,3$ are the three momenta of the van Hove singularities of the tight binding model $\hat{H}_0$ of Eq.\eqref{lattice}. The three form factors $f_a$ preserve the $C_3$ rotation symmetry and ensure that the expectation value of the order parameter in Eq. \eqref{eq:loopc} is purely imaginary. For example, it is fitted as: $f_{1}=2 \sin \left(k_{x} / 2\right) \sin \left(\sqrt{3} k_{y} / 2\right),f_{2,3}=\mp \cos \left(k_{x}\right) \pm \cos \left(\left(k_{x} \pm \sqrt{3} k_{y}\right) / 2\right)$ from the vertex obationed in fRG calculations \cite{PhysRevB.99.195120}.  Now we apply the flux insertion operator G, and the order parameter $\Delta_{\text{loop current}}$ becomes:
\begin{equation}
\begin{aligned}
    \Delta_{\phi=\frac{\pi}{3}}&=-\sum_{\bm{k},a,v}f_a(\bm{k})\hat{c}^{\dagger}_{v}(\bm{k}+\bm{M}_a+v \bm{K})c_{v}(\bm{k}+v \bm{K})\\
    &=-\sum_{\bm{k},a,v}f_a(\bm{k}+v \bm{K})\hat{c}^{\dagger}_{v}(\bm{k}+\bm{M}_a)c_{v}(\bm{k}),
    \end{aligned}
\end{equation}
where $v=\pm$ is the index of the valley. The order parameter $ \Delta_{\phi=\frac{\pi}{3}}$ is still the pure imaginary CDW order. The flux pattern at $\phi=\frac{\pi}{3}$ is shown in Fig. \ref{lc}.

As a result, the leading instability at $\phi=\frac{\pi}{3}$ is still the loop current phase with repulsive Hubbard and Heisenberg interaction near van Hove doping.  However, this is true with a nearly perfect nested Fermi surface. If we dope this loop current phase or induce a $t^{\prime}$ beyond a critical value to destroy the nesting, the loop current phase will become the degenerate chiral d-wave PDW order and uniform SC order. This phase transition corresponds to the transition from the loop current phase to chiral d-wave uniform SC order at $\phi=0$ \cite{PhysRevB.99.195120,PhysRevB.100.085136}.
\subsection{Bare susceptibilities}
The flux insertion argument above is specific to the lattice model like \eqref{lattice}. In cases when there lacks the information of the low Eq. energy lattice model, one can still apply the RG analysis to study the competing orders. The firs step is to identify 
the building blocks for parquet RG, which are various particle-hole and particle-particle susceptibilities:
\begin{equation}
    \begin{aligned} 
    \Pi_{\mathrm{ph}}(\bm{P},T)&=-T \sum_{n} \int \frac{d^{2} k}{(2 \pi)^{2}} G_{0}(\omega_n, k) G_{0}(\omega_n, \bm{P}+k) ,\\
    \Pi_{\mathrm{pp}}(\bm{P},T) &=T \sum_{n} \int \frac{d^{2} k}{(2 \pi)^{2}} G_{0}(\omega_n, k) G_{0}(-\omega_n, \bm{P}-k), 
    \end{aligned}
\end{equation}
where $G_{0}(\omega_n, q)=\frac{1}{i \omega_n-\epsilon(q)}$ and $\omega_n=(2n+1)\pi T$ is the fermion frequency. The leading divergent susceptibility is used as the flowing energy scale, with infrared limit $T\rightarrow 0$, in the parquet RG formalism.

We take the six patch model near the six van Hove singularities with the patch size $\Lambda$. There are three kinds of $\ln^2$ divergent susceptibilities:
\begin{equation}
    \begin{aligned}
    &\Pi_{pp}(0,T)=\frac{1}{4 \sqrt{3} \pi^{2} t} \ln \frac{\Lambda}{\max \{T, \mu\}} \ln \frac{\Lambda}{T},\\
      &\Pi_{pp}(\pm\bm{K},T)=\frac{1}{4 \sqrt{3} \pi^{2} t} \ln \frac{\Lambda}{\max \{T, \mu\}} \ln \frac{\Lambda}{T},\\
       &\Pi_{ph}(\bm{Q}_{\pm},T)=\frac{1}{8 \sqrt{3} \pi^{2} t} \ln \frac{\Lambda}{\max \{T, \mu\}} \ln \frac{\Lambda}{\max \left\{T, \mu, t^{\prime}\right\}}.
    \end{aligned}\label{eq:bare_six}
\end{equation}
Here the momentum $\bm{Q}_{\pm}$ are depicted in Fig.\ref{interaction_six}(a). We see that $\bm{Q}_+$ connects patches within the same valley, while $\bm{Q}_-$ connects patches from opposite valleys. 
The detailed calculations of Eq. \eqref{eq:bare_six} are presented in Appendix \ref{sec:calculations_of_the_bare_susceptibilities_in_the_six_patch_case}. This is also expected from the flux insertion argument. The only two $\ln^2$ divergent bare susceptibilities for $\phi=0$ are $\Pi_{pp}(0,T),\Pi_{ph}(\bm{M},T)$; after the flux insertion transformation G: $\phi=0\rightarrow\phi=\frac{\pi}{3}$ , the susceptibilities of the same valley and opposite valleys are mapped into those corresponding to different orders. Concretely, the particle-particle susceptibilities of the same and opposite valleys are mapped into $\Pi_{pp}(0,T)$ and $\Pi_{pp}(\pm\bm{K},T)$ respectively; the particle-hole susceptibilities of the same and opposite valleys are mapped into $\Pi_{ph}(\bm{Q}_+,T)$ and $\Pi_{ph}(\bm{Q}_-,T)$ respectively.

Note that even with a finite $t^{\prime}$ and/or finite doping (chemical potential), we still have $\Pi_{pp,\pm\pm}(\pm\bm{K},T)=\Pi_{pp}(0,T)$ and $\Pi_{ph}(\bm{Q}_+,T)=\Pi_{ph}(\bm{Q}_-,T) $. The comparison between cases with and without $t'$ is also shown in Fig.\ref{interaction_six}(a), which makes it clear that the particle-particle FS nesting is immune to the presence of a finite $t'$, consistent with Eq. \eqref{eq:bare_six}. This enables us to use a single parameter $d=\frac{\Pi_{ph}(\bm{Q}_+,T)}{\Pi_{pp}(0,T)}$ to characterize the degree of nesting in our following RG analysis. Note that for hexagonal lattices as we considered in the current work, the maximum value of $d$ is $1/2$, which is different from the square lattice case, where the maximum of $d$ is 1.

\subsection{Renormalization group analysis}
\begin{figure}
  \includegraphics[width=8.5cm]{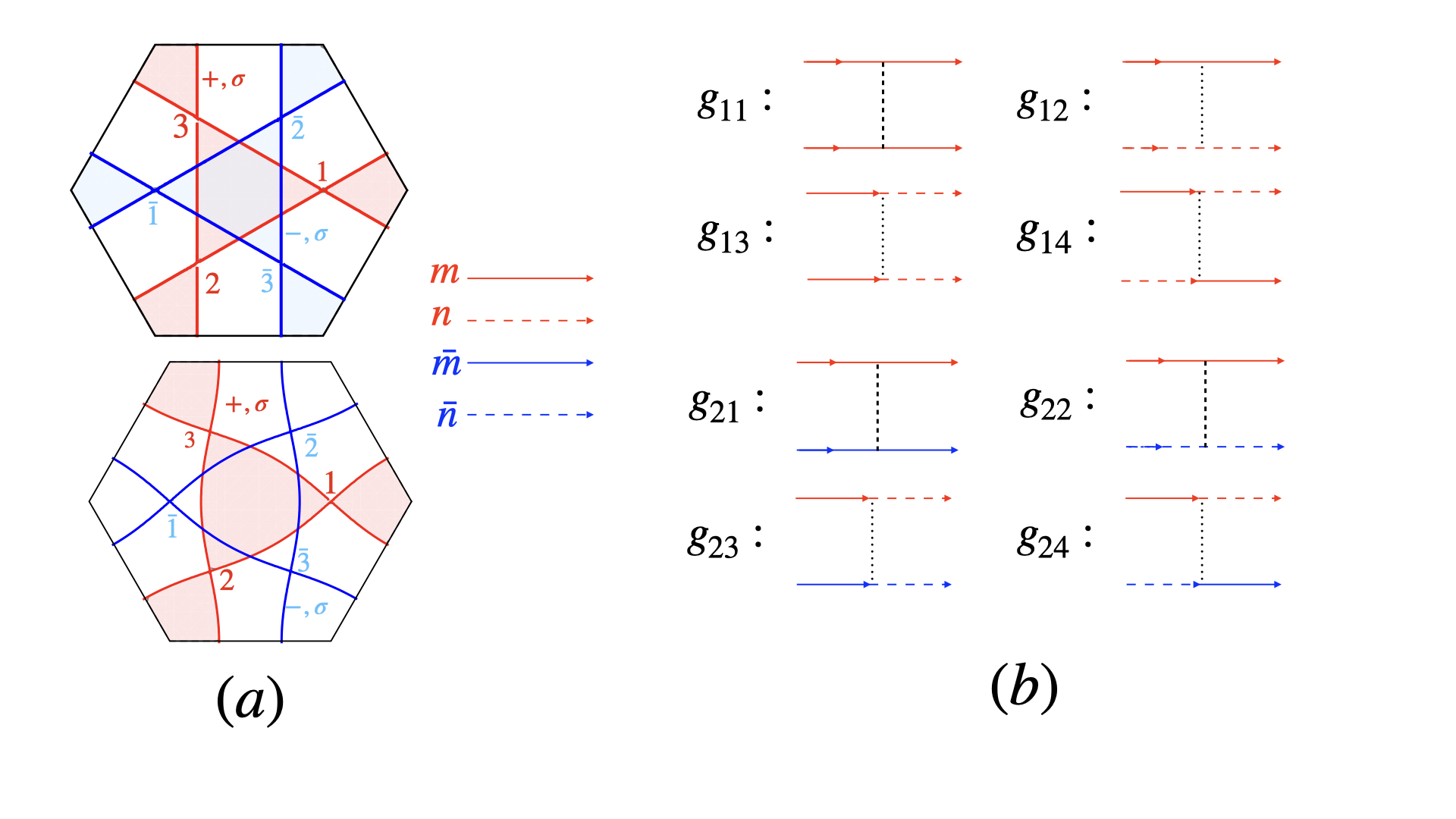}
  \caption{(a) The left panel is the Fermi surface at van Hove doping with $t^{\prime}=0$; the right panel is the Fermi surface at van Hove doping with $t^{\prime}=0.1$. The location of the van Hove singularities are not changed with nonzero $t^{\prime}$. The momentum vectors $\bm{Q}_+$ and $\bm{Q}_-$ are the particle-hole nesting vectors, at which the bare p-h susceptibilities diverge in $\ln^2$ manner. (b)All the twelve symmetry allowed independent four fermion interactions. The red and blue colors represent the opposite valleys and the solid and dashed lines represent different patches within the same valley. (c) Diagrammatic representation of the renormalization of the interaction $g_{13}$ by the particle-particle and particle-hole bubbles defined in Eq. \eqref{eq:bare_six}.}
  \label{interaction_six}
\end{figure}

There are twelve inequivalent symmetry-allowed four fermion interactions in the six patch case with $N_f=2$, which we show in Fig.\ref{interaction_six}(b). We label all the interactions as $g_{ij}$, where $i=1,2,4 $ and $ j=1,2,3,4$. Note $j=1,2$ represent forward scattering of the valley and patch degrees of freedom, while $j=3$ and $j=4$ represent umklapp and backward scattering respectively.  The valley umklapp interactions are forbidden by momentum conservation. But other than this, all the left twelve interactions are allowed. The three interactions $g_{13},g_{24}, g_{44}$ are umklapp scatterings, which show up only in the special case when $\phi=\pi/3$, and are absent for a general $\phi$ in the previous studies of similar systems \cite{PhysRevB.100.085136,PhysRevB.102.085103}. As a result, the stable PDW phases are absent in these studies. The valley preserving interactions $g_{1i},g_{2i}$ can arise from the $\text{SU(4)}$ symmetric lattice interactions such as Hubbard and Heisenberg interactions. The valley flipping interactions $g_{4i}$ can result from the  $\text{SU(4)}$ broken Hund couplings on the lattice. Meanwhile, the Heisenberg interaction also gives anisotropic initial values of $g_{1i}$ and  $g_{2i}$. Here, given that these different interactions are generally present in the system, we discuss the phase diagram from general initial values of the $g_{ij}$, instead of the original form of the interactions defined in Eq. \eqref{lat_int}. The projections of different lattice interactions $U, J, V_h$ and $K$ in Eq. \eqref{lat_int} to $g_{ij}$ are recorded in Appendix \ref{sec:projection_of_the_lattice_interaction}.

The one loop parquet RG equations for all the twelve interactions can be obtained using the diagrammatic technique. As an example, we show the renormalization of $g_{13}$ in Fig.\ref{interaction_six}(c); the renormalization for other interactions can be obtained similarly. Defining $\dot{g}_{ij}=d g_{ij}/dy$ where $y=\Pi_{pp}(0,T)$ and using $d=\frac{d \Pi_{ph}(\bm{Q}_{\pm},T)}{y}\approx d(y_c)$ where $y_c$ is the critical value at which at least one of $g_{ij}$ diverges, we arrive at
\begin{equation}
    \begin{aligned}
    \dot{g}_{11}=&-g_{11}^2-2g_{13}^2, \quad \dot{g}_{12}=d(g_{12}^2+g_{13}^2+g_{43}^2+g_{44}^2) ,\\
    \dot{g}_{13}=&-2g_{13}g_{11}-g_{13}^2+4d(g_{12}g_{13}-g_{23}g_{24})\\
    &+2d(g_{23}g_{44}+g_{24}g_{43}+g_{43}g_{44}-g_{13}g_{14}),\\
     \dot{g}_{14}=&2d(g_{12}g_{14}+g_{24}g_{44}+g_{23}g_{43}-g_{24}^2-g_{14}^2-g_{23}^2),\\
     \dot{g}_{21}=&-g_{21}^2-2g_{23}^2-g_{41}^2-2g_{43}^2,\quad \dot{g}_{22}=d(g_{22}^2+g_{23}^2),\\
    \dot{g}_{23}=&-2g_{23}g_{21}-g_{23}^2\\
    &+2d(g_{23}g_{22}+g_{12}g_{23}-2g_{23}g_{14}-g_{13}g_{24})\\
    &-(2g_{41}g_{43}+g_{43}^2)+2d(g_{13}g_{44}+g_{14}g_{43}),\\
    \dot{g}_{24}=&2d(g_{12}g_{24}+g_{14}g_{44}+g_{13}g_{43}-g_{13}g_{23}-2g_{24}g_{14}),\\
    \dot{g}_{41}=&-2(g_{21}g_{41}+2g_{23}g_{43}),\\
    \dot{g}_{42}=&2d(g_{22}g_{42}+g_{23}g_{43}-g_{43}^2-g_{42}^2),\\
    \dot{g}_{43}=&-2(g_{21}g_{43}+g_{23}g_{41}+g_{23}g_{43})\\
     &+2d(g_{12}g_{43}+g_{13}g_{44}+g_{22}g_{43}+g_{23}g_{42}-2g_{42}g_{43}),\\
    \dot{g}_{44}=&2d(g_{12}g_{44}+g_{13}g_{43}).
    \label{rg_six}
    \end{aligned}
\end{equation} 
We are interested in the stable strong coupling fixed points starting from different interactions, which correspond to the symmetry breaking ground states. Physically, the stable fixed points, or fixed trajectories mean that the corresponding ordered phases need no fine tuning of the interactions and exist in a wide parameter space. The asymptotic behavior of the fixed trajectories in the one-loop RG equations is 
\begin{equation}
    g_{ij}\approx \frac{G_{ij}}{y_c-y}
\end{equation}
Therefore, if $G_{ij}$ is nonzero, $g_{ij}$ diverges when $y$ approaches $y_c$ from below, i.e., it either flows to strong repulsion or strong attraction. We also notice that, for different $g_{ij}$, the critical value $y_c$ might not be the same. In cases we have different $y_c$, apparently the smallest one corresponds to the onset of instabilities. In that case, for those $g_{ij}$ which diverge at a larger $y_c$, their effective contributions vanish at the smallest $y_c$ where they are still small and can be neglected.
To examine the spontaneous symmetry breaking orders corresponding to the stable fixed trajectories , one needs to look into the order parameter vertices and susceptibilities under RG flow, which we discuss below.

\begin{figure}
  \includegraphics[width=8.5cm]{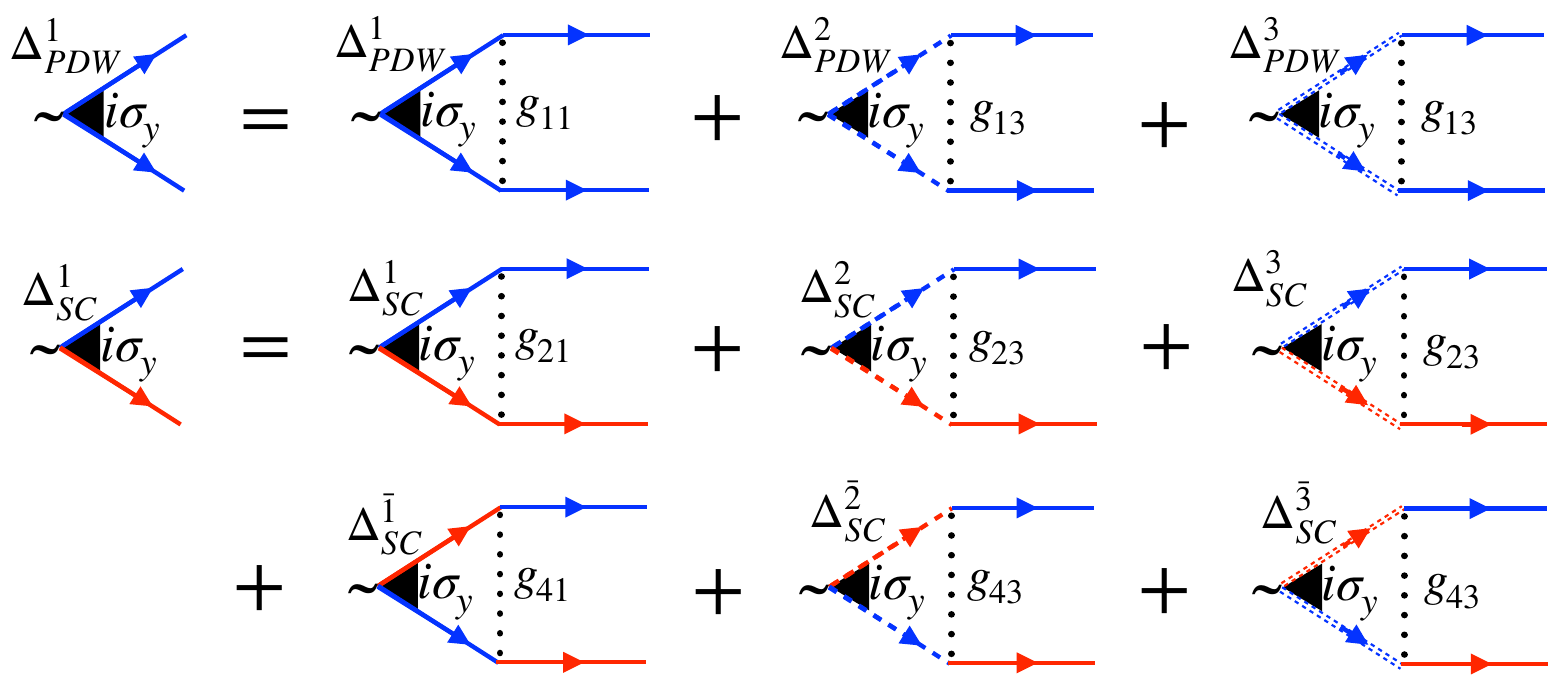}
 \caption{Diagrammatic representations of the one loop corrections to all the possible superconducting vertices of the six patch model. These include the PDW and uniform SC. The vertex of the uniform SC can be either spin singlet or triplet, and their diagrams are the same. As a result, the spin singlet and triplet pairings with $L_z=0$ are always degenerate.
 }
  \label{vertex_six}
\end{figure}

\subsection{RG enhanced susceptibilities and phases }
Having established the RG flow for the four fermion interactions $g_{ij}$, the competition among different orders can be readily identified. To this end, we need to write down the order parameters, and inspect how they flow when $g_{ij}$ changes with the energy scale. They can be well seen from calculating the triple vertex functions associated with each order parameter.

We start from the superconducting orders. The renormalizations of the corresponding vertcies are shown in Fig.\ref{vertex_six}. Note that the vertices of the PDW order have 3 components while the uniform superconducting order have $3\times2$ components in total. The three components correspond to three classes of paired patches, which are related with each other by $C_3$ rotation, and the additional double fold in the uniform SC comes from inversion. Corresponding to different superconducting form factors (such as s-wave, d-wave, etc.), the vertices can take different values on these different patches, and the leading instability is a combination from different patches \cite{Nandkishore2012}. 

In Fig.\ref{vertex_six}, the spin configuration is $i\sigma_y$ which corresponds to spin singlet pairing. This is because the PDW order can only be spin-singlet with total valley quantum number $|L_z|=1$ in our six patch model. This is due to that the vertex of spin-triplet pair with momentum $\bm{K}$ is actually zero: 
\begin{equation}
    \begin{aligned}
    &\Delta_{\text{PDW-triplet}}\sum_{\bm{k},a=1,2,3}c^{\dagger}_{a}(\bm{k})i\sigma_y\Vec{\sigma}[c^{\dagger}_{a}(-\bm{k})]^T+h.c.\\
    &=-\Delta_{\text{PDW-triplet}}\sum_{\bm{k},a=1,2,3}c^{\dagger}_{a}(-\bm{k})i\sigma_y\bm{\sigma}[c^{\dagger}_{a}(\bm{k})]^T+h.c.\\
    &=-\Delta_{\text{PDW-triplet}}\sum_{\bm{k},a=1,2,3}c^{\dagger}_{a}(\bm{k})i\sigma_y\bm{\sigma}[c^{\dagger}_{a}(-\bm{k})]^T+h.c.,
    \end{aligned}
\end{equation}
where $a$ is the patch label, and $\bm{k}$ is the momentum lies in a patch near each van Hove singularity. Meanwhile, the patch is inversion symmetric with the inversion center at the van Hove singularity. The two-component spinor operator $c^{\dagger}_a(\bm{k})$ is: $c^{\dagger}_a(\bm{k})=(c^{\dagger}_{a,\uparrow}(\bm{k}),c^{\dagger}_{a,\downarrow}(\bm{k}))$. Similarly, we can also prove that the spin triplet PDW vertex with nesting momentum $-\bm{K}$ is also zero. As a result, the spin triplet PDW order is expected to be energetically unfavored even if the whole Fermi surface is taken into consideration, as this order parameter cannot gap out the van Hove singularity. The vertices equations of uniform superconducting orders are more involved. The spin singlet and triplet uniform pairings are degenerate, and the SS (TT) pairing can mix with TS (ST) pairing in principle due to the broken $\text{SU(2)}_v$ valley symmetry.

We next identify the pairing symmetry of the leading instability. Explicitly, the RG equations corresponding to the diagrammatics in Fig.\ref{vertex_six} are:
\begin{equation}
    \begin{aligned}
            &
           \left(\begin{array}{c}
\dot{\Delta}_{\text{PDW}}^1\\ 
\dot{\Delta}_{\text{PDW}}^2\\
\dot{\Delta}_{\text{PDW}}^3
\end{array}\right)
=-\left(\begin{array}{ccc}
g_{11}&g_{13}&g_{13}\\ 
g_{13}&g_{11}&g_{13}\\
g_{13}&g_{13}&g_{11}
\end{array}\right)
\left(\begin{array}{c}
\Delta_{\text{PDW}}^1\\ 
\Delta_{\text{PDW}}^2\\
\Delta_{\text{PDW}}^3
\end{array}\right),\\
\end{aligned}\label{eq:PDW_order}
\end{equation}

\begin{equation}
    \begin{aligned}
           \left(\begin{array}{c}
\dot{\Delta}_{\text{SC}}^1\\ 
\dot{\Delta}_{\text{SC}}^2\\
\dot{\Delta}_{\text{SC}}^3\\
\dot{\Delta}_{\text{SC}}^{\bar{1}}\\ 
\dot{\Delta}_{\text{SC}}^{\bar{2}}\\
\dot{\Delta}_{\text{SC}}^{\bar{3}}
\end{array}\right)
=-\left(\begin{array}{cccccc}
g_{21}&g_{23}&g_{23}&g_{41}&g_{43}&g_{43}\\ 
g_{23}&g_{21}&g_{23}&g_{43}&g_{41}&g_{43}\\
g_{23}&g_{23}&g_{21}&g_{43}&g_{43}&g_{41}\\
g_{41}&g_{43}&g_{43}&g_{21}&g_{23}&g_{23}\\ 
g_{43}&g_{41}&g_{43}&g_{23}&g_{21}&g_{23}\\
g_{43}&g_{43}&g_{41}&g_{23}&g_{23}&g_{21}\\
\end{array}\right)
 \left(\begin{array}{c}
\Delta_{\text{SC}}^1\\ 
\Delta_{\text{SC}}^2\\
\Delta_{\text{SC}}^3\\
\Delta_{\text{SC}}^{\bar{1}}\\ 
\Delta_{\text{SC}}^{\bar{2}}\\
\Delta_{\text{SC}}^{\bar{3}}
\end{array}\right).\\
    \end{aligned}\label{eq:SC_order}
\end{equation}

We first discuss PDW order. If we diagonalize the three by three coefficient matrix in Eq.\eqref{eq:PDW_order}, we will arrive at three eigenfunctions corresponding to s-wave and two degnerate d-wave superconducting orders. Their eigen RG equations are: 
\begin{equation}
    \begin{aligned}
            &\Delta_{\text{PDW}}^s=(1,1,1):\quad  \dot{\Delta}_{\text{PDW}}^s=-(g_{11}+2g_{13})\Delta_{\text{PDW}}^s, \\
            &\Delta_{\text{PDW}}^{d_1}=(-1,0,1):\quad \dot{\Delta}_{\text{PDW}}^{d_1}=-(g_{11}-g_{13})\Delta_{\text{PDW}}^{d_1},\\
            &\Delta_{\text{PDW}}^{d_2}=(-1,1,0):\quad \dot{\Delta}_{\text{PDW}}^{d_2}=-(g_{11}-g_{13})\Delta_{\text{PDW}}^{d_2},\\
            \label{uni_d}
    \end{aligned}
\end{equation}
Note that the the two d-wave orders are degenerate, which is manifest in the coefficients $-(g_{11}-g_{13})$ in the RG equations of $\Delta_{\text{PDW}}^{d_{1,2}}$ in Eq. \eqref{uni_d}. The degeneracy is guaranteed by the underlying lattice symmetry $C_{3v}$, as the two $d$-wave orders belong to the same two-dimensional representation E. The ground state favours the chiral combination of these two components :$\Delta_{\text{PDW}}^{d_{1}}\pm i\Delta_{\text{PDW}}^{d_{2}}$, which can be verified through the Ginzburg-Landau free energy analysis similar to the three patch model in hexagonal systems\cite{Nandkishore2012}.

For the uniform SC order, we diagonalize the six by six coefficient matrix in Eq.\eqref{eq:SC_order}, and we arrive at different eigen pairing functions with different form factors and flavor quantum number:
\begin{equation}
    \begin{aligned}
   & \Delta_{\text{sc}}^s=(1,1,1,1,1,1):\\
    &\dot{\Delta}_{\text{sc}}^s=-(g_{21}+2g_{23}+g_{41}+2g_{43})\Delta_{\text{sc}}^s, \\
    &\\
   &\Delta_{\text{sc}}^f=(1,1,1,-1,-1,-1):\\
    &\dot{\Delta}_{\text{sc}}^f=-(g_{21}+2g_{23}-g_{41}-2g_{43})\Delta_{\text{sc}}^f, \\
        &\\
  &\Delta_{\text{sc}}^{d_1,d_2}=(-1,0,1,-1,0,1),(-1,1,0,-1,1,0):\\
    &\dot{\Delta}_{\text{sc}}^{d_1,d_2}=-(g_{21}-g_{23}+g_{41}-g_{43})\Delta_{\text{sc}}^{d_1,d_2}, \\
        &\\
    &\Delta_{\text{sc}}^{p_1,p_2}=(1,0,-1,-1,0,1),(1,-1,0,-1,1,0):\\
    &\dot{\Delta}_{\text{sc}}^{p_1,p_2}=-(g_{21}-g_{23}-g_{41}+g_{43})\Delta_{\text{sc}}^{p_1,p_2},
    \end{aligned}
\end{equation}
where $\Delta_{\text{sc}}^f$ is the vertex of f-wave uniform pairing. And  $\Delta_{\text{sc}}^{p_1,p_2}$ are the vertices of p-wave uniform pairing, which belong to the two dimensional E representation of $C_{3v}$. These two superconducting orders are TT or SS pairings with odd parity form factors. The remaining two vertices are TS or ST pairings with even parity form factors. $\Delta_{\text{sc}}^s$ is the vertex of s-wave uniform pairing. $\Delta_{\text{sc}}^{d_1,d_2}$ are the vertices of d-wave uniform pairing, which also belong to the E representation.

Now we move to the density wave vertices with momentum $\bm{Q}_{\pm}$. We will use $\text{CDW}^+/\text{SDW}^+$ to denote charge- and spin-density waves with momentum $\bm{Q}_+$, and use $\text{CDW}^-/\text{SDW}^-$ to denote charge- and spin-density waves with momentum $\bm{Q}_-$. The $\text{CDW}^+/\text{SDW}^+$ vertices contain both real and imaginary parts, each of which has two components, corresponding to two valleys with opposite fluxes. The one loop RG equations of the density wave orders can be obtained using similar diagrammatics shown in Fig.\ref{vertex_six}. The resulting equations are

\begin{equation}
    \begin{aligned}
            &
           \left(\begin{array}{c}
\dot{\Delta}_{\text{ReCDW}^+}^1\\ 
\dot{\Delta}_{\text{ReCDW}^+}^2\\
\end{array}\right)
=d\left(\begin{array}{cc}
a&b\\ 
b&a\\
\end{array}\right)
 \left(\begin{array}{c}
\Delta_{\text{ReCDW}^+}^1\\ 
\Delta_{\text{ReCDW}^+}^2\\
\end{array}\right),\\
&
          \left(\begin{array}{c}
\dot{\Delta}_{\text{ImCDW}^+}^1\\ 
\dot{\Delta}_{\text{ImCDW}^+}^2\\
\end{array}\right)
=d\left(\begin{array}{cc}
e&f\\ 
f&e\\
\end{array}\right)
 \left(\begin{array}{c}
\Delta_{\text{ImCDW}^+}^1\\ 
\Delta_{\text{ImCDW}^+}^2\\
\end{array}\right),
\end{aligned}\label{eq:cdw}
\end{equation}

\begin{equation}
    \begin{aligned}
 &\left(\begin{array}{c}
\dot{\Delta}_{\text{ReSDW}^+}^1\\ 
\dot{\Delta}_{\text{ReSDW}^+}^2\\
\end{array}\right)
=d\left(\begin{array}{cc}
g_{12}+g_{13}&g_{43}+g_{44}\\ 
g_{43}+g_{44}&g_{12}+g_{13}\\
\end{array}\right)
 \left(\begin{array}{c}
\Delta_{\text{ReSDW}^+}^1\\ 
\Delta_{\text{ReSDW}^+}^2\\
\end{array}\right),\\
&
           \left(\begin{array}{c}
\dot{\Delta}_{\text{ImSDW}^+}^1\\ 
\dot{\Delta}_{\text{ImSDW}^+}^2\\
\end{array}\right)
=d\left(\begin{array}{cc}
g_{12}-g_{13}&g_{43}-g_{44}\\ 
g_{43}-g_{44}&g_{12}-g_{13}\\
\end{array}\right)
 \left(\begin{array}{c}
\Delta_{\text{ImSDW}^+}^1\\ 
\Delta_{\text{ImSDW}^+}^2\\
\end{array}\right),
\end{aligned}
\end{equation}

\begin{equation}
    \begin{aligned}
 &
           \left(\begin{array}{c}
\dot{\Delta}_{\text{CDW}^-}^1\\ 
\dot{\Delta}_{\text{CDW}^-}^2\\
\end{array}\right)
=d\left(\begin{array}{cc}
g_{22}-2g_{42}&g_{23}-2g_{43}\\ 
g_{23}-2g_{43}&g_{22}-2g_{42}\\
\end{array}\right)
 \left(\begin{array}{c}
\Delta_{\text{CDW}^-}^1\\ 
\Delta_{\text{CDW}^-}^2\\
\end{array}\right),\\
&
           \left(\begin{array}{c}
\dot{\Delta}_{\text{SDW}^-}^1\\ 
\dot{\Delta}_{\text{SDW}^-}^2\\
\end{array}\right)
=d\left(\begin{array}{cc}
g_{22}&g_{23}\\ 
g_{23}&g_{22}\\
\end{array}\right)
 \left(\begin{array}{c}
{\Delta}_{\text{SDW}^-}^1\\ 
{\Delta}_{\text{SDW}^-}^2\\
\end{array}\right).
    \end{aligned}\label{eq:DW_order}
\end{equation}
In Eq.\eqref{eq:cdw} we have introduced the following quantities for brevity:
\begin{equation}
    \begin{aligned}
        &a=g_{12}-2g_{14}-g_{13},\quad b=-2g_{24}-2g_{23}+g_{43}+g_{44},\\
    &e=g_{12}-2g_{14}+g_{13},\quad f=2g_{24}-2g_{23}+g_{43}-g_{44}.
    \end{aligned}
    \end{equation}
Like in the superconducting case, here we also need to diagonalize all the two-by-two matrices in the vertex equations of density wave orders in particle-hole channel to find the leading instability configuration. Interestingly, despite of different order parameters, all these matrices contain only two distinct elements: the diagonal entry $h_{11}$ and the off-diagonal entry $h_{12}$, since the two diagonal (off-diagonal) entries have identical values. Matrix of this type has eigenvalues $h_{11}\pm h_{12}$, with the corresponding eigenfunctions being $\Delta_{\text{DW}}^{e,o}=\Delta_{\text{DW}}^1\pm\Delta_{\text{DW}}^2$ with e(o) standing for even (odd).

\begin{figure*}
  \includegraphics[width=18cm]{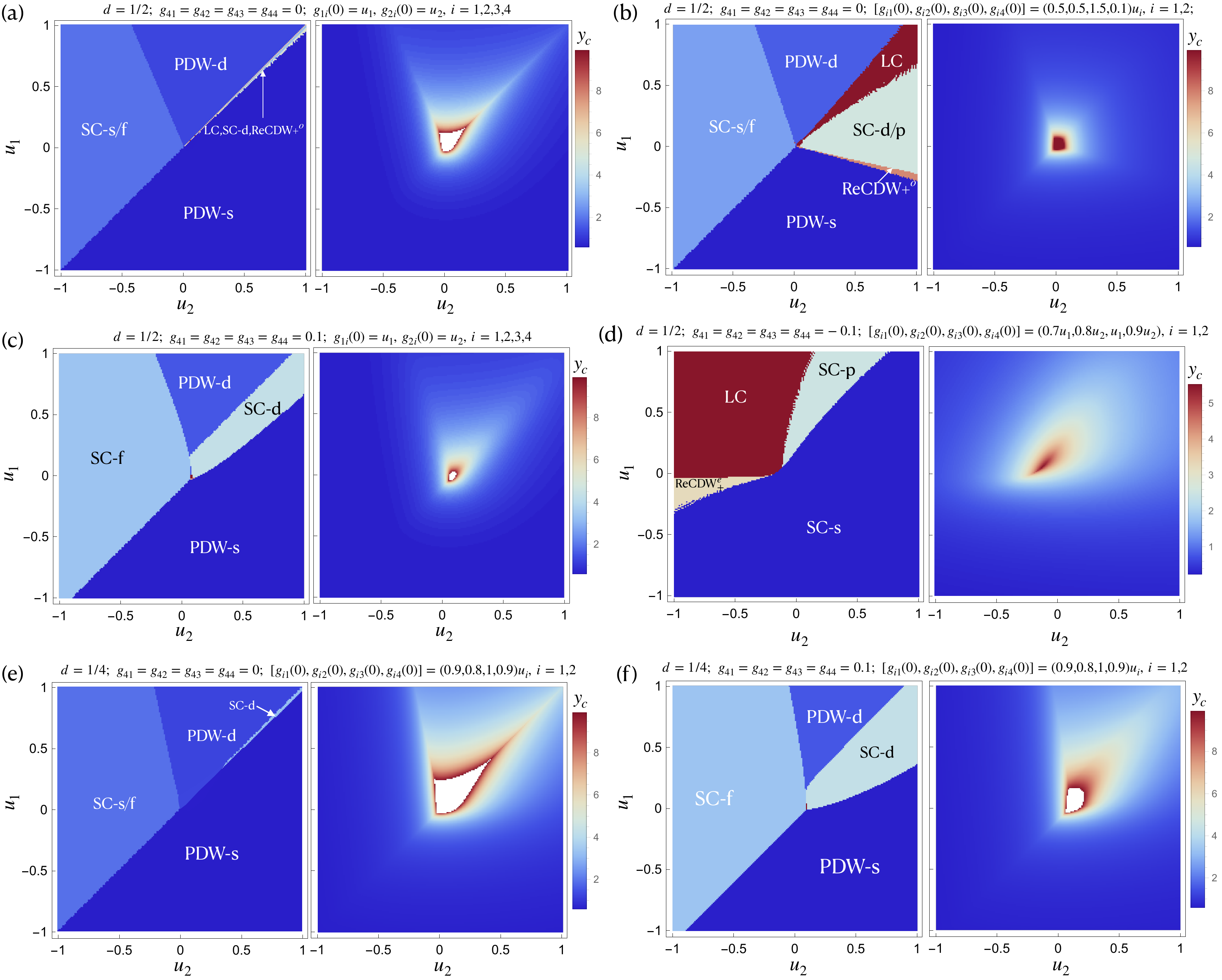}
  \caption{The phase diagram and divergent energy scale of the six patch model at $\phi=\frac{\pi}{3}$ Each sub-figure consists of  the phase diagram (the left panel) and the corresponding $y_c\sim \ln^2 (W/T_c)$ (the right panel) obtained from the nesting parameter d and initial values $g_{ij}$ labeled on the top. The color map only shows $y_c$ up to $y_c=10$, for $y_c>10$ the areas are denoted by white color. (a)The phase diagram from the initial values: $g_{1j}=u_1,g_{2j}=u_2,g_{4j}=0,j=1,2,3,4$ in the perfect nesting limit $d=\frac{1}{2}$. The chiral $d$-wave PDW, $s$-wave PDW and degenerate $s$- and $f$-wave uniform pairing donimates nearly all regimes in the phase diagram. The loop current (LC) order occurs only near the $\text{SU(4)}$ symmetric line with repulsive interactions: $g_{1i}=g_{2i}>0$, which is accompanied by chiral $d$-wave SC as well as parity-odd real CDW with momentum $\bm{Q}_+$. (b)The phase diagram from the anisotropic initial values of  $g_{1j}$ and $g_{2j}$ and zero $g_{4i}$ in the perfect nesting limit $d=\frac{1}{2}$. The loop current order is enhanced by the anisotropic initial values and occupies a finite area in this regime. (c)The phase diagram with repulsive $g_{4i}=0.1$. The loop current order is fragile to the repulsive valley flipping $g_{4i}$, but the other superconducting orders are robust to $g_{4i}$.  (d) The phase diagram with attractive $g_{4i}=-0.1$ and anisotropic initial values of $g_{1j}$ and $g_{2j}$. The chiral d-wave PDW is replaced with the loop current order. (e) The phase diagram with the nesting parameter $d=\frac{1}{4}$ and zero $g_{4i}$, which is away from the perfect nesting limit. (f)The phase diagram with $d=\frac{1}{4}$ and repulsive $g_{4i}$. The chiral d-wave PDW is robust to the valley flipping $g_{4i}$.}
  \label{alpha}
\end{figure*}

\begin{figure}
  \includegraphics[width=7.cm]{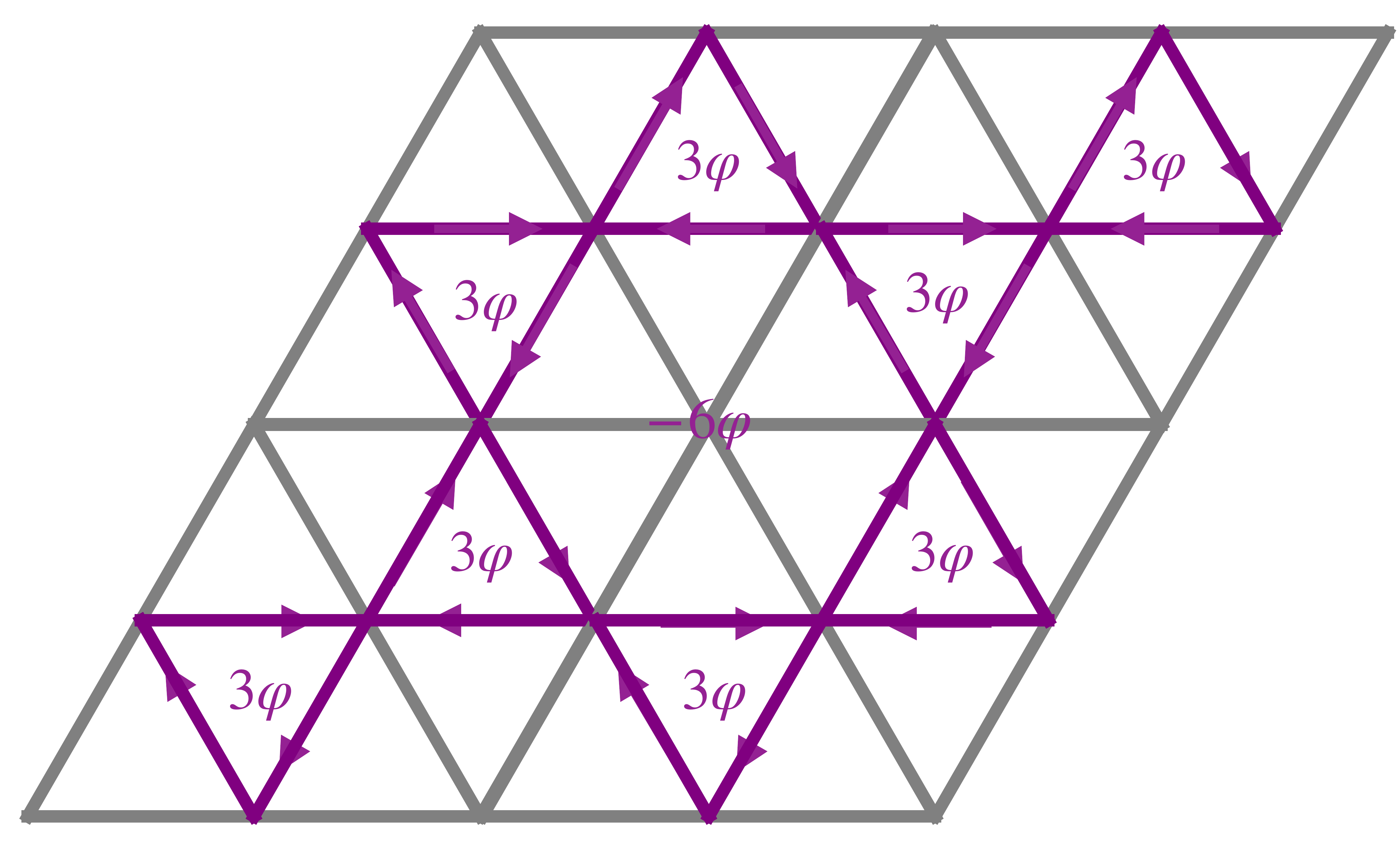}
  \caption{Kagome flux pattern of the loop current order at $\phi=\frac{\pi}{3}$. The flux of each triangle plaquette is $3\varphi$, and the flux of each hexagon plaquette is $-6\varphi$. And the pure imaginary CDW order here gives $\varphi=\frac{\pi}{2}$.}
  \label{lc}
\end{figure}

Finally, we are in the position to determine the leading order by calculating the renormalized susceptibilities of both the superconducting and the density wave order, which are governed by the following equations\cite{PhysRevB.100.085136,PhysRevX.6.041045}:
\begin{equation}
    \dot{\chi}_{SC}=|\Delta_{SC}|^2,~~ \dot{\chi}_{\text{DW}}=d|\Delta_{\text{DW}}|^2.
    \label{susce}
\end{equation}
Note here $\Delta_{SC}$ includes both PDW and uniform SC orders.
From the equations of vertices in the particle-particle and particle-hole channel, we can obtain the asymptotic solutions $\Delta_{SC/DW}(y)\approx (y_c-y)^{\beta_{SC/DW}}$, which is valid near $y_c$. Substituting this form into Eq.\eqref{susce} we obtain the asymptotic solution for $\chi$, which behaves like $\chi(y)\approx (y_c-y)^{\alpha}$ \cite{PhysRevX.6.041045,Cvetkovic,PhysRevB.100.085136,Binz2002} and the exponent can be obtained via
\begin{equation}
    \alpha_{SC,\text{DW}}=2\beta_{SC,\text{DW}}+1.\label{eq:alpha}
\end{equation}
If for some channel $\alpha<0$, the corresponding susceptibility diverges and signals an instability. 
The susceptibility with the most negative $\alpha$ is therefore the leading instability when reducing energy scale. It is obvious that $\alpha$ in the density wave channel depends on the nesting parameter $d$. If $d$ is small, $\beta_{DW}$ is suppressed and hence $\alpha_{DW}$ becomes positive and there is no onset of density wave order. This is expected since a smaller $d$ indicates that the FS nesting in the particle-hole channel is less important. The superconducting orders, however, are almost immune to this destruction.

In Fig. \ref{alpha} we present the phase diagram obtained for various initial interactions as well as nesting parameters. For each group of parameters, we show the phase boundary on the left panel, and the color map of $y_c$ on the right panel. 
We first focus on the the perfect nesting limit $d=\frac{1}{2}$, where not only the superconducting order, but also density wave orders are found. When the inter-patch interactions $g_{4j}$ are neglected, we find, in addition to PDW and SC orders, an imaginary CDW order with momentum $\bm{Q}_+$ near the $\text{SU(4)}$ symmetric line: $g_{1i}=g_{2i}>0$ [see Fig.\ref{alpha}(a) and (b)]. Since there are in fact 3 different $\bm{Q}_+$ related by $C_3$ rotation, this imaginary CDW is in the $3\bm{Q}_+$ state which gives rise to loop current order\cite{PhysRevB.93.115107}, and, similar to Haldane's model, can host quantum anomalous Hall effect. A real space configuration of this loop current is shown in Fig.\ref{lc}. The bond currents form a Kagome lattice pattern. This result is consistent with similar RG results at $\phi=0$\cite{PhysRevB.100.085136,PhysRevB.99.195120}. The loop current order can be enhanced by the anisotropic $g_{1i},g_{2i}$ and attractive valley flipping $g_{4i}$ [Fig.\ref{alpha} (b) and (d)]. The anisotropy of $g_{1i}$ and $g_{2i}$ can arise from the nearest-neighbour interactions on the lattice.

Apart from the loop current order, the phase diagram for the six patch case is almost dominated by PDW and uniform SC orders. Among those, the most interesting order is the $d$-wave PDW. In hexagonal lattices, the $d$-wave order parameters belong to the two-dimensional irreducible representation of the lattice group. Therefore, there are two degenerate $d$-wave state, and the true ground state must be obtained by comparing their Landau free energy. In most cases, the $d$-wave orders spontaneously break the time reversal symmetry to lower energy, leading to a chiral SC state. In our case, both the $d$-wave PDW and $d$-wave SC are chiral, with the order parameter being $\Delta_1+i\Delta_2$ or $\Delta_1-i\Delta_2$. It is worth to notice that, although the PDW we discussed here has both momentum $\bm{K}$ and $-\bm{K}$, the Larkin-Ovchinnikov(LO) state, in which the magnitude of the gap function oscillates in space and therefore has nodal lines, is not energetically favoured. This is because the FS of each valley at $\phi=\frac{\pi}{3}$ has only one nesting vector in the particle-particle channel. Thus, our PDW discussed here is similar to the original Fulde-Ferrell state. The degeneracy between opposite valleys can be lifted by a valley splitting field. The resulting valley polarized PDW has many interesting effects such as superconducting diode effect \cite{Noah2022,PhysRevLett.128.037001,Ando2020} and nonreciprocal Josephson effect \cite{Margarita}.

We have also confirmed that both the PDW and the uniform SC found here are indeed stable fixed points. To see this, one can include small perturbations around the fixed point and test whether the system flows away from this point. Formally this can be seen by examining the eigenvalues of the stability matrix discussed in Appendix \ref{sec:the_stability_of_the_fix_point}. Using this approach, we find that both the PDW and the uniform SC are stable against all kinds of symmetry allowed interactions. The fixed point corresponding to the loop-current order has two directions of relevant perturbations, which drive the RG flow to fixed points favouring nearby superconducting phases. But the loop current susceptibility is still the leading one until the interactions flow beyond the perturbative regime: $\text{max} |g_{ij}(y)| \gg 1$, in which case we have to stop the RG flow far before that energy scale. This means that the loop current order is still the stable phase in the regime of our phase diagram except in the weak coupling limit, where we can push the RG flow to the energy scale $y_c$ at which the interactions really diverge \cite{PhysRevB.102.115136}. 

To summarize, the orders that we find here, such as the loop-current, the PDW and the uniform SC, are all consistent with the analysis performed in a $\phi=0$ $\text{SU(4)}$ model, in the sense that they can be connected by the local transformation discussed in Sec.\ref{sec:flux} when a lattice model like Eq. \eqref{lattice} is available. However, our pRG results are quite general, and are applicable to the cases when there lacks the information of a lattice Hamiltonian. The key ingredients here are the presence of six CVHS, and the inclusion of the umklapp interactions $g_{13}, g_{24}$ and $g_{44}$.

\section{Higer order van Hove singularities: Two patch model } 
\label{sec:higer_order_van_hove_singularities_and_bare_susceptibilities}
\begin{figure}
  \includegraphics[width=8.5cm]{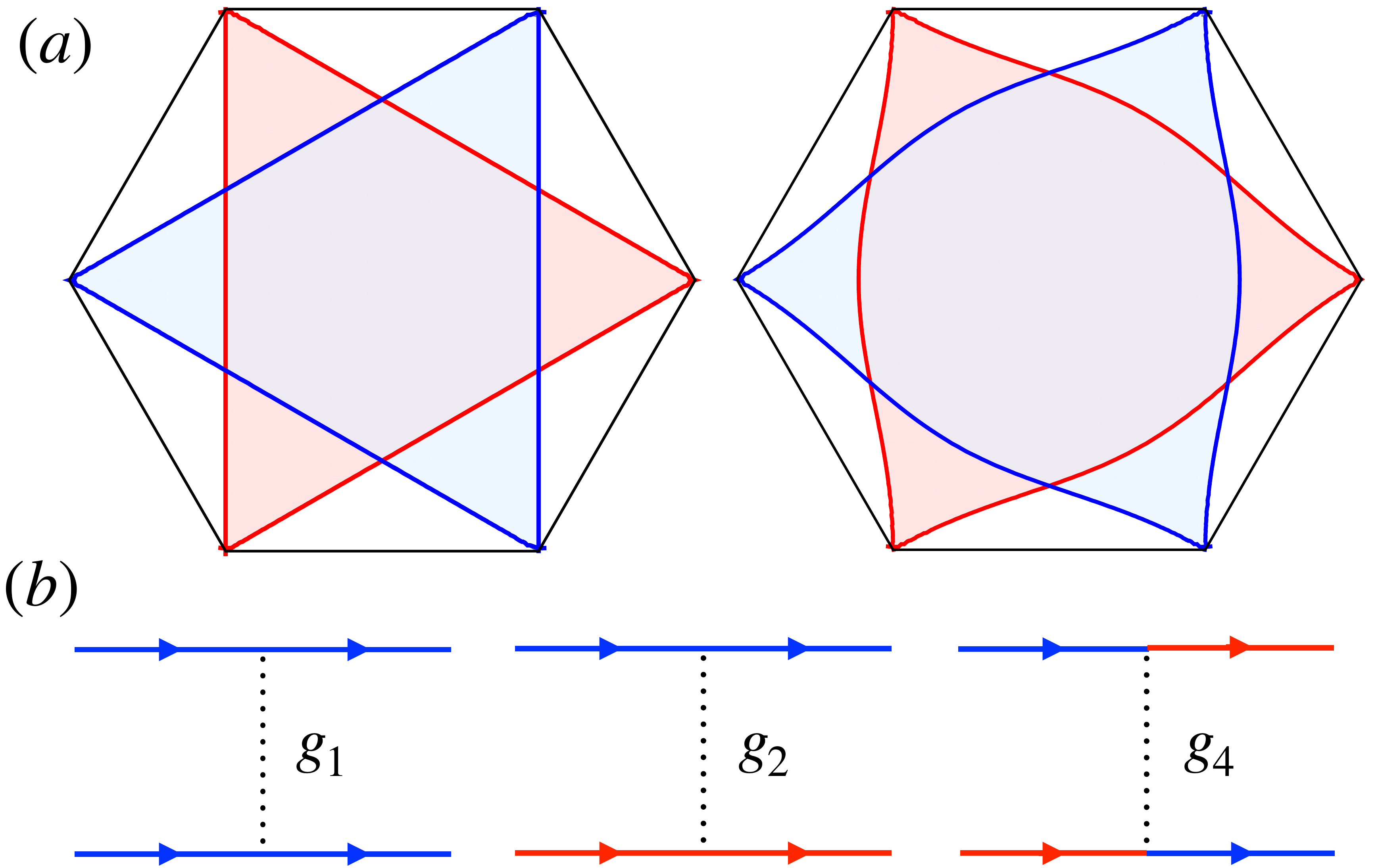}
  \caption{ (a)\quad  Fermi surface at higher-order van Hove doping with $t^{\prime}=0$ (left) and $t'=-0.1t$ (right). (b) The symmetry allowed four fermion interaction in the $\text{SU}(N_f) (N_f\geq 2)$ two patch model. Since the two patches are centered at $\pm \bm{K}$, there is no umklapp interaction.
  }\label{interaction}
\end{figure}

The band structure from Eq. \eqref{lattice} also hosts
 two higher-order van Hove singularities located at $\pm\bm{K}=\pm(\frac{4\pi}{3},0)$, if the condition  $\sin(\frac{\pi}{6}-\phi)=3t^{\prime}/t$ is satisfied. For $t'=0$, this happens when $\phi=\pi/6$. For a small but nonzero $t'$, this happens when $\phi$ is slightly below (above) $\pi/6$ for $t'>0$ ($<0$). The Fermi surface at this higher-order van Hove doping is illustrated in Fig.\ref{interaction}, which shows that a finite $t'$ breaks the perfect nesting. This can also be seen from the energy dispersion near these two HOVHS:
\begin{equation}
  \begin{aligned}
   \epsilon_1(\bm{k})=&\kappa_1(k_x^3-3k_xk_y^2)-\kappa_2(k_x^2+k_y^2)^2+\mu\\
    =&\kappa_1k^3\cos3\theta-\kappa_2k^4+\mu,\\
    \epsilon_2(\bm{k})=&-\kappa_1(k_x^3-3k_xk_y^2)-\kappa_2(k_x^2+k_y^2)^2+\mu\\
    =&-\kappa_1k^3\cos3\theta-\kappa_2k^4+\mu.\\
    \label{low}
  \end{aligned}
\end{equation}
where $\kappa_1=\sqrt{t^2-9t'^2}/4$, $\kappa_2=9t'/16$ and we have introduced $k$ and $\theta$ such that $k_x=k\cos\theta$ and $k_y=k\sin\theta$. Note that if $t'=0$ and $\mu=0$, only the cubic terms in these dispersions are present, and this corresponds to the perfect nesting case where $\epsilon_1(\bm{k})=-\epsilon_2(\bm{k})$. A finite $t'$ is associated with the $k^4$ term and hence spoils the FS nesting. A nonzero $\mu$ also has the effect of nesting breaking. As we shall see below, the effect of $t'$ is to diminish the divergence of the bare susceptibilities in different channels. $t^{\prime}$ and $\mu$ serve as tuning parameters in our model. 

Given the dispersions in Eq. \eqref{low}, the density of states near these two HOVHS can be obtained via $\nu_{1,2}(E)=\sum_{\bm{k}}\delta(E-\epsilon_{1,2}(\bm{k}))$. Since $\epsilon_1(\bm{k})=\epsilon_2(-\bm{k})$, $\nu_1(E)$ and $\nu_2(E)$ are in fact identical therefore we omit the subscript. The explicit expression of $\nu(E)$ is:
\begin{equation}
 \begin{aligned}
      \nu(E)&=\int\frac{k d k d\theta}{(2\pi)^2}\delta\left(E-\kappa_1 k^3\cos3\theta +\kappa_2k^4-\mu\right)\\
      &=\frac{2}{(2\pi)^2}\int_0^\infty \frac{d k}{|E'|}\text{Re}\left[\frac{k}{\sqrt{\kappa_1^2k^6/E'^2-(1+\kappa_2k^4/E')^2}}\right]
 \end{aligned}
\end{equation}
where $E'=E-\mu$. For the case when $t'=0$ and hence $\kappa_2=0$, the above integral can be evaluated and the result is 
\begin{equation}
    \begin{aligned}
        \nu(E)|_{t'=0}&=\frac{1}{|E-\mu|^{1/3}}\frac{2\kappa_1^{-2/3}}{(2\pi)^2}\int_1^\infty d x\frac{x}{\sqrt{x^6-1}}\\
            &=\frac{\nu_0}{|E-\mu|^{1/3}}\label{eq:DOS}
    \end{aligned}
\end{equation}
where $\nu_0=\kappa_1^{-2/3}\Gamma(7/6)/(2\Gamma(2/3)\pi^{3/2})\approx0.155 t^{-2/3}$, which is identical to that given in \cite{PhysRevB.104.195134}. 
With a nonzero $t'$, we have 
\begin{equation}
    \begin{aligned}
        \nu(E)=&\frac{\kappa_1^{-2/3}}{|E'|^{1/3}}\frac{2}{(2\pi)^2}\times\\
        &\int_0^\infty d x \text{Re}\frac{x}{\sqrt{x^6-(1-x^4|E'|^{1/3}\kappa_2/\kappa_1^{4/3})^2}}
    \end{aligned}
\end{equation}
Since we are interested in low-energy fermions in the vicinity near $\pm K$, we can make $E'$ small, and the leading divergent term in the above equation is the same with Eq. \eqref{eq:DOS}. Therefore, we anticipate that even for the non-perfect nesting case, $\nu(E)$ also have a power-law divergence. 

The divergence of the DOS near $\pm\bm{K}$ legitimizes our two patch approximation, in which we  consider fermions only near these two points, and apply pRG to investigate the competing electronic orders. In the following, we first discuss the building blocks (i.e. the bare susceptibilities) for our pRG analysis, and then we analyze the RG equations and identify the leading instability in various cases.

\subsection{Bare susceptibilities}

In the two patch model, Fermi surface nesting occurs with a nesting vector $\bm{Q}=2\bm{K}$ if $t'$ and $\mu$ are negligible. The nesting would result in a $\ln$ divergence for the particle-hole susceptibility which competes with superconductivity. However, because of the power-law divergence of the DOS in the presence of HOVHS, the $\ln$ divergence is less important here: the particle-hole instabilities compete with superconductivity in any case, regardless of the nesting effect. 

More interestingly, the HOVHSs located at $\pm\bm{K}$ are not time-reversal symmetric points and are dubbed as type-II HOVHS, in contrast to the type-I HOVHS where the dispersion has a form such as  $ak_y^2-bk_x^4$ \cite{PhysRevB.92.035132,PhysRevB.102.125141}. One remarkable feature of the type-II HOVHS is that, besides the divergent susceptibilities in particle-particle channel at zero momentum and particle-hole channel at $\bm{Q}$, those in particle-particle channel at $\bm{Q}$ and in particle-hole channel at zero momentum also diverge in similar manner. In other words, the four channels are comparable in low energy limit and one has to treat all of them on equal footing. This leads to a competition among the superconductivity, the finite momentum pairing, the density wave and the Pomeranchuck instability.

Consequently, we need the following four bare susceptibilities as our RG building blocks:
\begin{equation}
\begin{aligned}
  \Pi_{pp}(0)&=\int \frac{d^2\bm{k}}{(2\pi)^2}\frac{1-n_F[\epsilon_1(\bm{k})]-n_F[\epsilon_2(-\bm{k})]}{\epsilon_{1}(\bm{k})+\epsilon_{2}(-\bm{k})}\\
  \Pi_{pp}(\bm{Q})&=\int \frac{d^2\bm{k}}{(2\pi)^2}\frac{1-n_F[\epsilon_1(\bm{k})]-n_F[\epsilon_1(-\bm{k})]}{\epsilon_{1}(\bm{k})+\epsilon_{1}(-\bm{k})}\\
  \Pi_{ph}(\bm{Q})&=-\int \frac{d^2\bm{k}}{(2\pi)^2}\frac{n_F[\epsilon_1(\bm{k})]-n_F[\epsilon_2(\bm{k})]}{\epsilon_{1}(\bm{k})-\epsilon_{2}(\bm{k})}\\
  \Pi_{ph}(0)&=-\int \frac{d^2\bm{k}}{(2\pi)^2}\frac{\partial n_F(\epsilon)}{\partial \epsilon}=\int \frac{d^2\bm{k}}{(2\pi)^2}\frac{\beta}{4\cosh^2[\beta\epsilon_1(\bm{k})/2]}\\  
  \end{aligned}\label{eq:bubbles}
\end{equation}
where $\epsilon_1(\bm{k})$ and $\epsilon_2(\bm{k})$ are given in Eq. \eqref{low}. In the special case when $t'=0$ and $\mu=0$, one can make use of Eq. \eqref{eq:DOS} and $\int d^2\bm{k}/(2\pi)^2=\int d E\nu(E)$ to obtain the low energy behavior of Eq. \eqref{eq:bubbles}. After evaluating the factors numerically we have
\begin{equation}
   \begin{aligned}
        \Pi_{pp}(0)=\Pi_{ph}(\bm{Q})&\approx0.527\frac{t^{-2/3}}{T^{1/3}}\\
        \Pi_{pp}(\bm{Q})=\Pi_{ph}(0)&\approx0.177\frac{t^{-2/3}}{T^{1/3}}\\
   \end{aligned}\label{eq:tp0}
\end{equation}
If $\mu\neq0$, the above scaling behaviors still hold (but with different numerical factors) when $T$ is much larger than $\mu$. However, if $\mu$ becomes the largest, these bare susceptibilities no longer have a power-law divergence with $T$. Instead, it is easy to see in this case,
\begin{equation}
    \begin{aligned}
        &\Pi_{pp}(0)\sim \frac{1}{|\mu|^{1/3}}\log\frac{\Lambda}{T},\\
        &\Pi_{pp}(\bm{Q}),\Pi_{ph}(0),\Pi_{ph}(\bm{Q})\sim\frac{1}{|\mu|^{1/3}}.
    \end{aligned}
\end{equation}
Thus, in low temperature limit with a finite $\mu$, only the uniform SC channel has the potential instability. Below we will disregard this case, by assuming we are in the limit $T\gg|\mu|$ such that the scaling behaviors in Eq. \eqref{eq:tp0} persist to the lowest $T$ of our interest.

\begin{figure}
    \includegraphics[width=8.cm]{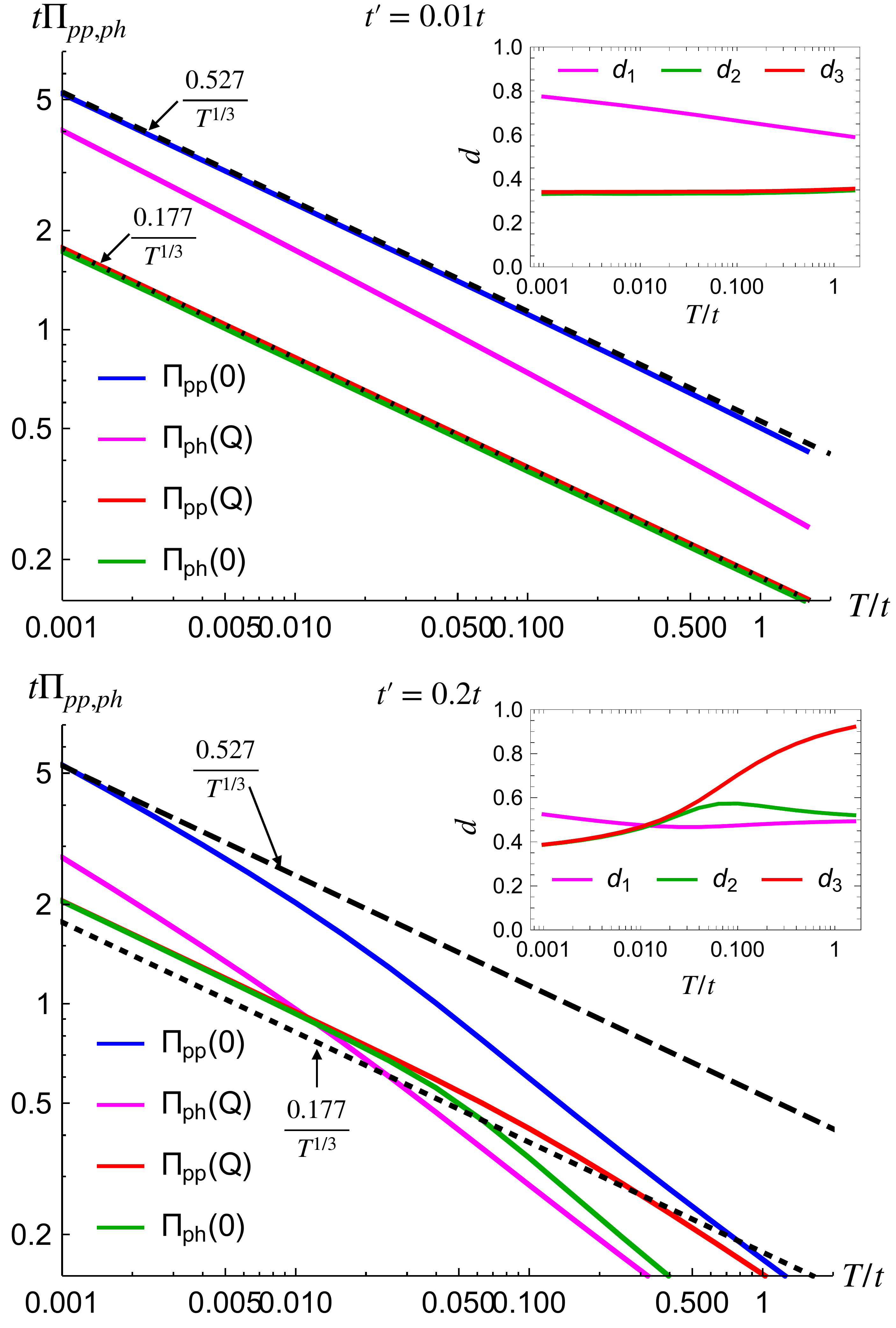}
    \caption{Various susceptibilities as a function of temperature $T$, obtained using $t'=0.01t$ (upper panel) and $t'=0.1t$ (lower panel) and in the limit $T\gg|\mu|$. The insets show temperature dependence of the nesting parameters $d_1$, $d_2$ and $d_3$. For comparison we also show $\Pi_{pp}(0)$ and $\Pi_{pp}(\bm{Q})$ at $t'=\mu=0$ as the dashed and dotted lines. }\label{fig:bubbles}
\end{figure}

For $t'\neq0$, it's rather difficult to obtain a relation as simple as Eq. \eqref{eq:tp0}, but a direct numerical calculation from Eq. \eqref{eq:bubbles} is feasible. In Fig.\ref{fig:bubbles} we plot the numerical results of the four bare susceptibilities as a function of $T$ with $t'=0.01t$ and $t'= 0.2 t$ in the upper and lower panels, respectively. For comparison, the result in Eq. \eqref{eq:tp0} at $t'=0$ is plotted as the dashed and dotted lines. In both cases, all the four bare susceptibilities scale as $1/T^{1/3}$ when $T$ becomes small enough. We clearly see that the finite $t'$ has little effect on the small $T$ behavior of $\Pi_{pp}(0)$, while it reduces the prefactor in $\Pi_{ph}(\bm{Q})$ significantly and enhances  $\Pi_{pp}(\bm{Q})$ and $\Pi_{ph}(0)$ slightly. As a result, $\Pi_{ph}(\bm{Q})$, once identical to $\Pi_{pp}(0)$ when $t'=0$ [see Eq.\eqref{eq:tp0}], now becomes smaller. In the insets of Fig. \ref{fig:bubbles}, we show the temperature dependence of the nesting parameteres, defined as $d_1=\Pi_{ph}(\bm{Q})/\Pi_{pp}(0)$, $d_2=\Pi_{ph}(0)/\Pi_{pp}(0)$ and $d_3=\Pi_{pp}(\bm{Q})/\Pi_{pp}(0)$. In the ideal case when $t'=0$, we have $d_1=1$, $d_2=d_3=0.336\approx1/3$. With a nonzero $t'$, all these parameters becomes $T$-dependent, but have weak $T$-dependence in $T\to0$ limit. Moreover, we now have $d_1$ significantly reduced, while $d_2\approx d_3$ almost intact. These results legitimize our following RG analysis, in which we take all the three nesting parameters as constant in low $T$ limit.

\subsection{Renormalization group analysis} 
\label{sec:renormalization_group_analysis}
As in the six patch model, here we also consider the system with spin $\text{SU(2)}$ symmetry. Unlike the six patch model, the symmetry allowed interactions are much fewer. In Fig.\ref{interaction}(b) show all the three interactions. Note there is no Umklapp interaction.

The one-loop RG equations for these interactions can be obtained in a similar way as in the six patch model, but here the running parameter $y=\Pi_{pp}(0)$ scales as $1/T^{1/3}$ instead of $\ln^2T$. The results are:
\begin{equation}
  \begin{aligned}
    \dot{g_1}=&\left[\left(3-N_f\right)d_2-d_3\right]g_1^2-N_fd_2g_2^2+d_2g_4^2+2d_2g_2g_4,\\
    \dot{g_2}=&(d_1-1)g_2^2-g_4^2+2d_2g_1g_4+(2-2N_f)d_2g_1g_2,\\
    \dot{g_4}=&2(d_1-1)g_2g_4+2d_2g_1g_4-N_fd_1g_4^2,
    \label{rg}
  \end{aligned}
\end{equation}
where $N_f$ is the number of fermion flavors for each valley component. We take $N_f=2$ in our following discussion.  The nesting parameters  $d_i$ at low energy limit are approximated by constant values and defined via $d_1=\Pi_{ph}(\bm{Q})/y$, $d_2=\Pi_{ph}(0)/y$ and $d_3=\Pi_{pp}(\bm{Q})/y$. This is justified by the numerical results plotted in Fig.\ref{fig:bubbles}. Similar to the six patch model analysis, the interactions $g_i$ can also flow to some strong coupling fixed point at some critical value $y_c$. We thus can assume the scaling form $g_i=G_i/(y_c-y)$ near $y_c$ and then confirm it. 

\begin{figure}
  \includegraphics[width=8.5cm]{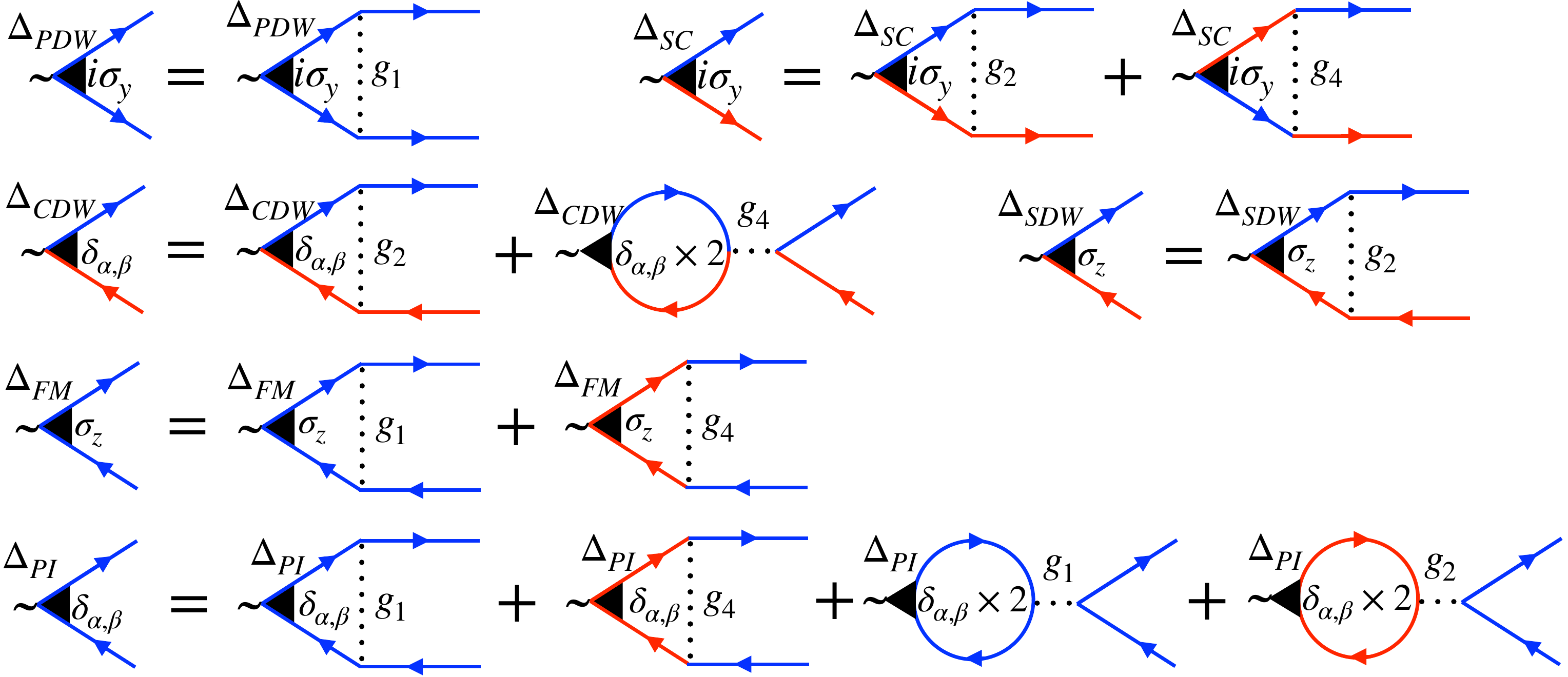}
  \caption{Diagrammatic representation of the one-loop RG equations for various order parameters in the two patch case.}\label{2porder}
\end{figure}
Because of the fewer interactions in the two patch case, competing orders are also fewer. We consider PDW, uniform SC, CDW, SDW, ferromagnetism (FM) and charge Pomeranchuck instability (PI). We list all the possible order parameters and their RG equations diagrammatically in Fig.\ref{2porder}. The corresponding RG equations for these orders are explicitly given by 
\begin{equation}
    \begin{aligned}
        &\dot{\Delta}_{\text{PDW}}=-d_3g_1 \Delta_{\text{PDW}}\\
        &\dot{\Delta}_{\text{SC}}^s=-(g_2+g_4) \Delta_{\text{SC}}^s,~~\dot{\Delta}_{\text{SC}}^p=-(g_2-g_4) \Delta_{\text{SC}}^p \\
        &\dot{\Delta}_{\text{CDW}}=d_1(g_2-2 g_4) \Delta_{\text{CDW}}\\
         &\dot{\Delta}_{\text{SDW}}=d_1g_2 \Delta_{\text{SDW}}\\
          &\dot{\Delta}_{\text{FM1}}=d_2(g_1+g_4) \Delta_{\text{FM}},\\
          &\dot{\Delta}_{\text{FM2}}=d_2(g_1-g_4) \Delta_{\text{FM}},\\
          &\dot{\Delta}_{\text{PI1}}=d_2(-g_1-2g_2+g_4) \Delta_{\text{PI1}}\\
          &\dot{\Delta}_{\text{PI2}}=d_2(-g_1+2g_2-g_4) \Delta_{\text{PI2}}\\
    \end{aligned}\label{eq:2pDelta}
\end{equation}

Here the $s$-wave SC order parameter has the same sign at both patches, while the $p$-wave order parameter changes sign between patches. Similarly, both $\Delta_{FM1}$ and $\Delta_{PI1}$ preserve sign when changing patches, while $\Delta_{FM2}$ and $\Delta_{PI2}$ do not.

 The possible order is associated with a divergent susceptibility, for which the behavior close to $y_c$ can also be expressed as $\chi\sim(y_c-y)^\alpha$ with $\alpha<0$. Like in the six patch model, we can determine $\alpha$ from the RG equations for $\chi$, and the resulting $\alpha$ is given in terms of $G_i$, i.e. the same as Eq. \eqref{eq:alpha}.

We first look into the case of perfect nesting, where $d_1=1$, $d_2=d_3=1/3$. Under this condition, the RG equation has a fixed point at $G_1=-G_2<0$ and $G_4=0$. This fixed point indicates a degenerate gound state among $\Delta_{\text{CDW}}$, $\Delta_{\text{SDW}}$ and $\Delta_{\text{PI2}}$. This can be directly seen from Eq. \eqref{eq:2pDelta}: the flow equation for these three orders are the same if $g_4$ vanishes. The PDW order parameter is the subleading one, due to the fact that $d_3<d_1$ and $g_1\to-\infty$. However it cannot be ordered upon deceasing $T$ because the corresponding $\alpha_{PDW}>0$. 

\begin{figure*}
    \centering
    \includegraphics[width=16cm]{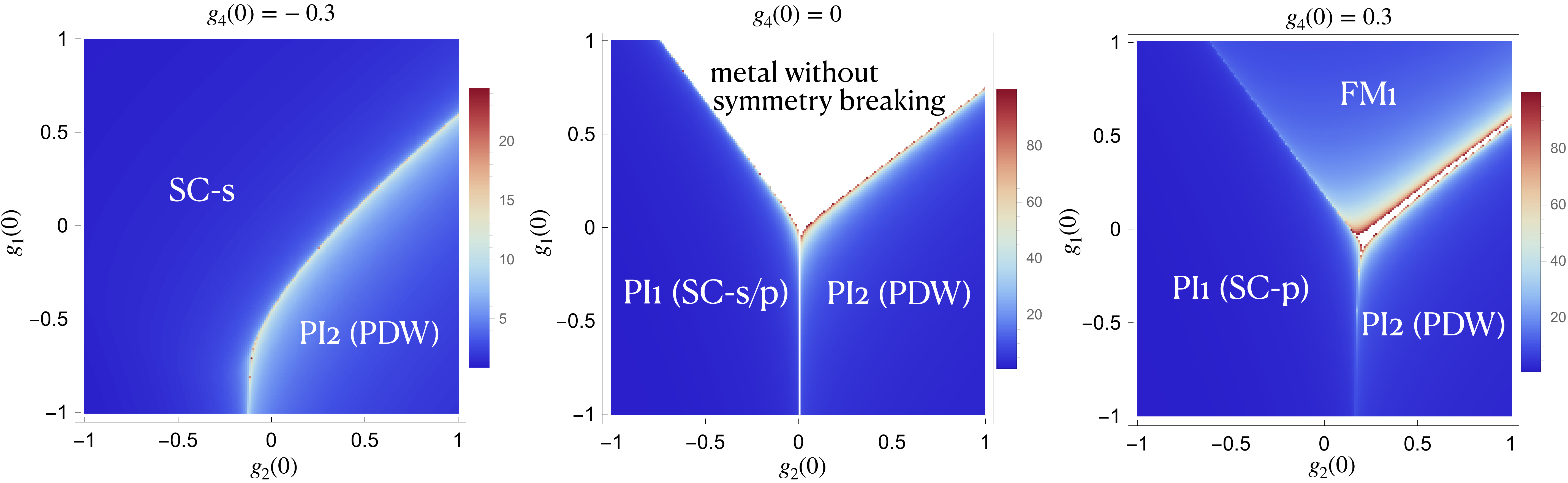}
    \caption{Phase diagram in the two patch HOVHS case for different initial values of $g_1$, $g_2$ and $g_4$. Again all the interactions are measured in units of $t$. The RG equation is solved with $t'=0.2t$, $\mu=0$. The leading orders are marked, while those inside the parenthesis are subleading.}
    \label{fig:2p_phase}
\end{figure*}

There are other fixed points of the RG equations with a nonzero $t'$, which can be reached by tuning the initial values of the interactions. As a result, a finite $t^{\prime}$ gives a richer phase diagram. In Fig.\ref{fig:2p_phase} we present the phase diagram for the two patch model with $t'=0.2t$, obtained from various initial interactions. These phase diagrams are shown in the plane of $g_1(0)$-$g_2(0)$, and we choose three different $g_4(0)$ from the left panel to the right panel of Fig. \ref{fig:2p_phase}, which correspond to $g_4(0)=-0.3,0,0.3$ respectively. Firstly, if $g_4(0)=0$, $g_4$ stays zero under the RG flow. In this case, depending on whether $g_2(0)$ is repulsive or attractive, the leading instabilities are $\Delta_{\text{PI2}}$ and $\Delta_{\text{PI1}}$ respectively. We also note that for repulsive $g_2$, the subleading order is PDW, while for attractive $g_2$, the subleading order is degenerate $s$- and $p$-wave uniform SC. If $g_1(0)$ becomes large and $g_2(0)$ stays small, both $g_1$ and $g_2$ flows to zero, for which there is no onset of instability. We term this case as metal without symmetry breaking. Similar behavior has been found in a model with a single HOVHS in the weak coupling regime in Ref.\cite{Isobe2019}, which names the gapless metallic state as the `supermetal', and also in the same two patch model but with $N_f=1$ instead \cite{PhysRevB.104.195134,https://doi.org/10.48550/arxiv.2203.05480}. Whether the ground state is interacting or non-interacting can be justified by the Wilsonian RG which includes the competition between tree level and one loop contribution.

Secondly, if we start from a repulsive initial $g_4(0)>0$, the charge Pomeranchuck instabilities $\Delta_{\text{PI1}}$ and $\Delta_{\text{PI2}}$ are stable with respect to this perturbation. What's more, the degeneracy between the subleading $s$-wave and $p$-wave SC when $g_4=0$ is lifted by the nonzero $g_4$. And the $p$-wave SC wins over the $s$-wave SC with a repulsive $g_4(0)>0$. In the metal phase regime when $g_4(0)=0$, we now have $\Delta_{\text{FM1}}$ as the leading order. Near the boundary between $\text{FM}1$ and $\text{PI}2$ orders, the critical value $y_c$ is progressively larger than in other areas, which means the boundary between FM1 and PI2 is still a gapless metal without symmetry breaking. 
Finally, if we start from an attractive $g_4(0)<0$, we have another fixed point where the $s$-wave uniform SC becomes the leading order (see the left panel in Fig.\ref{fig:2p_phase}).

In all cases, the PDW phase is subleading, which indicates a strong tendency of the long-range ordering of PDW. 
We close this section by noticing that, PDW can be the leading order once we introduce an valley Zeeman term splitting term in the system \cite{PhysRevB.99.205150}. For example, one can differentiate the phase factor $\phi$ in Eq. \eqref{lattice} for different valleys by setting $\phi_+\neq\phi_-$. This valley Zeeman breaks the time-reversal symmetry between $\bm{K}$ and $-\bm{K}$, and as a result the two HOVHS do not show up at the same energy. Then we effectively obtain a one-patch model, located either around $\bm{K}$ or $-\bm{K}$. The inter-patch interactions $g_2$ and $g_4$ are therefore absent. The only allowed interaction is $g_1$, governed by the RG equation
\begin{equation}
    \dot{g}_1=[(3-N_f)d-1]g_1^2
\end{equation}
where we use $\Pi_{pp}(\bm{Q})$ as the running parameter and $d=\Pi_{ph}(0)/\Pi_{pp}(\bm{Q})$. Once $g_1$ flows to strong attraction and $d<1$, the leading order is then a valley polarized PDW.

\section{Discussion and conclusion} 
\label{sec:discussion}
In this work we analyzed, based on an unbiased weak coupling RG approach, various competing orders of two models with different types of van Hove singularities. In the first case, the system hosts six different conventional van Hove singularities where the DOS diverges logrithmically. In the other case, we discuss a system with two type-II higher order van Hove singularities where the DOS diverges in a power-law manner. In both cases, the bare PDW susceptibilities are enhanced to the same order of the most divergent bare BCS susceptibility, and thus becomes one of many competing orders including the uniform SC and density waves in particle-hole channel. We find that, with certain initial interactions and nesting parameters, the PDW order wins over all other orders and becomes the leading instability.  

We argue the both cases considered here can be realized in moir\'e systems such as twisted bilayer graphene, twisted double bilayer graphene and twisted bilayer TMD,  for which the CVHS and HOVHS can be achieved by tuning moir\'e band structure through an applied out-of-plane displacement field. Another crucial ingredient for realizing the PDW order is the necessity of keeping both spin and valley degrees of freedom. This can be seen by comparing the $N_f=1$ and $N_f=2$ models. In the former case, the system becomes effectively spinless with a nonzero $\phi$. Then the number of symmetry allowed interactions is reduced. For example, in the two patch spinless model at higher-order van Hove filling, only $g_2$ is allowed. As a result, the PDW ground state cannot be realized in the $N_f=1$ two patch model \cite{PhysRevB.104.195134,https://doi.org/10.48550/arxiv.2203.05480}. 

Keeping both spin and valley degrees of freedom results in an $\text{SU(4)}$ symmetry at $\phi=0$. Pairing at this case can be classified based how the spin and valley form doublet as we discussed in Sec.\ref{sec:class}. We also discussed the local transformation that connects the order parameters at $\phi=0$ and those at $\phi=\frac{\pi}{3}$. From the view of the local transformation, we argued that only the valley triplet pairing at $\phi=0$ can be mapped into the PDW order at $\phi=\frac{\pi}{3}$. We find that the pairing symmetry must be either $s$-wave or $d$-wave. In the $d$-wave situation, the ground state is most likely to be a $d+id$ configuration in order to be fully gaped. It will be interesting to further investigate this chiral PDW phase. 

The PDW order obtained in the six patch case has finite momentum $\pm\bm{K}$. However we note this is different from the unidirectional PDW, which has both $\bm{Q}$ and $-\bm{Q}$ pairing and the spatial configuration is a stripe SC. Here the momentum is associated with valley index such that the PDW is like Fulde-Ferrell state for each valley. Since the FS of each valley has only one nesting vector in the particle-particle channel, a natural way to arrive at the LO state is to add a small coupling between different valleys. However, we find that a small valley coupling also changes the pairing momentum in the six patch model, such that the PDW becomes incommensurate.

In the six patch case, the degeneracy between different valleys can be lifted by including some valley splitting field, giving rise to valley polarized PDW. In the two patch case, such valley splitting field is necessary for realizing the PDW order. In both cases, we obtain a valley polarized PDW. Although this PDW order does not have spatial nodal lines, it has other interesting properties such as superconducting diode effect \cite{Ando2020,PhysRevLett.128.037001,Noah2022}, in which the critical current is nonreciprocal, i.e. it is larger in one direction but smaller in the opposite direction. Other interesting effect such as dc-Josephson effect has also been discussed when this PDW order develops\cite{Margarita}.  

Finally, we notice it is possible to realize some exotic orders through partial melting of the chiral PDW found in this paper. These include, chiral nematic order $(\Delta_1(\bm{K})+i\Delta_2(\bm{K}))(\Delta^*_1(\bm{K})+i\Delta^*_2(\bm{K}))$ which breaks lattice rotation and time-reversal symmetry, charge-$4e$ PDW $(\Delta_1(\bm{K})+i\Delta_2(\bm{K}))(\Delta_1(\bm{K})-i\Delta_2(\bm{K}))$, and even chiral charge-$6e$ uniform SC $(\Delta_1(\bm{K})+i\Delta_2(\bm{K}))^3$. Identifying the instabilities towards these orders and revealing about their physical properties require additional work which we leave for future studies. 

\acknowledgements 
We thank Hong Yao, Sri Raghu, Mengxing Ye and in particular Andrey Chubukov for useful discussions. YMW is grateful to the support of Shuimu Fellow Foundation at Tsinghua University. F.W. is supported by National Key Research and Development Program of China (Grant No. 2021YFA1401300) and start-up funding of Wuhan University.

\appendix

\section{Calculations of the bare susceptibilities in the six patch case} 
\label{sec:calculations_of_the_bare_susceptibilities_in_the_six_patch_case}
The dispersion near the six conventional van Hove singularities are:
\begin{equation}
 \begin{aligned}
     &\epsilon_{1}(\Vec{k})=\epsilon_{\bar{1}}(\Vec{k})=-\frac{1}{2}(t+9t_1)k_x^2+\frac{3}{2}(t+t_1)k_y^2,\\
   &\epsilon_{2,3}(\Vec{k})=\epsilon_{\bar{2},\bar{3}}(\Vec{k})=tk_x^2-3t_1k_y^2\pm\sqrt{3}(t+t_1)k_xk_y,
       \label{dispersion}
      \end{aligned}
  \end{equation}
 We start from the simple case with perfect nested Fermi surface ($t^{\prime}=0$). We introduce two parameters: $a_{\pm}=\sqrt{\frac{t}{2}}(k_x\pm\sqrt{3}k_y)$, and the dispersion in Eq.\eqref{dispersion} can be rewritten as: $\epsilon_1(\Vec{k})=-a_+a_-,\epsilon_{2,3}(\Vec{k})=a_{\pm}(a_++a_-)$. And the bare susceptibilities are:
\begin{equation}
         \begin{aligned} 
        &\Pi_{ph}(\bm{Q}_{\pm},T)=\\
        &-h T \sum_{n} \int_{-\sqrt{\Lambda}}^{\sqrt{\Lambda}}  \frac{d a_{+} d a_{-}}{\left(i \omega_{n}-a_{+} a_{-}\right)\left[i \omega_{n}+a_{+}\left(a_{+}+a_{-}\right)\right]}, \\
         & \Pi_{pp}(0,T)=\Pi_{pp}(\pm\bm{K},T)=\\
         &-h T \sum_{n} \int_{-\sqrt{\Lambda}}^{\sqrt{\Lambda}}  \frac{d a_{+} d a_{-}}{\left(i \omega_{n}-a_{+} a_{-}\right)\left[i \omega_{n}+a_{+}\left(a_{+}+a_{-}\right)\right]}, 
         \end{aligned}
         \label{bare}
 \end{equation}
 where $\Lambda$ is the UV energy cutoff. We note the expressions in Eq.\eqref{bare} have been evaluated in Ref \cite{PhysRevB.100.085136}, and the results are:
 \begin{equation}
    \begin{aligned}
    &\Pi_{pp}(0,T)=\Pi_{pp}(\pm\bm{K},T)=\frac{1}{4 \sqrt{3} \pi^{2} t} \ln^2 \frac{\Lambda}{T},\\
       &\Pi_{ph}(\bm{Q}_{\pm},T)=\frac{1}{8 \sqrt{3} \pi^{2} t} \ln^2 \frac{\Lambda}{T}.
    \end{aligned}
\end{equation}
When a finite next-nearest neighbour hopping $t^{\prime}$ or/and chemical potential is introduced, all the log divergence will be replaced by $\ln \frac{\Lambda}{\text{max}\{T,\mu,t^{\prime}\}}$ except the Cooper logarithm in $\Pi_{pp}(0,T)$ and $\Pi_{pp}(\pm\bm{K},T)$ \cite{Nandkishore2012,Furukawa}.

\section{Projection of the lattice interaction} 
\label{sec:projection_of_the_lattice_interaction}

In this section, we project the lattice interaction in Eq. \eqref{lat_int} to $g_{ij}$ of the six patch model. The lattice interaction consists the Hubbard interaction, spin Hund interaction, orbital Hund interaction and Heisenberg interaction. The Hubbard interaction gives the same initial values for all the $g_{1i}$ and $g_{2i}$.  We start from the spin Hund interaction:
\begin{equation}
    \begin{aligned}
    V_h \sum_{i} (c_i^{\dagger}\vec{S}c_i)^2&=V_h \sum_{i,\alpha,\beta,a,b}[2c_{i,\alpha,a}^{\dagger}c_{i,\beta,a}c_{i,\beta,b}^{\dagger}c_{i,\alpha,b}-n_i^2]\\
    &=-V_h\sum_{i,a,b}[2c_{i,a}^{\dagger}c_{i,b}c_{i,b}^{\dagger}c_{i,a}+n_i^2],
    \end{aligned}
\end{equation}
where $\alpha,\beta$ are the spin indices, $a,b$ are the valley indices and the spin summation is implied in the last line. We have also neglected the quadratic term of fermion operators going from the first line to the second line, which is just the chemical potential. The spin Hund coupling gives the initial values of $g_{ij}$ as: $g_{1i}=-3V_h,g_{2i}=-V_h,g_{4i}=-2V_h$.

Similarly, the projection of orbital Hund interaction to the low energy $g_{ij}$ interactions is:
\begin{equation}
    \begin{aligned}
   & K \sum_{i} (c_i^{\dagger}\vec{L}c_i)^2=K \sum_{i,\alpha,\beta,a,b}[(c_{i,+}^{\dagger}c_{i,-}+c_{i,-}^{\dagger}c_{i,+})^2\\
    &-(c_{i,+}^{\dagger}c_{i,-}-c_{i,-}^{\dagger}c_{i,+})^2+(c_{i,+}^{\dagger}c_{i,+}-c_{i,-}^{\dagger}c_{i,-})^2]\\
    &=K\sum_{i}[2c_{i,+}^{\dagger}c_{i,-}c_{i,-}^{\dagger}c_{i,+}\\
    &+2c_{i,-}^{\dagger}c_{i,+}c_{i,+}^{\dagger}c_{i,-}+(c_{i,+}^{\dagger}c_{i,+}-c_{i,-}^{\dagger}c_{i,-})^2],
    \end{aligned}
\end{equation}
which give the initial values of $g_{ij}$ as: $g_{1i}=K,g_{2i}=-K,g_{4i}=2K$.

Finally, the Heisenberg interaction consists three parts: the nearest-neighbour spin Hund coupling, orbital Hund coupling and spin-orbital Hund coupling:
\begin{equation}
    \begin{aligned}
    &J \sum_{\braket{ij},n} c_i^{\dagger}T^nc_ic_j^{\dagger}T^nc_j\\
    &=J\sum_{<ij>}(c_i^{\dagger}\vec{\sigma}c_i) (c^{\dagger}_j\vec{\sigma}c_j)+J\sum_{<ij>}(c_i^{\dagger}\vec{\tau}c_i) (c^{\dagger}_j\vec{\tau}c_j)\\
    &+J\sum_{<ij>}[c_i^{\dagger}\vec{\tau}\otimes\vec{\sigma}c_i] [c^{\dagger}_j\vec{\tau}\otimes\vec{\sigma}c_j].
    \end{aligned}
\end{equation}
We project each part of the Heisenberg interaction to the low energy degrees of freedom separately. We start from the simplest nearest-neighbor orbital Hund coupling which does not flip the spin explicitly:
\begin{equation}
    \begin{aligned}
    &J\sum_{\braket{ij}}(c_i^{\dagger}\vec{\tau}c_i) (c^{\dagger}_j\vec{\tau}c_j)\\
    &=J\sum_{<ij>}(2c_{i,+}^{\dagger}c_{i,-}c_{j,-}^{\dagger}c_{j,+}+2c_{i,-}^{\dagger}c_{i,+}c_{j,+}^{\dagger}c_{j,-}\\
    &+c_{i,+}^{\dagger}c_{i,+}c_{j,+}^{\dagger}c_{j,+}+c_{i,-}^{\dagger}c_{i,-}c_{j,-}^{\dagger}c_{j,-}\\
    &-c_{i,+}^{\dagger}c_{i,+}c_{j,-}^{\dagger}c_{j,-}-c_{i,-}^{\dagger}c_{i,-}c_{j,+}^{\dagger}c_{j,+}
    ).
    \end{aligned}
\end{equation}
$g_{ij}$ acquire anisotropic initial values from this term as:
\begin{equation}
\begin{aligned}
   & g_{11}=g_{12}=J,\quad  g_{13}=g_{14}=J\sum_{\hat{a}_i}\cos(\bm{Q}_+\cdot\hat{a}_i)=-J,\\
   & g_{21}=g_{22}=-J,\quad  g_{23}=g_{24}=-J\sum_{\hat{a}_i}\cos(\bm{Q}_+\cdot\hat{a}_i)=J,\\
   & g_{41}=g_{44}=2J\sum_{\hat{a}_i}\cos(\bm{K}\cdot\hat{a}_i)=-3J,\\
   &g_{42}=g_{43}=2J\sum_{\hat{a}_i}\cos(\bm{Q}_-\cdot\hat{a}_i)=J.
    \end{aligned}
\end{equation}

Next we consider the spin Hund coupling, to which we apply the $\text{SU(2)}$ Fierz identity to bring it into the spin preserving form:
\begin{equation}
    \begin{aligned}
    &J\sum_{\braket{ij}} (c_i^{\dagger}\vec{S}c_i)(c_j^{\dagger}\vec{S}c_j)\\
    &=J \sum_{i,\alpha,\beta,a,b}(2c_{i,\alpha,a}^{\dagger}c_{i,\beta,a}c_{j,\beta,b}^{\dagger}c_{j,\alpha,b}-n_in_j)\\
    &=-J\sum_{\braket{ij}}(2c_{i,+}^{\dagger}c_{j,+}c_{j,+}^{\dagger}c_{i,+}+2c_{i,-}^{\dagger}c_{j,-}c_{j,-}^{\dagger}c_{i,-}\\
    &+2c_{i,+}^{\dagger}c_{j,-}c_{j,-}^{\dagger}c_{i,+}+2c_{i,-}^{\dagger}c_{j,+}c_{j,+}^{\dagger}c_{i,-}+n_in_j),
    \end{aligned}
\end{equation}
$g_{ij}$ acquire the following initial values from this term as:
\begin{equation}
\begin{aligned}
   & g_{11}=-3J,\quad g_{12}=-2J\sum_{\hat{a}_i}\cos(\bm{Q}_+\cdot\hat{a}_i)-J=J\\
   & g_{13}=-3J\sum_{\hat{a}_i}\cos(\bm{Q}_+\cdot\hat{a}_i)=3J,\\
   & g_{14}=-2J-J\sum_{\hat{a}_i}\cos(\bm{Q}_+\cdot\hat{a}_i)=-J,\\
   & g_{21}=g_{22}=-J,\quad g_{23}=g_{24}=-J\sum_{\hat{a}_i}\cos(\bm{Q}_+\cdot\hat{a}_i)=J\\
   & g_{41}=g_{42}=-2J,\quad g_{43}=g_{44}=-2J\sum_{\hat{a}_i}\cos(\bm{Q}_+\cdot\hat{a}_i)=2J.
    \end{aligned}
\end{equation}

Finally, the projection of the spin-orbital Hund coupling coupling is:
\begin{equation}
    \begin{aligned}
    &J\sum_{<ij>}[c_i^{\dagger}\vec{\tau}\otimes\vec{\sigma}c_i] [c^{\dagger}_j\vec{\tau}\otimes\vec{\sigma}c_j]\\
    &=J[2c_{i,+}^{\dagger}\vec{\sigma}c_{i,-}c_{j,-}^{\dagger}\vec{\sigma}c_{j,+}+2c_{i,-}^{\dagger}\vec{\sigma}c_{i,+}c_{j,+}^{\dagger}\vec{\sigma}c_{j,-}\\
    &+(c_{i,+}^{\dagger}\vec{\sigma}c_{i,+})\cdot(c_{j,+}^{\dagger}\vec{\sigma}c_{j,+})+(c_{i,-}^{\dagger}\vec{\sigma}c_{i,-})\cdot(c_{j,-}^{\dagger}\vec{\sigma}c_{j,-})\\
    &-(c_{i,+}^{\dagger}\vec{\sigma}c_{i,+})\cdot(c_{j,-}^{\dagger}\vec{\sigma}c_{j,-})-(c_{i,-}^{\dagger}\vec{\sigma}c_{i,-})\cdot(c_{j,+}^{\dagger}\vec{\sigma}c_{j,+})]\\
    &=J\sum_{\braket{ij}}(-4c_{i,+}^{\dagger}c_{j,+}c_{j,-}^{\dagger}c_{i,-}-2c_{i,+}^{\dagger}c_{i,-}c_{j,-}^{\dagger}c_{j,+}\\
    &-2c_{i,+}^{\dagger}c_{j,+}c_{j,+}^{\dagger}c_{i,+}-c_{i,+}^{\dagger}c_{i,+}c_{j,+}^{\dagger}c_{j,+}+2c_{i,+}^{\dagger}c_{j,-}c_{j,-}^{\dagger}c_{i,+}\\
    &+c_{i,+}^{\dagger}c_{i,+}c_{j,-}^{\dagger}c_{j,-})+\quad (+\leftrightarrow-)
    \end{aligned}
\end{equation}
$g_{ij}$ acquire the following initial values from this term as:
\begin{equation}
\begin{aligned}
& g_{11}=-2J-J=-3J,\quad g_{12}=-2J\sum_{\hat{a}_i}\cos(\bm{Q}_+\cdot\hat{a}_i)-J=J,\\
   &g_{13}=-2J\sum_{\hat{a}_i}\cos(\bm{Q}_+\cdot\hat{a}_i)-J\sum_{\hat{a}_i}\cos(\bm{Q}_+\cdot\hat{a}_i)=3J,\\
   &g_{14}=-2J-J\sum_{\hat{a}_i}\cos(\bm{Q}_+\cdot\hat{a}_i)=-J,\\
   & g_{21}=-4J\sum_{\hat{a}_i}\cos(\bm{K}\cdot\hat{a}_i)+J=7J,\\
   & g_{22}=-4J\sum_{\hat{a}_i}\cos(\bm{Q}_-\cdot\hat{a}_i)+J=-J,\\
   &g_{23}=-4J\sum_{\hat{a}_i}\cos(\bm{Q}_-\cdot\hat{a}_i)+J\sum_{\hat{a}_i}\cos(\bm{Q}_+\cdot\hat{a}_i)=-3J,\\
   &g_{24}=-4J\sum_{\hat{a}_i}\cos(\bm{K}\cdot\hat{a}_i)+J\sum_{\hat{a}_i}\cos(\bm{Q}_+\cdot\hat{a}_i)=5J,\\
   & g_{41}=2J-2J\sum_{\hat{a}_i}\cos(\bm{K}\cdot\hat{a}_i)=5J,\\   & g_{42}=2J-2J\sum_{\hat{a}_i}\cos(\bm{Q}_-\cdot\hat{a}_i)=J,\\
   &g_{43}=2J\sum_{\hat{a}_i}\cos(\bm{Q}_+\cdot\hat{a}_i)-2J\sum_{\hat{a}_i}\cos(\bm{Q}_-\cdot\hat{a}_i)=-3J,\\
   &g_{44}=2J\sum_{\hat{a}_i}\cos(\bm{Q}_+\cdot\hat{a}_i)-2J\sum_{\hat{a}_i}\cos(\bm{K}\cdot\hat{a}_i)=J
    \end{aligned}
\end{equation}
The total contribution of the Heisenberg interaction to the initial values of $g_{ij}$ are:
\begin{equation}
\begin{aligned}
    &g_{11}=-5J,
    \quad g_{12}=3J,\quad  g_{13}=5J,\quad  g_{14}=-3J,\\
    &g_{21}=5J,\quad g_{22}=-3J,\quad g_{23}=-J,\quad g_{24}=7J,\\
    &g_{4i}=0
    \end{aligned}
\end{equation}

\section{The stability of the fix point} 
\label{sec:the_stability_of_the_fix_point}
The asymptotic behavior of strong coupling fixed points in the one-loop RG equations are : $g_{ij}\approx \frac{G_{ij}}{y_c-y} , i=1,2,\quad j=1,2,3,4$. The stability of fixed trajectories toward strong coupling is analyzed through the stability matrices of the ratios of coupling constants, or 'rays'\cite{PhysRevB.81.041401,PhysRevB.82.205106,Cvetkovic, PhysRevB.100.085136}. Since the interaction $g_{12},g_{22}$ generally flow to $+\infty$, as we can see from the RG equations of the six patch model. We can use either of them as a new running parameter, and define the rays as : $x_{ij}=g_{ij}/g_{12}$ or $x_{ij}=g_{ij}/g_{22}$ . 
 
 We take a typical nesting parameter $d=\frac{1}{2}$. The $g_{22}$ interaction flows to $+\infty$ in the strong coupling fix point which favour the chiral d-wave PDW, and we use $g_{22}$ as the new flow parameter to analyze the stability of this ray. The fixed poitn ratios are:

The RG equations can be cast into 
  \begin{equation}
   \begin{aligned}
        &\frac{d x_{11}}{d\ln g_{22}}=-x_{11}-\frac{x_{11}^2+2x_{13}^2}{d(1+x_{23}^2)},\\
        &\frac{d x_{12}}{d\ln g_{22}}=-x_{12}+\frac{x_{12}^2+x_{13}^2+x_{43}^2+x_{44}^2}{1+x_{23}^2},\\
        &\frac{d x_{13}}{d\ln g_{22}}=-x_{13}-\frac{2x_{11}x_{13}+x_{13}^2}{d(1+x_{23}^2)}+2\times\\
        &\frac{2x_{12}x_{13}-2x_{23}x_{24}+x_{23}x_{44}+x_{24}x_{43}+x_{43}x_{44}-x_{13}x_{14}}{1+x_{23}^2},\\
        &\frac{d x_{14}}{d\ln g_{22}}=-x_{14}+2\times\\
        &\frac{x_{12}x_{14}+x_{24}x_{44}+x_{23}x_{43}-x_{24}^2-x_{14}^2-x_{23}^2}{1+x_{23}^2},\\
        &\frac{d x_{21}}{d\ln g_{22}}=-x_{21}-\frac{x_{21}^2+2x_{23}^2+x_{41}^2+2x_{43}^2}{d(1+x_{23}^2)},\\
        &\frac{d x_{23}}{d\ln g_{22}}=-x_{23}+\\
        &\frac{-2x_{21}x_{23}-x_{23}^2+2d(x_{23}+x_{12}x_{23}-2x_{23}x_{14}-x_{13}x_{24})}{d(1+x_{23}^2)},\\
        &+\frac{-2x_{41}x_{43}-x_{43}^2+2d(x_{13}x_{44}+x_{14}x_{43})}{d(1+x_{23}^2)},\\
        &\frac{d x_{24}}{d\ln g_{22}}=-x_{24}\\
        &+2\frac{x_{12}x_{24}+x_{14}x_{44}+x_{13}x_{43}-x_{13}x_{23}-2x_{14}x_{24}}{1+x_{23}^2},\\
        &\frac{d x_{41}}{d\ln g_{22}}=-x_{41}-2\frac{x_{21}x_{41}+2x_{23}x_{43}}{d(1+x_{23}^2)},\\
        &\frac{d x_{42}}{d\ln g_{22}}=-x_{42}+2\frac{x_{42}+x_{23}x_{43}-x_{43}^2-x_{42}^2}{1+x_{23}^2},\\
        &\frac{d x_{43}}{d\ln g_{22}}=-x_{43}-2\frac{x_{21}x_{43}+x_{23}x_{41}+x_{23}x_{43}}{d(1+x_{23}^2)}\\
        &+2\frac{x_{12}x_{43}+x_{13}x_{44}+x_{43}+x_{23}x_{42}-2x_{42}x_{43}}{1+x_{23}^2},\\
        &\frac{d x_{44}}{d\ln g_{22}}=-x_{44}+2\frac{x_{12}x_{44}+x_{13}x_{43}}{1+x_{23}^2}.
        \label{ray}
   \end{aligned}
 \end{equation}
Similar equations can also be obtained if we choose $g_{12}$ as the new flow parameter. For convenience
we use the eleven component vector $\Vec{x}$ to compactly label the eleven ratios on the right hand side of the Eq.\eqref{ray}, and use $f_i(\Vec{x}),\quad i=1,2,3...7$ to label the expressions on the left hand side of the Eq.\eqref{ray}.  Then we do small perturbations to the stable ray, which is equivalent to linearize the Eq.\eqref{ray} around the stable ray:
\begin{equation}
     \begin{array}{l}
\frac{d\delta x_i}{d \ln g_{22}}
=\frac{\partial f_{i}(\Vec{x})}{\partial x_{j}}|_{\Vec{x}=\Vec{x}^*}
\delta x_j=M_{ij}\delta x_j
\end{array}
\end{equation}
If the stability matrix $M_{ij}=\partial f_{i}(\rho_1,\rho_4)/\partial \rho_{j}$ has eigenvalues which are all negative, then the fixed point is stable. Otherwise it has positive eigenvalue(s), then the fixed point is not stable. We have testified that for the PDW and uniform SC orders, all the eigenvalues of their corresponding stability matrix are negative. 

In the two patch model, similar analysis can be applied. For example, when the interactions flow to a fixed point at which $g_1\to -\infty$, we can introduce $\ln|g_1|$ as a new running parameter around the fixed point to see if this is stable. Defining $x_2=-g_2/g_1$ and $x_4=-g_4/g_1$, the RG equations for $g_2$ and $g_4$ can be cast into
\begin{equation}
    \begin{aligned}
    \frac{d x_2}{d \ln |g_1|}&=-x_2-\frac{(d_1-1)x_2^2-x_4^2-2d_2x_4-(2-2N_f)d_2x_2}{[(3-N_f)d_2-d_3]-N_fd_2x_2^2+d_2x_4^2+2d_2x_2x_4}\\
     \frac{d x_4}{d \ln |g_1|}&=-x_4-\frac{2(d_1-1)x_2x_4-2d_2x_4-N_fd_1x_4^2}{[(3-N_f)d_2-d_3]-N_fd_2x_2^2+d_2x_4^2+2d_2x_2x_4}\\
    \end{aligned}
\end{equation}
Again we can define the corresponding stability matrix, and we find that all the leading orders shown in Fig.\ref{fig:2p_phase} are stable.

\bibliography{PDW}

\begin{thebibliography}{87}%
\makeatletter
\providecommand \@ifxundefined [1]{%
 \@ifx{#1\undefined}
}%
\providecommand \@ifnum [1]{%
 \ifnum #1\expandafter \@firstoftwo
 \else \expandafter \@secondoftwo
 \fi
}%
\providecommand \@ifx [1]{%
 \ifx #1\expandafter \@firstoftwo
 \else \expandafter \@secondoftwo
 \fi
}%
\providecommand \natexlab [1]{#1}%
\providecommand \enquote  [1]{``#1''}%
\providecommand \bibnamefont  [1]{#1}%
\providecommand \bibfnamefont [1]{#1}%
\providecommand \citenamefont [1]{#1}%
\providecommand \href@noop [0]{\@secondoftwo}%
\providecommand \href [0]{\begingroup \@sanitize@url \@href}%
\providecommand \@href[1]{\@@startlink{#1}\@@href}%
\providecommand \@@href[1]{\endgroup#1\@@endlink}%
\providecommand \@sanitize@url [0]{\catcode `\\12\catcode `\$12\catcode
  `\&12\catcode `\#12\catcode `\^12\catcode `\_12\catcode `\%12\relax}%
\providecommand \@@startlink[1]{}%
\providecommand \@@endlink[0]{}%
\providecommand \url  [0]{\begingroup\@sanitize@url \@url }%
\providecommand \@url [1]{\endgroup\@href {#1}{\urlprefix }}%
\providecommand \urlprefix  [0]{URL }%
\providecommand \Eprint [0]{\href }%
\providecommand \doibase [0]{https://doi.org/}%
\providecommand \selectlanguage [0]{\@gobble}%
\providecommand \bibinfo  [0]{\@secondoftwo}%
\providecommand \bibfield  [0]{\@secondoftwo}%
\providecommand \translation [1]{[#1]}%
\providecommand \BibitemOpen [0]{}%
\providecommand \bibitemStop [0]{}%
\providecommand \bibitemNoStop [0]{.\EOS\space}%
\providecommand \EOS [0]{\spacefactor3000\relax}%
\providecommand \BibitemShut  [1]{\csname bibitem#1\endcsname}%
\let\auto@bib@innerbib\@empty
\bibitem [{\citenamefont {Agterberg}\ \emph {et~al.}(2020)\citenamefont
  {Agterberg}, \citenamefont {Davis}, \citenamefont {Edkins}, \citenamefont
  {Fradkin}, \citenamefont {Van~Harlingen}, \citenamefont {Kivelson},
  \citenamefont {Lee}, \citenamefont {Radzihovsky}, \citenamefont {Tranquada},\
  and\ \citenamefont {Wang}}]{Agterberg}%
  \BibitemOpen
  \bibfield  {author} {\bibinfo {author} {\bibfnamefont {D.~F.}\ \bibnamefont
  {Agterberg}}, \bibinfo {author} {\bibfnamefont {J.~S.}\ \bibnamefont
  {Davis}}, \bibinfo {author} {\bibfnamefont {S.~D.}\ \bibnamefont {Edkins}},
  \bibinfo {author} {\bibfnamefont {E.}~\bibnamefont {Fradkin}}, \bibinfo
  {author} {\bibfnamefont {D.~J.}\ \bibnamefont {Van~Harlingen}}, \bibinfo
  {author} {\bibfnamefont {S.~A.}\ \bibnamefont {Kivelson}}, \bibinfo {author}
  {\bibfnamefont {P.~A.}\ \bibnamefont {Lee}}, \bibinfo {author} {\bibfnamefont
  {L.}~\bibnamefont {Radzihovsky}}, \bibinfo {author} {\bibfnamefont {J.~M.}\
  \bibnamefont {Tranquada}},\ and\ \bibinfo {author} {\bibfnamefont
  {Y.}~\bibnamefont {Wang}},\ }\bibfield  {title} {\bibinfo {title} {The
  physics of pair-density waves: Cuprate superconductors and beyond},\ }\href
  {https://doi.org/10.1146/annurev-conmatphys-031119-050711} {\bibfield
  {journal} {\bibinfo  {journal} {Annual Review of Condensed Matter Physics}\
  }\textbf {\bibinfo {volume} {11}},\ \bibinfo {pages} {231} (\bibinfo {year}
  {2020})}\BibitemShut {NoStop}%
\bibitem [{\citenamefont {Fulde}\ and\ \citenamefont {Ferrell}(1964)}]{FF}%
  \BibitemOpen
  \bibfield  {author} {\bibinfo {author} {\bibfnamefont {P.}~\bibnamefont
  {Fulde}}\ and\ \bibinfo {author} {\bibfnamefont {R.~A.}\ \bibnamefont
  {Ferrell}},\ }\bibfield  {title} {\bibinfo {title} {Superconductivity in a
  strong spin-exchange field},\ }\href
  {https://doi.org/10.1103/PhysRev.135.A550} {\bibfield  {journal} {\bibinfo
  {journal} {Phys. Rev.}\ }\textbf {\bibinfo {volume} {135}},\ \bibinfo {pages}
  {A550} (\bibinfo {year} {1964})}\BibitemShut {NoStop}%
\bibitem [{\citenamefont {Larkin}\ and\ \citenamefont
  {Ovchinnikov}(1965)}]{LO}%
  \BibitemOpen
  \bibfield  {author} {\bibinfo {author} {\bibfnamefont {A.~I.}\ \bibnamefont
  {Larkin}}\ and\ \bibinfo {author} {\bibfnamefont {Y.~N.}\ \bibnamefont
  {Ovchinnikov}},\ }\bibfield  {title} {\bibinfo {title} {Nonuniform state of
  superconductors},\ }\href {https://www.osti.gov/biblio/4653415} {\bibfield
  {journal} {\bibinfo  {journal} {Sov. Phys. JETP}\ }\textbf {\bibinfo {volume}
  {20}},\ \bibinfo {pages} {762} (\bibinfo {year} {1965})}\BibitemShut
  {NoStop}%
\bibitem [{\citenamefont {Agosta}(2018)}]{cryst8070285}%
  \BibitemOpen
  \bibfield  {author} {\bibinfo {author} {\bibfnamefont {C.~C.}\ \bibnamefont
  {Agosta}},\ }\bibfield  {title} {\bibinfo {title} {Inhomogeneous
  superconductivity in organic and related superconductors},\ }\href
  {https://doi.org/10.3390/cryst8070285} {\bibfield  {journal} {\bibinfo
  {journal} {Crystals}\ }\textbf {\bibinfo {volume} {8}},\ \bibinfo {pages}
  {285} (\bibinfo {year} {2018})}\BibitemShut {NoStop}%
\bibitem [{\citenamefont {Matsuda}\ and\ \citenamefont
  {Shimahara}(2007)}]{Matsuda2007}%
  \BibitemOpen
  \bibfield  {author} {\bibinfo {author} {\bibfnamefont {Y.}~\bibnamefont
  {Matsuda}}\ and\ \bibinfo {author} {\bibfnamefont {H.}~\bibnamefont
  {Shimahara}},\ }\bibfield  {title} {\bibinfo {title}
  {Fulde–ferrell–larkin–ovchinnikov state in heavy fermion
  superconductors},\ }\href {https://doi.org/10.1143/JPSJ.76.051005} {\bibfield
   {journal} {\bibinfo  {journal} {Journal of the Physical Society of Japan}\
  }\textbf {\bibinfo {volume} {76}},\ \bibinfo {pages} {051005} (\bibinfo
  {year} {2007})}\BibitemShut {NoStop}%
\bibitem [{\citenamefont {Gurevich}(2010)}]{Gurevich}%
  \BibitemOpen
  \bibfield  {author} {\bibinfo {author} {\bibfnamefont {A.}~\bibnamefont
  {Gurevich}},\ }\bibfield  {title} {\bibinfo {title} {Upper critical field and
  the fulde-ferrel-larkin-ovchinnikov transition in multiband
  superconductors},\ }\href {https://doi.org/10.1103/PhysRevB.82.184504}
  {\bibfield  {journal} {\bibinfo  {journal} {Phys. Rev. B}\ }\textbf {\bibinfo
  {volume} {82}},\ \bibinfo {pages} {184504} (\bibinfo {year}
  {2010})}\BibitemShut {NoStop}%
\bibitem [{\citenamefont {Cho}\ \emph {et~al.}(2017)\citenamefont {Cho},
  \citenamefont {Yang}, \citenamefont {Yuan}, \citenamefont {Shen},
  \citenamefont {Wolf},\ and\ \citenamefont {Lortz}}]{FeSC}%
  \BibitemOpen
  \bibfield  {author} {\bibinfo {author} {\bibfnamefont {C.-w.}\ \bibnamefont
  {Cho}}, \bibinfo {author} {\bibfnamefont {J.~H.}\ \bibnamefont {Yang}},
  \bibinfo {author} {\bibfnamefont {N.~F.~Q.}\ \bibnamefont {Yuan}}, \bibinfo
  {author} {\bibfnamefont {J.}~\bibnamefont {Shen}}, \bibinfo {author}
  {\bibfnamefont {T.}~\bibnamefont {Wolf}},\ and\ \bibinfo {author}
  {\bibfnamefont {R.}~\bibnamefont {Lortz}},\ }\bibfield  {title} {\bibinfo
  {title} {Thermodynamic evidence for the fulde-ferrell-larkin-ovchinnikov
  state in the ${\mathrm{kfe}}_{2}{\mathrm{as}}_{2}$ superconductor},\ }\href
  {https://doi.org/10.1103/PhysRevLett.119.217002} {\bibfield  {journal}
  {\bibinfo  {journal} {Phys. Rev. Lett.}\ }\textbf {\bibinfo {volume} {119}},\
  \bibinfo {pages} {217002} (\bibinfo {year} {2017})}\BibitemShut {NoStop}%
\bibitem [{\citenamefont {Berg}\ \emph {et~al.}(2007)\citenamefont {Berg},
  \citenamefont {Fradkin}, \citenamefont {Kim}, \citenamefont {Kivelson},
  \citenamefont {Oganesyan}, \citenamefont {Tranquada},\ and\ \citenamefont
  {Zhang}}]{Berg2007}%
  \BibitemOpen
  \bibfield  {author} {\bibinfo {author} {\bibfnamefont {E.}~\bibnamefont
  {Berg}}, \bibinfo {author} {\bibfnamefont {E.}~\bibnamefont {Fradkin}},
  \bibinfo {author} {\bibfnamefont {E.-A.}\ \bibnamefont {Kim}}, \bibinfo
  {author} {\bibfnamefont {S.~A.}\ \bibnamefont {Kivelson}}, \bibinfo {author}
  {\bibfnamefont {V.}~\bibnamefont {Oganesyan}}, \bibinfo {author}
  {\bibfnamefont {J.~M.}\ \bibnamefont {Tranquada}},\ and\ \bibinfo {author}
  {\bibfnamefont {S.~C.}\ \bibnamefont {Zhang}},\ }\bibfield  {title} {\bibinfo
  {title} {Dynamical layer decoupling in a stripe-ordered high-${T}_{c}$
  superconductor},\ }\href {https://doi.org/10.1103/PhysRevLett.99.127003}
  {\bibfield  {journal} {\bibinfo  {journal} {Phys. Rev. Lett.}\ }\textbf
  {\bibinfo {volume} {99}},\ \bibinfo {pages} {127003} (\bibinfo {year}
  {2007})}\BibitemShut {NoStop}%
\bibitem [{\citenamefont {Wang}\ \emph
  {et~al.}(2015{\natexlab{a}})\citenamefont {Wang}, \citenamefont {Agterberg},\
  and\ \citenamefont {Chubukov}}]{Wang2015a}%
  \BibitemOpen
  \bibfield  {author} {\bibinfo {author} {\bibfnamefont {Y.}~\bibnamefont
  {Wang}}, \bibinfo {author} {\bibfnamefont {D.~F.}\ \bibnamefont
  {Agterberg}},\ and\ \bibinfo {author} {\bibfnamefont {A.}~\bibnamefont
  {Chubukov}},\ }\bibfield  {title} {\bibinfo {title} {Coexistence of
  charge-density-wave and pair-density-wave orders in underdoped cuprates},\
  }\href {https://doi.org/10.1103/PhysRevLett.114.197001} {\bibfield  {journal}
  {\bibinfo  {journal} {Phys. Rev. Lett.}\ }\textbf {\bibinfo {volume} {114}},\
  \bibinfo {pages} {197001} (\bibinfo {year} {2015}{\natexlab{a}})}\BibitemShut
  {NoStop}%
\bibitem [{\citenamefont {Wang}\ \emph
  {et~al.}(2015{\natexlab{b}})\citenamefont {Wang}, \citenamefont {Agterberg},\
  and\ \citenamefont {Chubukov}}]{Wang2015b}%
  \BibitemOpen
  \bibfield  {author} {\bibinfo {author} {\bibfnamefont {Y.}~\bibnamefont
  {Wang}}, \bibinfo {author} {\bibfnamefont {D.~F.}\ \bibnamefont
  {Agterberg}},\ and\ \bibinfo {author} {\bibfnamefont {A.}~\bibnamefont
  {Chubukov}},\ }\bibfield  {title} {\bibinfo {title} {Interplay between pair-
  and charge-density-wave orders in underdoped cuprates},\ }\href
  {https://doi.org/10.1103/PhysRevB.91.115103} {\bibfield  {journal} {\bibinfo
  {journal} {Phys. Rev. B}\ }\textbf {\bibinfo {volume} {91}},\ \bibinfo
  {pages} {115103} (\bibinfo {year} {2015}{\natexlab{b}})}\BibitemShut
  {NoStop}%
\bibitem [{\citenamefont {Wang}\ \emph {et~al.}(2018)\citenamefont {Wang},
  \citenamefont {Edkins}, \citenamefont {Hamidian}, \citenamefont {Davis},
  \citenamefont {Fradkin},\ and\ \citenamefont {Kivelson}}]{Wang2018}%
  \BibitemOpen
  \bibfield  {author} {\bibinfo {author} {\bibfnamefont {Y.}~\bibnamefont
  {Wang}}, \bibinfo {author} {\bibfnamefont {S.~D.}\ \bibnamefont {Edkins}},
  \bibinfo {author} {\bibfnamefont {M.~H.}\ \bibnamefont {Hamidian}}, \bibinfo
  {author} {\bibfnamefont {J.~C.~S.}\ \bibnamefont {Davis}}, \bibinfo {author}
  {\bibfnamefont {E.}~\bibnamefont {Fradkin}},\ and\ \bibinfo {author}
  {\bibfnamefont {S.~A.}\ \bibnamefont {Kivelson}},\ }\bibfield  {title}
  {\bibinfo {title} {Pair density waves in superconducting vortex halos},\
  }\href {https://doi.org/10.1103/PhysRevB.97.174510} {\bibfield  {journal}
  {\bibinfo  {journal} {Phys. Rev. B}\ }\textbf {\bibinfo {volume} {97}},\
  \bibinfo {pages} {174510} (\bibinfo {year} {2018})}\BibitemShut {NoStop}%
\bibitem [{\citenamefont {Tranquada}\ \emph {et~al.}(1995)\citenamefont
  {Tranquada}, \citenamefont {Sternlieb}, \citenamefont {Axe}, \citenamefont
  {Nakamura},\ and\ \citenamefont {Uchida}}]{Tranquada1995}%
  \BibitemOpen
  \bibfield  {author} {\bibinfo {author} {\bibfnamefont {J.~M.}\ \bibnamefont
  {Tranquada}}, \bibinfo {author} {\bibfnamefont {B.~J.}\ \bibnamefont
  {Sternlieb}}, \bibinfo {author} {\bibfnamefont {J.~D.}\ \bibnamefont {Axe}},
  \bibinfo {author} {\bibfnamefont {Y.}~\bibnamefont {Nakamura}},\ and\
  \bibinfo {author} {\bibfnamefont {S.}~\bibnamefont {Uchida}},\ }\bibfield
  {title} {\bibinfo {title} {Evidence for stripe correlations of spins and
  holes in copper oxide superconductors},\ }\href
  {https://doi.org/10.1038/375561a0} {\bibfield  {journal} {\bibinfo  {journal}
  {Nature}\ }\textbf {\bibinfo {volume} {375}},\ \bibinfo {pages} {561}
  (\bibinfo {year} {1995})}\BibitemShut {NoStop}%
\bibitem [{\citenamefont {Fujita}\ \emph {et~al.}(2004)\citenamefont {Fujita},
  \citenamefont {Goka}, \citenamefont {Yamada}, \citenamefont {Tranquada},\
  and\ \citenamefont {Regnault}}]{Fujita2004}%
  \BibitemOpen
  \bibfield  {author} {\bibinfo {author} {\bibfnamefont {M.}~\bibnamefont
  {Fujita}}, \bibinfo {author} {\bibfnamefont {H.}~\bibnamefont {Goka}},
  \bibinfo {author} {\bibfnamefont {K.}~\bibnamefont {Yamada}}, \bibinfo
  {author} {\bibfnamefont {J.~M.}\ \bibnamefont {Tranquada}},\ and\ \bibinfo
  {author} {\bibfnamefont {L.~P.}\ \bibnamefont {Regnault}},\ }\bibfield
  {title} {\bibinfo {title} {Stripe order, depinning, and fluctuations in
  ${\mathrm{la}}_{1.875}{\mathrm{ba}}_{0.125}{\mathrm{cuo}}_{4}$ and
  ${\mathrm{la}}_{1.875}{\mathrm{ba}}_{0.075}{\mathrm{sr}}_{0.050}{\mathrm{cuo}}_{4}$},\
  }\href {https://doi.org/10.1103/PhysRevB.70.104517} {\bibfield  {journal}
  {\bibinfo  {journal} {Phys. Rev. B}\ }\textbf {\bibinfo {volume} {70}},\
  \bibinfo {pages} {104517} (\bibinfo {year} {2004})}\BibitemShut {NoStop}%
\bibitem [{\citenamefont {H\"ucker}\ \emph {et~al.}(2011)\citenamefont
  {H\"ucker}, \citenamefont {v.~Zimmermann}, \citenamefont {Gu}, \citenamefont
  {Xu}, \citenamefont {Wen}, \citenamefont {Xu}, \citenamefont {Kang},
  \citenamefont {Zheludev},\ and\ \citenamefont {Tranquada}}]{Zimmermann2011}%
  \BibitemOpen
  \bibfield  {author} {\bibinfo {author} {\bibfnamefont {M.}~\bibnamefont
  {H\"ucker}}, \bibinfo {author} {\bibfnamefont {M.}~\bibnamefont
  {v.~Zimmermann}}, \bibinfo {author} {\bibfnamefont {G.~D.}\ \bibnamefont
  {Gu}}, \bibinfo {author} {\bibfnamefont {Z.~J.}\ \bibnamefont {Xu}}, \bibinfo
  {author} {\bibfnamefont {J.~S.}\ \bibnamefont {Wen}}, \bibinfo {author}
  {\bibfnamefont {G.}~\bibnamefont {Xu}}, \bibinfo {author} {\bibfnamefont
  {H.~J.}\ \bibnamefont {Kang}}, \bibinfo {author} {\bibfnamefont
  {A.}~\bibnamefont {Zheludev}},\ and\ \bibinfo {author} {\bibfnamefont
  {J.~M.}\ \bibnamefont {Tranquada}},\ }\bibfield  {title} {\bibinfo {title}
  {Stripe order in superconducting
  la${}_{2\ensuremath{-}x}$ba${}_{x}$cuo${}_{4}$
  ($0.095\ensuremath{\leqslant}x\ensuremath{\leqslant}0.155$)},\ }\href
  {https://doi.org/10.1103/PhysRevB.83.104506} {\bibfield  {journal} {\bibinfo
  {journal} {Phys. Rev. B}\ }\textbf {\bibinfo {volume} {83}},\ \bibinfo
  {pages} {104506} (\bibinfo {year} {2011})}\BibitemShut {NoStop}%
\bibitem [{\citenamefont {Berg}\ \emph
  {et~al.}(2009{\natexlab{a}})\citenamefont {Berg}, \citenamefont {Fradkin},
  \citenamefont {Kivelson},\ and\ \citenamefont {Tranquada}}]{Berg_2009}%
  \BibitemOpen
  \bibfield  {author} {\bibinfo {author} {\bibfnamefont {E.}~\bibnamefont
  {Berg}}, \bibinfo {author} {\bibfnamefont {E.}~\bibnamefont {Fradkin}},
  \bibinfo {author} {\bibfnamefont {S.~A.}\ \bibnamefont {Kivelson}},\ and\
  \bibinfo {author} {\bibfnamefont {J.~M.}\ \bibnamefont {Tranquada}},\
  }\bibfield  {title} {\bibinfo {title} {Striped superconductors: how spin,
  charge and superconducting orders intertwine in the cuprates},\ }\href
  {https://doi.org/10.1088/1367-2630/11/11/115004} {\bibfield  {journal}
  {\bibinfo  {journal} {New Journal of Physics}\ }\textbf {\bibinfo {volume}
  {11}},\ \bibinfo {pages} {115004} (\bibinfo {year}
  {2009}{\natexlab{a}})}\BibitemShut {NoStop}%
\bibitem [{\citenamefont {Lee}(2014)}]{PALee2014}%
  \BibitemOpen
  \bibfield  {author} {\bibinfo {author} {\bibfnamefont {P.~A.}\ \bibnamefont
  {Lee}},\ }\bibfield  {title} {\bibinfo {title} {Amperean pairing and the
  pseudogap phase of cuprate superconductors},\ }\href
  {https://doi.org/10.1103/PhysRevX.4.031017} {\bibfield  {journal} {\bibinfo
  {journal} {Phys. Rev. X}\ }\textbf {\bibinfo {volume} {4}},\ \bibinfo {pages}
  {031017} (\bibinfo {year} {2014})}\BibitemShut {NoStop}%
\bibitem [{\citenamefont {Agterberg}\ and\ \citenamefont
  {Tsunetsugu}(2008)}]{Agterberg2008}%
  \BibitemOpen
  \bibfield  {author} {\bibinfo {author} {\bibfnamefont {D.~F.}\ \bibnamefont
  {Agterberg}}\ and\ \bibinfo {author} {\bibfnamefont {H.}~\bibnamefont
  {Tsunetsugu}},\ }\bibfield  {title} {\bibinfo {title} {Dislocations and
  vortices in pair-density-wave superconductors},\ }\href
  {https://doi.org/10.1038/nphys999} {\bibfield  {journal} {\bibinfo  {journal}
  {Nature Physics}\ }\textbf {\bibinfo {volume} {4}},\ \bibinfo {pages} {639}
  (\bibinfo {year} {2008})}\BibitemShut {NoStop}%
\bibitem [{\citenamefont {Nie}\ \emph {et~al.}(2014)\citenamefont {Nie},
  \citenamefont {Tarjus},\ and\ \citenamefont {Kivelson}}]{Nie2014}%
  \BibitemOpen
  \bibfield  {author} {\bibinfo {author} {\bibfnamefont {L.}~\bibnamefont
  {Nie}}, \bibinfo {author} {\bibfnamefont {G.}~\bibnamefont {Tarjus}},\ and\
  \bibinfo {author} {\bibfnamefont {S.~A.}\ \bibnamefont {Kivelson}},\
  }\bibfield  {title} {\bibinfo {title} {Quenched disorder and vestigial
  nematicity in the pseudogap regime of the cuprates},\ }\href
  {https://doi.org/10.1073/pnas.1406019111} {\bibfield  {journal} {\bibinfo
  {journal} {Proceedings of the National Academy of Sciences}\ }\textbf
  {\bibinfo {volume} {111}},\ \bibinfo {pages} {7980} (\bibinfo {year}
  {2014})}\BibitemShut {NoStop}%
\bibitem [{\citenamefont {Fradkin}\ \emph {et~al.}(2015)\citenamefont
  {Fradkin}, \citenamefont {Kivelson},\ and\ \citenamefont
  {Tranquada}}]{Fradkin2015}%
  \BibitemOpen
  \bibfield  {author} {\bibinfo {author} {\bibfnamefont {E.}~\bibnamefont
  {Fradkin}}, \bibinfo {author} {\bibfnamefont {S.~A.}\ \bibnamefont
  {Kivelson}},\ and\ \bibinfo {author} {\bibfnamefont {J.~M.}\ \bibnamefont
  {Tranquada}},\ }\bibfield  {title} {\bibinfo {title} {Colloquium: Theory of
  intertwined orders in high temperature superconductors},\ }\href
  {https://doi.org/10.1103/RevModPhys.87.457} {\bibfield  {journal} {\bibinfo
  {journal} {Rev. Mod. Phys.}\ }\textbf {\bibinfo {volume} {87}},\ \bibinfo
  {pages} {457} (\bibinfo {year} {2015})}\BibitemShut {NoStop}%
\bibitem [{\citenamefont {Berg}\ \emph
  {et~al.}(2009{\natexlab{b}})\citenamefont {Berg}, \citenamefont {Fradkin},\
  and\ \citenamefont {Kivelson}}]{Berg2009}%
  \BibitemOpen
  \bibfield  {author} {\bibinfo {author} {\bibfnamefont {E.}~\bibnamefont
  {Berg}}, \bibinfo {author} {\bibfnamefont {E.}~\bibnamefont {Fradkin}},\ and\
  \bibinfo {author} {\bibfnamefont {S.~A.}\ \bibnamefont {Kivelson}},\
  }\bibfield  {title} {\bibinfo {title} {Charge-4e superconductivity from
  pair-density-wave order in certain high-temperature superconductors},\ }\href
  {https://doi.org/10.1038/nphys1389} {\bibfield  {journal} {\bibinfo
  {journal} {Nature Physics}\ }\textbf {\bibinfo {volume} {5}},\ \bibinfo
  {pages} {830} (\bibinfo {year} {2009}{\natexlab{b}})}\BibitemShut {NoStop}%
\bibitem [{\citenamefont {Agterberg}\ \emph {et~al.}(2011)\citenamefont
  {Agterberg}, \citenamefont {Geracie},\ and\ \citenamefont
  {Tsunetsugu}}]{Agterberg2011}%
  \BibitemOpen
  \bibfield  {author} {\bibinfo {author} {\bibfnamefont {D.~F.}\ \bibnamefont
  {Agterberg}}, \bibinfo {author} {\bibfnamefont {M.}~\bibnamefont {Geracie}},\
  and\ \bibinfo {author} {\bibfnamefont {H.}~\bibnamefont {Tsunetsugu}},\
  }\bibfield  {title} {\bibinfo {title} {Conventional and charge-six
  superfluids from melting hexagonal fulde-ferrell-larkin-ovchinnikov phases in
  two dimensions},\ }\href {https://doi.org/10.1103/PhysRevB.84.014513}
  {\bibfield  {journal} {\bibinfo  {journal} {Phys. Rev. B}\ }\textbf {\bibinfo
  {volume} {84}},\ \bibinfo {pages} {014513} (\bibinfo {year}
  {2011})}\BibitemShut {NoStop}%
\bibitem [{\citenamefont {Ge}\ \emph {et~al.}(2022)\citenamefont {Ge},
  \citenamefont {Wang}, \citenamefont {Xing}, \citenamefont {Yin},
  \citenamefont {Lei}, \citenamefont {Wang},\ and\ \citenamefont
  {Wang}}]{ge2022}%
  \BibitemOpen
  \bibfield  {author} {\bibinfo {author} {\bibfnamefont {J.}~\bibnamefont
  {Ge}}, \bibinfo {author} {\bibfnamefont {P.}~\bibnamefont {Wang}}, \bibinfo
  {author} {\bibfnamefont {Y.}~\bibnamefont {Xing}}, \bibinfo {author}
  {\bibfnamefont {Q.}~\bibnamefont {Yin}}, \bibinfo {author} {\bibfnamefont
  {H.}~\bibnamefont {Lei}}, \bibinfo {author} {\bibfnamefont {Z.}~\bibnamefont
  {Wang}},\ and\ \bibinfo {author} {\bibfnamefont {J.}~\bibnamefont {Wang}},\
  }\bibfield  {title} {\bibinfo {title} {Discovery of charge-4e and charge-6e
  superconductivity in kagome superconductor csv3sb5}\ }\href
  {https://doi.org/10.48550/ARXIV.2201.10352} {10.48550/ARXIV.2201.10352}
  (\bibinfo {year} {2022})\BibitemShut {NoStop}%
\bibitem [{\citenamefont {Li}\ \emph {et~al.}(2010)\citenamefont {Li},
  \citenamefont {Luican}, \citenamefont {Lopes~dos Santos}, \citenamefont
  {Castro~Neto}, \citenamefont {Reina}, \citenamefont {Kong},\ and\
  \citenamefont {Andrei}}]{Li2010}%
  \BibitemOpen
  \bibfield  {author} {\bibinfo {author} {\bibfnamefont {G.}~\bibnamefont
  {Li}}, \bibinfo {author} {\bibfnamefont {A.}~\bibnamefont {Luican}}, \bibinfo
  {author} {\bibfnamefont {J.~M.~B.}\ \bibnamefont {Lopes~dos Santos}},
  \bibinfo {author} {\bibfnamefont {A.~H.}\ \bibnamefont {Castro~Neto}},
  \bibinfo {author} {\bibfnamefont {A.}~\bibnamefont {Reina}}, \bibinfo
  {author} {\bibfnamefont {J.}~\bibnamefont {Kong}},\ and\ \bibinfo {author}
  {\bibfnamefont {E.~Y.}\ \bibnamefont {Andrei}},\ }\bibfield  {title}
  {\bibinfo {title} {Observation of van hove singularities in twisted graphene
  layers},\ }\href {https://doi.org/10.1038/nphys1463} {\bibfield  {journal}
  {\bibinfo  {journal} {Nature Physics}\ }\textbf {\bibinfo {volume} {6}},\
  \bibinfo {pages} {109} (\bibinfo {year} {2010})}\BibitemShut {NoStop}%
\bibitem [{\citenamefont {Brihuega}\ \emph {et~al.}(2012)\citenamefont
  {Brihuega}, \citenamefont {Mallet}, \citenamefont {Gonz\'alez-Herrero},
  \citenamefont {Trambly~de Laissardi\`ere}, \citenamefont {Ugeda},
  \citenamefont {Magaud}, \citenamefont {G\'omez-Rodr\'{\i}guez}, \citenamefont
  {Yndur\'ain},\ and\ \citenamefont {Veuillen}}]{Brihuega2012}%
  \BibitemOpen
  \bibfield  {author} {\bibinfo {author} {\bibfnamefont {I.}~\bibnamefont
  {Brihuega}}, \bibinfo {author} {\bibfnamefont {P.}~\bibnamefont {Mallet}},
  \bibinfo {author} {\bibfnamefont {H.}~\bibnamefont {Gonz\'alez-Herrero}},
  \bibinfo {author} {\bibfnamefont {G.}~\bibnamefont {Trambly~de
  Laissardi\`ere}}, \bibinfo {author} {\bibfnamefont {M.~M.}\ \bibnamefont
  {Ugeda}}, \bibinfo {author} {\bibfnamefont {L.}~\bibnamefont {Magaud}},
  \bibinfo {author} {\bibfnamefont {J.~M.}\ \bibnamefont
  {G\'omez-Rodr\'{\i}guez}}, \bibinfo {author} {\bibfnamefont {F.}~\bibnamefont
  {Yndur\'ain}},\ and\ \bibinfo {author} {\bibfnamefont {J.-Y.}\ \bibnamefont
  {Veuillen}},\ }\bibfield  {title} {\bibinfo {title} {Unraveling the intrinsic
  and robust nature of van hove singularities in twisted bilayer graphene by
  scanning tunneling microscopy and theoretical analysis},\ }\href
  {https://doi.org/10.1103/PhysRevLett.109.196802} {\bibfield  {journal}
  {\bibinfo  {journal} {Phys. Rev. Lett.}\ }\textbf {\bibinfo {volume} {109}},\
  \bibinfo {pages} {196802} (\bibinfo {year} {2012})}\BibitemShut {NoStop}%
\bibitem [{\citenamefont {Wu}\ \emph {et~al.}(2021)\citenamefont {Wu},
  \citenamefont {Zhang}, \citenamefont {Watanabe}, \citenamefont {Taniguchi},\
  and\ \citenamefont {Andrei}}]{Wu2021}%
  \BibitemOpen
  \bibfield  {author} {\bibinfo {author} {\bibfnamefont {S.}~\bibnamefont
  {Wu}}, \bibinfo {author} {\bibfnamefont {Z.}~\bibnamefont {Zhang}}, \bibinfo
  {author} {\bibfnamefont {K.}~\bibnamefont {Watanabe}}, \bibinfo {author}
  {\bibfnamefont {T.}~\bibnamefont {Taniguchi}},\ and\ \bibinfo {author}
  {\bibfnamefont {E.~Y.}\ \bibnamefont {Andrei}},\ }\bibfield  {title}
  {\bibinfo {title} {Chern insulators, van hove singularities and topological
  flat bands in magic-angle twisted bilayer graphene},\ }\href
  {https://doi.org/10.1038/s41563-020-00911-2} {\bibfield  {journal} {\bibinfo
  {journal} {Nature Materials}\ }\textbf {\bibinfo {volume} {20}},\ \bibinfo
  {pages} {488} (\bibinfo {year} {2021})}\BibitemShut {NoStop}%
\bibitem [{\citenamefont {Furukawa}\ \emph {et~al.}(1998)\citenamefont
  {Furukawa}, \citenamefont {Rice},\ and\ \citenamefont
  {Salmhofer}}]{Furukawa}%
  \BibitemOpen
  \bibfield  {author} {\bibinfo {author} {\bibfnamefont {N.}~\bibnamefont
  {Furukawa}}, \bibinfo {author} {\bibfnamefont {T.~M.}\ \bibnamefont {Rice}},\
  and\ \bibinfo {author} {\bibfnamefont {M.}~\bibnamefont {Salmhofer}},\
  }\bibfield  {title} {\bibinfo {title} {Truncation of a two-dimensional fermi
  surface due to quasiparticle gap formation at the saddle points},\ }\href
  {https://doi.org/10.1103/PhysRevLett.81.3195} {\bibfield  {journal} {\bibinfo
   {journal} {Phys. Rev. Lett.}\ }\textbf {\bibinfo {volume} {81}},\ \bibinfo
  {pages} {3195} (\bibinfo {year} {1998})}\BibitemShut {NoStop}%
\bibitem [{\citenamefont {Nandkishore}\ \emph {et~al.}(2012)\citenamefont
  {Nandkishore}, \citenamefont {Levitov},\ and\ \citenamefont
  {Chubukov}}]{Nandkishore2012}%
  \BibitemOpen
  \bibfield  {author} {\bibinfo {author} {\bibfnamefont {R.}~\bibnamefont
  {Nandkishore}}, \bibinfo {author} {\bibfnamefont {L.~S.}\ \bibnamefont
  {Levitov}},\ and\ \bibinfo {author} {\bibfnamefont {A.~V.}\ \bibnamefont
  {Chubukov}},\ }\bibfield  {title} {\bibinfo {title} {Chiral superconductivity
  from repulsive interactions in doped graphene},\ }\href
  {https://doi.org/10.1038/nphys2208} {\bibfield  {journal} {\bibinfo
  {journal} {Nature Physics}\ }\textbf {\bibinfo {volume} {8}},\ \bibinfo
  {pages} {158} (\bibinfo {year} {2012})}\BibitemShut {NoStop}%
\bibitem [{\citenamefont {Isobe}\ \emph {et~al.}(2018)\citenamefont {Isobe},
  \citenamefont {Yuan},\ and\ \citenamefont {Fu}}]{PhysRevX.8.041041}%
  \BibitemOpen
  \bibfield  {author} {\bibinfo {author} {\bibfnamefont {H.}~\bibnamefont
  {Isobe}}, \bibinfo {author} {\bibfnamefont {N.~F.~Q.}\ \bibnamefont {Yuan}},\
  and\ \bibinfo {author} {\bibfnamefont {L.}~\bibnamefont {Fu}},\ }\bibfield
  {title} {\bibinfo {title} {Unconventional superconductivity and density waves
  in twisted bilayer graphene},\ }\href
  {https://doi.org/10.1103/PhysRevX.8.041041} {\bibfield  {journal} {\bibinfo
  {journal} {Phys. Rev. X}\ }\textbf {\bibinfo {volume} {8}},\ \bibinfo {pages}
  {041041} (\bibinfo {year} {2018})}\BibitemShut {NoStop}%
\bibitem [{\citenamefont {Lin}\ and\ \citenamefont
  {Nandkishore}(2019)}]{PhysRevB.100.085136}%
  \BibitemOpen
  \bibfield  {author} {\bibinfo {author} {\bibfnamefont {Y.-P.}\ \bibnamefont
  {Lin}}\ and\ \bibinfo {author} {\bibfnamefont {R.~M.}\ \bibnamefont
  {Nandkishore}},\ }\bibfield  {title} {\bibinfo {title} {Chiral twist on the
  high-${T}_{c}$ phase diagram in moir\'e heterostructures},\ }\href
  {https://doi.org/10.1103/PhysRevB.100.085136} {\bibfield  {journal} {\bibinfo
   {journal} {Phys. Rev. B}\ }\textbf {\bibinfo {volume} {100}},\ \bibinfo
  {pages} {085136} (\bibinfo {year} {2019})}\BibitemShut {NoStop}%
\bibitem [{\citenamefont {Hsu}\ \emph {et~al.}(2020)\citenamefont {Hsu},
  \citenamefont {Wu},\ and\ \citenamefont {Das~Sarma}}]{PhysRevB.102.085103}%
  \BibitemOpen
  \bibfield  {author} {\bibinfo {author} {\bibfnamefont {Y.-T.}\ \bibnamefont
  {Hsu}}, \bibinfo {author} {\bibfnamefont {F.}~\bibnamefont {Wu}},\ and\
  \bibinfo {author} {\bibfnamefont {S.}~\bibnamefont {Das~Sarma}},\ }\bibfield
  {title} {\bibinfo {title} {Topological superconductivity, ferromagnetism, and
  valley-polarized phases in moir\'e systems: Renormalization group analysis
  for twisted double bilayer graphene},\ }\href
  {https://doi.org/10.1103/PhysRevB.102.085103} {\bibfield  {journal} {\bibinfo
   {journal} {Phys. Rev. B}\ }\textbf {\bibinfo {volume} {102}},\ \bibinfo
  {pages} {085103} (\bibinfo {year} {2020})}\BibitemShut {NoStop}%
\bibitem [{\citenamefont {Hsu}\ \emph {et~al.}(2021)\citenamefont {Hsu},
  \citenamefont {Wu},\ and\ \citenamefont {Das~Sarma}}]{PhysRevB.104.195134}%
  \BibitemOpen
  \bibfield  {author} {\bibinfo {author} {\bibfnamefont {Y.-T.}\ \bibnamefont
  {Hsu}}, \bibinfo {author} {\bibfnamefont {F.}~\bibnamefont {Wu}},\ and\
  \bibinfo {author} {\bibfnamefont {S.}~\bibnamefont {Das~Sarma}},\ }\bibfield
  {title} {\bibinfo {title} {Spin-valley locked instabilities in moir\'e
  transition metal dichalcogenides with conventional and higher-order van hove
  singularities},\ }\href {https://doi.org/10.1103/PhysRevB.104.195134}
  {\bibfield  {journal} {\bibinfo  {journal} {Phys. Rev. B}\ }\textbf {\bibinfo
  {volume} {104}},\ \bibinfo {pages} {195134} (\bibinfo {year}
  {2021})}\BibitemShut {NoStop}%
\bibitem [{\citenamefont {Yuan}\ \emph {et~al.}(2019)\citenamefont {Yuan},
  \citenamefont {Isobe},\ and\ \citenamefont {Fu}}]{Yuan2019}%
  \BibitemOpen
  \bibfield  {author} {\bibinfo {author} {\bibfnamefont {N.~F.~Q.}\
  \bibnamefont {Yuan}}, \bibinfo {author} {\bibfnamefont {H.}~\bibnamefont
  {Isobe}},\ and\ \bibinfo {author} {\bibfnamefont {L.}~\bibnamefont {Fu}},\
  }\bibfield  {title} {\bibinfo {title} {Magic of high-order van hove
  singularity},\ }\href {https://doi.org/10.1038/s41467-019-13670-9} {\bibfield
   {journal} {\bibinfo  {journal} {Nature Communications}\ }\textbf {\bibinfo
  {volume} {10}},\ \bibinfo {pages} {5769} (\bibinfo {year}
  {2019})}\BibitemShut {NoStop}%
\bibitem [{\citenamefont {Wu}\ and\ \citenamefont
  {Das~Sarma}(2020)}]{Wufengcheng2020}%
  \BibitemOpen
  \bibfield  {author} {\bibinfo {author} {\bibfnamefont {F.}~\bibnamefont
  {Wu}}\ and\ \bibinfo {author} {\bibfnamefont {S.}~\bibnamefont {Das~Sarma}},\
  }\bibfield  {title} {\bibinfo {title} {Ferromagnetism and superconductivity
  in twisted double bilayer graphene},\ }\href
  {https://doi.org/10.1103/PhysRevB.101.155149} {\bibfield  {journal} {\bibinfo
   {journal} {Phys. Rev. B}\ }\textbf {\bibinfo {volume} {101}},\ \bibinfo
  {pages} {155149} (\bibinfo {year} {2020})}\BibitemShut {NoStop}%
\bibitem [{\citenamefont {Guerci}\ \emph {et~al.}(2022)\citenamefont {Guerci},
  \citenamefont {Simon},\ and\ \citenamefont {Mora}}]{Guerci}%
  \BibitemOpen
  \bibfield  {author} {\bibinfo {author} {\bibfnamefont {D.}~\bibnamefont
  {Guerci}}, \bibinfo {author} {\bibfnamefont {P.}~\bibnamefont {Simon}},\ and\
  \bibinfo {author} {\bibfnamefont {C.}~\bibnamefont {Mora}},\ }\bibfield
  {title} {\bibinfo {title} {Higher-order van hove singularity in magic-angle
  twisted trilayer graphene},\ }\href
  {https://doi.org/10.1103/PhysRevResearch.4.L012013} {\bibfield  {journal}
  {\bibinfo  {journal} {Phys. Rev. Research}\ }\textbf {\bibinfo {volume}
  {4}},\ \bibinfo {pages} {L012013} (\bibinfo {year} {2022})}\BibitemShut
  {NoStop}%
\bibitem [{\citenamefont {Diatlov}\ \emph {et~al.}(1957)\citenamefont
  {Diatlov}, \citenamefont {Sudakov},\ and\ \citenamefont
  {Ter-Martirosian}}]{osti_4338008}%
  \BibitemOpen
  \bibfield  {author} {\bibinfo {author} {\bibfnamefont {I.~T.}\ \bibnamefont
  {Diatlov}}, \bibinfo {author} {\bibfnamefont {V.~V.}\ \bibnamefont
  {Sudakov}},\ and\ \bibinfo {author} {\bibfnamefont {K.~A.}\ \bibnamefont
  {Ter-Martirosian}},\ }\bibfield  {title} {\bibinfo {title} {Asymptotic
  meson-meson scattering theory},\ }\href {https://www.osti.gov/biblio/4338008}
  {\bibfield  {journal} {\bibinfo  {journal} {Soviet Phys. JETP}\ }\textbf
  {\bibinfo {volume} {5}},\ \bibinfo {pages} {631} (\bibinfo {year}
  {1957})}\BibitemShut {NoStop}%
\bibitem [{\citenamefont {Sólyom}(1979)}]{oneD1979}%
  \BibitemOpen
  \bibfield  {author} {\bibinfo {author} {\bibfnamefont {J.}~\bibnamefont
  {Sólyom}},\ }\bibfield  {title} {\bibinfo {title} {The fermi gas model of
  one-dimensional conductors},\ }\href
  {https://doi.org/10.1080/00018737900101375} {\bibfield  {journal} {\bibinfo
  {journal} {Advances in Physics}\ }\textbf {\bibinfo {volume} {28}},\ \bibinfo
  {pages} {201} (\bibinfo {year} {1979})}\BibitemShut {NoStop}%
\bibitem [{\citenamefont {Zheleznyak}\ \emph {et~al.}(1997)\citenamefont
  {Zheleznyak}, \citenamefont {Yakovenko},\ and\ \citenamefont
  {Dzyaloshinskii}}]{Dzyaloshinskii1997}%
  \BibitemOpen
  \bibfield  {author} {\bibinfo {author} {\bibfnamefont {A.~T.}\ \bibnamefont
  {Zheleznyak}}, \bibinfo {author} {\bibfnamefont {V.~M.}\ \bibnamefont
  {Yakovenko}},\ and\ \bibinfo {author} {\bibfnamefont {I.~E.}\ \bibnamefont
  {Dzyaloshinskii}},\ }\bibfield  {title} {\bibinfo {title} {Parquet solution
  for a flat fermi surface},\ }\href {https://doi.org/10.1103/PhysRevB.55.3200}
  {\bibfield  {journal} {\bibinfo  {journal} {Phys. Rev. B}\ }\textbf {\bibinfo
  {volume} {55}},\ \bibinfo {pages} {3200} (\bibinfo {year}
  {1997})}\BibitemShut {NoStop}%
\bibitem [{\citenamefont {Chubukov}\ \emph {et~al.}(2008)\citenamefont
  {Chubukov}, \citenamefont {Efremov},\ and\ \citenamefont
  {Eremin}}]{Chubukov2008}%
  \BibitemOpen
  \bibfield  {author} {\bibinfo {author} {\bibfnamefont {A.~V.}\ \bibnamefont
  {Chubukov}}, \bibinfo {author} {\bibfnamefont {D.~V.}\ \bibnamefont
  {Efremov}},\ and\ \bibinfo {author} {\bibfnamefont {I.}~\bibnamefont
  {Eremin}},\ }\bibfield  {title} {\bibinfo {title} {Magnetism,
  superconductivity, and pairing symmetry in iron-based superconductors},\
  }\href {https://doi.org/10.1103/PhysRevB.78.134512} {\bibfield  {journal}
  {\bibinfo  {journal} {Phys. Rev. B}\ }\textbf {\bibinfo {volume} {78}},\
  \bibinfo {pages} {134512} (\bibinfo {year} {2008})}\BibitemShut {NoStop}%
\bibitem [{\citenamefont {Wu}\ \emph {et~al.}(2018)\citenamefont {Wu},
  \citenamefont {Lovorn}, \citenamefont {Tutuc},\ and\ \citenamefont
  {MacDonald}}]{Wufc2018}%
  \BibitemOpen
  \bibfield  {author} {\bibinfo {author} {\bibfnamefont {F.}~\bibnamefont
  {Wu}}, \bibinfo {author} {\bibfnamefont {T.}~\bibnamefont {Lovorn}}, \bibinfo
  {author} {\bibfnamefont {E.}~\bibnamefont {Tutuc}},\ and\ \bibinfo {author}
  {\bibfnamefont {A.~H.}\ \bibnamefont {MacDonald}},\ }\bibfield  {title}
  {\bibinfo {title} {Hubbard model physics in transition metal dichalcogenide
  moir\'e bands},\ }\href {https://doi.org/10.1103/PhysRevLett.121.026402}
  {\bibfield  {journal} {\bibinfo  {journal} {Phys. Rev. Lett.}\ }\textbf
  {\bibinfo {volume} {121}},\ \bibinfo {pages} {026402} (\bibinfo {year}
  {2018})}\BibitemShut {NoStop}%
\bibitem [{\citenamefont {Wu}\ \emph {et~al.}(2019)\citenamefont {Wu},
  \citenamefont {Lovorn}, \citenamefont {Tutuc}, \citenamefont {Martin},\ and\
  \citenamefont {MacDonald}}]{Wufc2019}%
  \BibitemOpen
  \bibfield  {author} {\bibinfo {author} {\bibfnamefont {F.}~\bibnamefont
  {Wu}}, \bibinfo {author} {\bibfnamefont {T.}~\bibnamefont {Lovorn}}, \bibinfo
  {author} {\bibfnamefont {E.}~\bibnamefont {Tutuc}}, \bibinfo {author}
  {\bibfnamefont {I.}~\bibnamefont {Martin}},\ and\ \bibinfo {author}
  {\bibfnamefont {A.~H.}\ \bibnamefont {MacDonald}},\ }\bibfield  {title}
  {\bibinfo {title} {Topological insulators in twisted transition metal
  dichalcogenide homobilayers},\ }\href
  {https://doi.org/10.1103/PhysRevLett.122.086402} {\bibfield  {journal}
  {\bibinfo  {journal} {Phys. Rev. Lett.}\ }\textbf {\bibinfo {volume} {122}},\
  \bibinfo {pages} {086402} (\bibinfo {year} {2019})}\BibitemShut {NoStop}%
\bibitem [{\citenamefont {Zhang}\ \emph {et~al.}(2020)\citenamefont {Zhang},
  \citenamefont {Wang}, \citenamefont {Watanabe}, \citenamefont {Taniguchi},
  \citenamefont {Ueno}, \citenamefont {Tutuc},\ and\ \citenamefont
  {LeRoy}}]{Zhang2020}%
  \BibitemOpen
  \bibfield  {author} {\bibinfo {author} {\bibfnamefont {Z.}~\bibnamefont
  {Zhang}}, \bibinfo {author} {\bibfnamefont {Y.}~\bibnamefont {Wang}},
  \bibinfo {author} {\bibfnamefont {K.}~\bibnamefont {Watanabe}}, \bibinfo
  {author} {\bibfnamefont {T.}~\bibnamefont {Taniguchi}}, \bibinfo {author}
  {\bibfnamefont {K.}~\bibnamefont {Ueno}}, \bibinfo {author} {\bibfnamefont
  {E.}~\bibnamefont {Tutuc}},\ and\ \bibinfo {author} {\bibfnamefont {B.~J.}\
  \bibnamefont {LeRoy}},\ }\bibfield  {title} {\bibinfo {title} {Flat bands in
  twisted bilayer transition metal dichalcogenides},\ }\href
  {https://doi.org/10.1038/s41567-020-0958-x} {\bibfield  {journal} {\bibinfo
  {journal} {Nature Physics}\ }\textbf {\bibinfo {volume} {16}},\ \bibinfo
  {pages} {1093} (\bibinfo {year} {2020})}\BibitemShut {NoStop}%
\bibitem [{\citenamefont {Shabani}\ \emph {et~al.}(2021)\citenamefont
  {Shabani}, \citenamefont {Halbertal}, \citenamefont {Wu}, \citenamefont
  {Chen}, \citenamefont {Liu}, \citenamefont {Hone}, \citenamefont {Yao},
  \citenamefont {Basov}, \citenamefont {Zhu},\ and\ \citenamefont
  {Pasupathy}}]{Shabani2021}%
  \BibitemOpen
  \bibfield  {author} {\bibinfo {author} {\bibfnamefont {S.}~\bibnamefont
  {Shabani}}, \bibinfo {author} {\bibfnamefont {D.}~\bibnamefont {Halbertal}},
  \bibinfo {author} {\bibfnamefont {W.}~\bibnamefont {Wu}}, \bibinfo {author}
  {\bibfnamefont {M.}~\bibnamefont {Chen}}, \bibinfo {author} {\bibfnamefont
  {S.}~\bibnamefont {Liu}}, \bibinfo {author} {\bibfnamefont {J.}~\bibnamefont
  {Hone}}, \bibinfo {author} {\bibfnamefont {W.}~\bibnamefont {Yao}}, \bibinfo
  {author} {\bibfnamefont {D.~N.}\ \bibnamefont {Basov}}, \bibinfo {author}
  {\bibfnamefont {X.}~\bibnamefont {Zhu}},\ and\ \bibinfo {author}
  {\bibfnamefont {A.~N.}\ \bibnamefont {Pasupathy}},\ }\bibfield  {title}
  {\bibinfo {title} {Deep moir{\'e} potentials in twisted transition metal
  dichalcogenide bilayers},\ }\href
  {https://doi.org/10.1038/s41567-021-01174-7} {\bibfield  {journal} {\bibinfo
  {journal} {Nature Physics}\ }\textbf {\bibinfo {volume} {17}},\ \bibinfo
  {pages} {720} (\bibinfo {year} {2021})}\BibitemShut {NoStop}%
\bibitem [{\citenamefont {Weston}\ \emph {et~al.}(2020)\citenamefont {Weston},
  \citenamefont {Zou}, \citenamefont {Enaldiev}, \citenamefont {Summerfield},
  \citenamefont {Clark}, \citenamefont {Z{\'o}lyomi}, \citenamefont {Graham},
  \citenamefont {Yelgel}, \citenamefont {Magorrian}, \citenamefont {Zhou},
  \citenamefont {Zultak}, \citenamefont {Hopkinson}, \citenamefont {Barinov},
  \citenamefont {Bointon}, \citenamefont {Kretinin}, \citenamefont {Wilson},
  \citenamefont {Beton}, \citenamefont {Fal'ko}, \citenamefont {Haigh},\ and\
  \citenamefont {Gorbachev}}]{Weston2020}%
  \BibitemOpen
  \bibfield  {author} {\bibinfo {author} {\bibfnamefont {A.}~\bibnamefont
  {Weston}}, \bibinfo {author} {\bibfnamefont {Y.}~\bibnamefont {Zou}},
  \bibinfo {author} {\bibfnamefont {V.}~\bibnamefont {Enaldiev}}, \bibinfo
  {author} {\bibfnamefont {A.}~\bibnamefont {Summerfield}}, \bibinfo {author}
  {\bibfnamefont {N.}~\bibnamefont {Clark}}, \bibinfo {author} {\bibfnamefont
  {V.}~\bibnamefont {Z{\'o}lyomi}}, \bibinfo {author} {\bibfnamefont
  {A.}~\bibnamefont {Graham}}, \bibinfo {author} {\bibfnamefont
  {C.}~\bibnamefont {Yelgel}}, \bibinfo {author} {\bibfnamefont
  {S.}~\bibnamefont {Magorrian}}, \bibinfo {author} {\bibfnamefont
  {M.}~\bibnamefont {Zhou}}, \bibinfo {author} {\bibfnamefont {J.}~\bibnamefont
  {Zultak}}, \bibinfo {author} {\bibfnamefont {D.}~\bibnamefont {Hopkinson}},
  \bibinfo {author} {\bibfnamefont {A.}~\bibnamefont {Barinov}}, \bibinfo
  {author} {\bibfnamefont {T.~H.}\ \bibnamefont {Bointon}}, \bibinfo {author}
  {\bibfnamefont {A.}~\bibnamefont {Kretinin}}, \bibinfo {author}
  {\bibfnamefont {N.~R.}\ \bibnamefont {Wilson}}, \bibinfo {author}
  {\bibfnamefont {P.~H.}\ \bibnamefont {Beton}}, \bibinfo {author}
  {\bibfnamefont {V.~I.}\ \bibnamefont {Fal'ko}}, \bibinfo {author}
  {\bibfnamefont {S.~J.}\ \bibnamefont {Haigh}},\ and\ \bibinfo {author}
  {\bibfnamefont {R.}~\bibnamefont {Gorbachev}},\ }\bibfield  {title} {\bibinfo
  {title} {Atomic reconstruction in twisted bilayers of transition metal
  dichalcogenides},\ }\href {https://doi.org/10.1038/s41565-020-0682-9}
  {\bibfield  {journal} {\bibinfo  {journal} {Nature Nanotechnology}\ }\textbf
  {\bibinfo {volume} {15}},\ \bibinfo {pages} {592} (\bibinfo {year}
  {2020})}\BibitemShut {NoStop}%
\bibitem [{\citenamefont {Devakul}\ \emph {et~al.}(2021)\citenamefont
  {Devakul}, \citenamefont {Cr{\'e}pel}, \citenamefont {Zhang},\ and\
  \citenamefont {Fu}}]{Devakul2021}%
  \BibitemOpen
  \bibfield  {author} {\bibinfo {author} {\bibfnamefont {T.}~\bibnamefont
  {Devakul}}, \bibinfo {author} {\bibfnamefont {V.}~\bibnamefont {Cr{\'e}pel}},
  \bibinfo {author} {\bibfnamefont {Y.}~\bibnamefont {Zhang}},\ and\ \bibinfo
  {author} {\bibfnamefont {L.}~\bibnamefont {Fu}},\ }\bibfield  {title}
  {\bibinfo {title} {Magic in twisted transition metal dichalcogenide
  bilayers},\ }\href {https://doi.org/10.1038/s41467-021-27042-9} {\bibfield
  {journal} {\bibinfo  {journal} {Nature Communications}\ }\textbf {\bibinfo
  {volume} {12}},\ \bibinfo {pages} {6730} (\bibinfo {year}
  {2021})}\BibitemShut {NoStop}%
\bibitem [{\citenamefont {Zhang}\ \emph {et~al.}(2021)\citenamefont {Zhang},
  \citenamefont {Liu},\ and\ \citenamefont {Fu}}]{ZhangYang2021}%
  \BibitemOpen
  \bibfield  {author} {\bibinfo {author} {\bibfnamefont {Y.}~\bibnamefont
  {Zhang}}, \bibinfo {author} {\bibfnamefont {T.}~\bibnamefont {Liu}},\ and\
  \bibinfo {author} {\bibfnamefont {L.}~\bibnamefont {Fu}},\ }\bibfield
  {title} {\bibinfo {title} {Electronic structures, charge transfer, and charge
  order in twisted transition metal dichalcogenide bilayers},\ }\href
  {https://doi.org/10.1103/PhysRevB.103.155142} {\bibfield  {journal} {\bibinfo
   {journal} {Phys. Rev. B}\ }\textbf {\bibinfo {volume} {103}},\ \bibinfo
  {pages} {155142} (\bibinfo {year} {2021})}\BibitemShut {NoStop}%
\bibitem [{\citenamefont {Angeli}\ and\ \citenamefont
  {MacDonald}(2021)}]{doi:10.1073/pnas.2021826118}%
  \BibitemOpen
  \bibfield  {author} {\bibinfo {author} {\bibfnamefont {M.}~\bibnamefont
  {Angeli}}\ and\ \bibinfo {author} {\bibfnamefont {A.~H.}\ \bibnamefont
  {MacDonald}},\ }\bibfield  {title} {\bibinfo {title} {Gamma valley transition
  metal dichalcogenide moir\'e bands},\ }\href
  {https://doi.org/10.1073/pnas.2021826118} {\bibfield  {journal} {\bibinfo
  {journal} {Proceedings of the National Academy of Sciences}\ }\textbf
  {\bibinfo {volume} {118}},\ \bibinfo {pages} {e2021826118} (\bibinfo {year}
  {2021})}\BibitemShut {NoStop}%
\bibitem [{\citenamefont {Tran}\ \emph {et~al.}(2020)\citenamefont {Tran},
  \citenamefont {Choi},\ and\ \citenamefont {Singh}}]{Tran_2020}%
  \BibitemOpen
  \bibfield  {author} {\bibinfo {author} {\bibfnamefont {K.}~\bibnamefont
  {Tran}}, \bibinfo {author} {\bibfnamefont {J.}~\bibnamefont {Choi}},\ and\
  \bibinfo {author} {\bibfnamefont {A.}~\bibnamefont {Singh}},\ }\bibfield
  {title} {\bibinfo {title} {Moir{\'{e}} and beyond in transition metal
  dichalcogenide twisted bilayers},\ }\href
  {https://doi.org/10.1088/2053-1583/abd3e7} {\bibfield  {journal} {\bibinfo
  {journal} {2D Materials}\ }\textbf {\bibinfo {volume} {8}},\ \bibinfo {pages}
  {022002} (\bibinfo {year} {2020})}\BibitemShut {NoStop}%
\bibitem [{\citenamefont {Vitale}\ \emph {et~al.}(2021)\citenamefont {Vitale},
  \citenamefont {Atalar}, \citenamefont {Mostofi},\ and\ \citenamefont
  {Lischner}}]{Vitale_2021}%
  \BibitemOpen
  \bibfield  {author} {\bibinfo {author} {\bibfnamefont {V.}~\bibnamefont
  {Vitale}}, \bibinfo {author} {\bibfnamefont {K.}~\bibnamefont {Atalar}},
  \bibinfo {author} {\bibfnamefont {A.~A.}\ \bibnamefont {Mostofi}},\ and\
  \bibinfo {author} {\bibfnamefont {J.}~\bibnamefont {Lischner}},\ }\bibfield
  {title} {\bibinfo {title} {Flat band properties of twisted transition metal
  dichalcogenide homo- and heterobilayers of {MoS}2, {MoSe}2, {WS}2 and
  {WSe}2},\ }\href {https://doi.org/10.1088/2053-1583/ac15d9} {\bibfield
  {journal} {\bibinfo  {journal} {2D Materials}\ }\textbf {\bibinfo {volume}
  {8}},\ \bibinfo {pages} {045010} (\bibinfo {year} {2021})}\BibitemShut
  {NoStop}%
\bibitem [{\citenamefont {Bi}\ and\ \citenamefont {Fu}(2021)}]{Bi2021}%
  \BibitemOpen
  \bibfield  {author} {\bibinfo {author} {\bibfnamefont {Z.}~\bibnamefont
  {Bi}}\ and\ \bibinfo {author} {\bibfnamefont {L.}~\bibnamefont {Fu}},\
  }\bibfield  {title} {\bibinfo {title} {Excitonic density wave and spin-valley
  superfluid in bilayer transition metal dichalcogenide},\ }\href
  {https://doi.org/10.1038/s41467-020-20802-z} {\bibfield  {journal} {\bibinfo
  {journal} {Nature Communications}\ }\textbf {\bibinfo {volume} {12}},\
  \bibinfo {pages} {642} (\bibinfo {year} {2021})}\BibitemShut {NoStop}%
\bibitem [{\citenamefont {Scherer}\ \emph {et~al.}(2021)\citenamefont
  {Scherer}, \citenamefont {Kennes},\ and\ \citenamefont
  {Classen}}]{Scherer2021}%
  \BibitemOpen
  \bibfield  {author} {\bibinfo {author} {\bibfnamefont {M.~M.}\ \bibnamefont
  {Scherer}}, \bibinfo {author} {\bibfnamefont {D.~M.}\ \bibnamefont
  {Kennes}},\ and\ \bibinfo {author} {\bibfnamefont {L.}~\bibnamefont
  {Classen}},\ }\bibfield  {title} {\bibinfo {title} {$\mathcal{N}=4$ chiral
  superconductivity in moiré transition metal dichalcogenides}\ }\href
  {https://doi.org/10.48550/ARXIV.2108.11406} {10.48550/ARXIV.2108.11406}
  (\bibinfo {year} {2021})\BibitemShut {NoStop}%
\bibitem [{\citenamefont {Schrade}\ and\ \citenamefont
  {Fu}(2019)}]{Schrade2019}%
  \BibitemOpen
  \bibfield  {author} {\bibinfo {author} {\bibfnamefont {C.}~\bibnamefont
  {Schrade}}\ and\ \bibinfo {author} {\bibfnamefont {L.}~\bibnamefont {Fu}},\
  }\bibfield  {title} {\bibinfo {title} {Spin-valley density wave in moir\'e
  materials},\ }\href {https://doi.org/10.1103/PhysRevB.100.035413} {\bibfield
  {journal} {\bibinfo  {journal} {Phys. Rev. B}\ }\textbf {\bibinfo {volume}
  {100}},\ \bibinfo {pages} {035413} (\bibinfo {year} {2019})}\BibitemShut
  {NoStop}%
\bibitem [{\citenamefont {Zhang}\ and\ \citenamefont
  {Senthil}(2019)}]{PhysRevB.99.205150}%
  \BibitemOpen
  \bibfield  {author} {\bibinfo {author} {\bibfnamefont {Y.-H.}\ \bibnamefont
  {Zhang}}\ and\ \bibinfo {author} {\bibfnamefont {T.}~\bibnamefont
  {Senthil}},\ }\bibfield  {title} {\bibinfo {title} {Bridging hubbard model
  physics and quantum hall physics in trilayer
  $\text{graphene}/h\ensuremath{-}\mathrm{BN}$ moir\'e superlattice},\ }\href
  {https://doi.org/10.1103/PhysRevB.99.205150} {\bibfield  {journal} {\bibinfo
  {journal} {Phys. Rev. B}\ }\textbf {\bibinfo {volume} {99}},\ \bibinfo
  {pages} {205150} (\bibinfo {year} {2019})}\BibitemShut {NoStop}%
\bibitem [{\citenamefont {Zhang}\ and\ \citenamefont
  {Mao}(2020)}]{PhysRevB.101.035122}%
  \BibitemOpen
  \bibfield  {author} {\bibinfo {author} {\bibfnamefont {Y.-H.}\ \bibnamefont
  {Zhang}}\ and\ \bibinfo {author} {\bibfnamefont {D.}~\bibnamefont {Mao}},\
  }\bibfield  {title} {\bibinfo {title} {Spin liquids and pseudogap metals in
  the su(4) hubbard model in a moir\'e superlattice},\ }\href
  {https://doi.org/10.1103/PhysRevB.101.035122} {\bibfield  {journal} {\bibinfo
   {journal} {Phys. Rev. B}\ }\textbf {\bibinfo {volume} {101}},\ \bibinfo
  {pages} {035122} (\bibinfo {year} {2020})}\BibitemShut {NoStop}%
\bibitem [{\citenamefont {Zhang}\ \emph {et~al.}(2019)\citenamefont {Zhang},
  \citenamefont {Mao}, \citenamefont {Cao}, \citenamefont {Jarillo-Herrero},\
  and\ \citenamefont {Senthil}}]{PhysRevB.99.075127}%
  \BibitemOpen
  \bibfield  {author} {\bibinfo {author} {\bibfnamefont {Y.-H.}\ \bibnamefont
  {Zhang}}, \bibinfo {author} {\bibfnamefont {D.}~\bibnamefont {Mao}}, \bibinfo
  {author} {\bibfnamefont {Y.}~\bibnamefont {Cao}}, \bibinfo {author}
  {\bibfnamefont {P.}~\bibnamefont {Jarillo-Herrero}},\ and\ \bibinfo {author}
  {\bibfnamefont {T.}~\bibnamefont {Senthil}},\ }\bibfield  {title} {\bibinfo
  {title} {Nearly flat chern bands in moir\'e superlattices},\ }\href
  {https://doi.org/10.1103/PhysRevB.99.075127} {\bibfield  {journal} {\bibinfo
  {journal} {Phys. Rev. B}\ }\textbf {\bibinfo {volume} {99}},\ \bibinfo
  {pages} {075127} (\bibinfo {year} {2019})}\BibitemShut {NoStop}%
\bibitem [{\citenamefont {Chebrolu}\ \emph {et~al.}(2019)\citenamefont
  {Chebrolu}, \citenamefont {Chittari},\ and\ \citenamefont
  {Jung}}]{PhysRevB.99.235417}%
  \BibitemOpen
  \bibfield  {author} {\bibinfo {author} {\bibfnamefont {N.~R.}\ \bibnamefont
  {Chebrolu}}, \bibinfo {author} {\bibfnamefont {B.~L.}\ \bibnamefont
  {Chittari}},\ and\ \bibinfo {author} {\bibfnamefont {J.}~\bibnamefont
  {Jung}},\ }\bibfield  {title} {\bibinfo {title} {Flat bands in twisted double
  bilayer graphene},\ }\href {https://doi.org/10.1103/PhysRevB.99.235417}
  {\bibfield  {journal} {\bibinfo  {journal} {Phys. Rev. B}\ }\textbf {\bibinfo
  {volume} {99}},\ \bibinfo {pages} {235417} (\bibinfo {year}
  {2019})}\BibitemShut {NoStop}%
\bibitem [{\citenamefont {Koshino}(2019)}]{PhysRevB.99.235406}%
  \BibitemOpen
  \bibfield  {author} {\bibinfo {author} {\bibfnamefont {M.}~\bibnamefont
  {Koshino}},\ }\bibfield  {title} {\bibinfo {title} {Band structure and
  topological properties of twisted double bilayer graphene},\ }\href
  {https://doi.org/10.1103/PhysRevB.99.235406} {\bibfield  {journal} {\bibinfo
  {journal} {Phys. Rev. B}\ }\textbf {\bibinfo {volume} {99}},\ \bibinfo
  {pages} {235406} (\bibinfo {year} {2019})}\BibitemShut {NoStop}%
\bibitem [{\citenamefont {Liu}\ \emph {et~al.}(2019)\citenamefont {Liu},
  \citenamefont {Ma}, \citenamefont {Gao},\ and\ \citenamefont
  {Dai}}]{PhysRevX.9.031021}%
  \BibitemOpen
  \bibfield  {author} {\bibinfo {author} {\bibfnamefont {J.}~\bibnamefont
  {Liu}}, \bibinfo {author} {\bibfnamefont {Z.}~\bibnamefont {Ma}}, \bibinfo
  {author} {\bibfnamefont {J.}~\bibnamefont {Gao}},\ and\ \bibinfo {author}
  {\bibfnamefont {X.}~\bibnamefont {Dai}},\ }\bibfield  {title} {\bibinfo
  {title} {Quantum valley hall effect, orbital magnetism, and anomalous hall
  effect in twisted multilayer graphene systems},\ }\href
  {https://doi.org/10.1103/PhysRevX.9.031021} {\bibfield  {journal} {\bibinfo
  {journal} {Phys. Rev. X}\ }\textbf {\bibinfo {volume} {9}},\ \bibinfo {pages}
  {031021} (\bibinfo {year} {2019})}\BibitemShut {NoStop}%
\bibitem [{\citenamefont {Lee}\ \emph {et~al.}(2019)\citenamefont {Lee},
  \citenamefont {Khalaf}, \citenamefont {Liu}, \citenamefont {Liu},
  \citenamefont {Hao}, \citenamefont {Kim},\ and\ \citenamefont
  {Vishwanath}}]{Lee2019}%
  \BibitemOpen
  \bibfield  {author} {\bibinfo {author} {\bibfnamefont {J.~Y.}\ \bibnamefont
  {Lee}}, \bibinfo {author} {\bibfnamefont {E.}~\bibnamefont {Khalaf}},
  \bibinfo {author} {\bibfnamefont {S.}~\bibnamefont {Liu}}, \bibinfo {author}
  {\bibfnamefont {X.}~\bibnamefont {Liu}}, \bibinfo {author} {\bibfnamefont
  {Z.}~\bibnamefont {Hao}}, \bibinfo {author} {\bibfnamefont {P.}~\bibnamefont
  {Kim}},\ and\ \bibinfo {author} {\bibfnamefont {A.}~\bibnamefont
  {Vishwanath}},\ }\bibfield  {title} {\bibinfo {title} {Theory of correlated
  insulating behaviour and spin-triplet superconductivity in twisted double
  bilayer graphene},\ }\href {https://doi.org/10.1038/s41467-019-12981-1}
  {\bibfield  {journal} {\bibinfo  {journal} {Nature Communications}\ }\textbf
  {\bibinfo {volume} {10}},\ \bibinfo {pages} {5333} (\bibinfo {year}
  {2019})}\BibitemShut {NoStop}%
\bibitem [{\citenamefont {Haddadi}\ \emph {et~al.}(2020)\citenamefont
  {Haddadi}, \citenamefont {Wu}, \citenamefont {Kruchkov},\ and\ \citenamefont
  {Yazyev}}]{Haddadi2020}%
  \BibitemOpen
  \bibfield  {author} {\bibinfo {author} {\bibfnamefont {F.}~\bibnamefont
  {Haddadi}}, \bibinfo {author} {\bibfnamefont {Q.}~\bibnamefont {Wu}},
  \bibinfo {author} {\bibfnamefont {A.~J.}\ \bibnamefont {Kruchkov}},\ and\
  \bibinfo {author} {\bibfnamefont {O.~V.}\ \bibnamefont {Yazyev}},\ }\bibfield
   {title} {\bibinfo {title} {Moir{\'e} flat bands in twisted double bilayer
  graphene},\ }\href {https://doi.org/10.1021/acs.nanolett.9b05117} {\bibfield
  {journal} {\bibinfo  {journal} {Nano Letters}\ }\textbf {\bibinfo {volume}
  {20}},\ \bibinfo {pages} {2410} (\bibinfo {year} {2020})}\BibitemShut
  {NoStop}%
\bibitem [{\citenamefont {Koshino}\ \emph {et~al.}(2018)\citenamefont
  {Koshino}, \citenamefont {Yuan}, \citenamefont {Koretsune}, \citenamefont
  {Ochi}, \citenamefont {Kuroki},\ and\ \citenamefont
  {Fu}}]{PhysRevX.8.031087}%
  \BibitemOpen
  \bibfield  {author} {\bibinfo {author} {\bibfnamefont {M.}~\bibnamefont
  {Koshino}}, \bibinfo {author} {\bibfnamefont {N.~F.~Q.}\ \bibnamefont
  {Yuan}}, \bibinfo {author} {\bibfnamefont {T.}~\bibnamefont {Koretsune}},
  \bibinfo {author} {\bibfnamefont {M.}~\bibnamefont {Ochi}}, \bibinfo {author}
  {\bibfnamefont {K.}~\bibnamefont {Kuroki}},\ and\ \bibinfo {author}
  {\bibfnamefont {L.}~\bibnamefont {Fu}},\ }\bibfield  {title} {\bibinfo
  {title} {Maximally localized wannier orbitals and the extended hubbard model
  for twisted bilayer graphene},\ }\href
  {https://doi.org/10.1103/PhysRevX.8.031087} {\bibfield  {journal} {\bibinfo
  {journal} {Phys. Rev. X}\ }\textbf {\bibinfo {volume} {8}},\ \bibinfo {pages}
  {031087} (\bibinfo {year} {2018})}\BibitemShut {NoStop}%
\bibitem [{\citenamefont {Yuan}\ and\ \citenamefont
  {Fu}(2018)}]{PhysRevB.98.045103}%
  \BibitemOpen
  \bibfield  {author} {\bibinfo {author} {\bibfnamefont {N.~F.~Q.}\
  \bibnamefont {Yuan}}\ and\ \bibinfo {author} {\bibfnamefont {L.}~\bibnamefont
  {Fu}},\ }\bibfield  {title} {\bibinfo {title} {Model for the metal-insulator
  transition in graphene superlattices and beyond},\ }\href
  {https://doi.org/10.1103/PhysRevB.98.045103} {\bibfield  {journal} {\bibinfo
  {journal} {Phys. Rev. B}\ }\textbf {\bibinfo {volume} {98}},\ \bibinfo
  {pages} {045103} (\bibinfo {year} {2018})}\BibitemShut {NoStop}%
\bibitem [{\citenamefont {Kang}\ and\ \citenamefont
  {Vafek}(2018)}]{PhysRevX.8.031088}%
  \BibitemOpen
  \bibfield  {author} {\bibinfo {author} {\bibfnamefont {J.}~\bibnamefont
  {Kang}}\ and\ \bibinfo {author} {\bibfnamefont {O.}~\bibnamefont {Vafek}},\
  }\bibfield  {title} {\bibinfo {title} {Symmetry, maximally localized wannier
  states, and a low-energy model for twisted bilayer graphene narrow bands},\
  }\href {https://doi.org/10.1103/PhysRevX.8.031088} {\bibfield  {journal}
  {\bibinfo  {journal} {Phys. Rev. X}\ }\textbf {\bibinfo {volume} {8}},\
  \bibinfo {pages} {031088} (\bibinfo {year} {2018})}\BibitemShut {NoStop}%
\bibitem [{\citenamefont {Chichinadze}\ \emph {et~al.}(2022)\citenamefont
  {Chichinadze}, \citenamefont {Classen}, \citenamefont {Wang},\ and\
  \citenamefont {Chubukov}}]{Dmitry2022}%
  \BibitemOpen
  \bibfield  {author} {\bibinfo {author} {\bibfnamefont {D.~V.}\ \bibnamefont
  {Chichinadze}}, \bibinfo {author} {\bibfnamefont {L.}~\bibnamefont
  {Classen}}, \bibinfo {author} {\bibfnamefont {Y.}~\bibnamefont {Wang}},\ and\
  \bibinfo {author} {\bibfnamefont {A.~V.}\ \bibnamefont {Chubukov}},\
  }\bibfield  {title} {\bibinfo {title} {Su(4) symmetry in twisted bilayer
  graphene: An itinerant perspective},\ }\href
  {https://doi.org/10.1103/PhysRevLett.128.227601} {\bibfield  {journal}
  {\bibinfo  {journal} {Phys. Rev. Lett.}\ }\textbf {\bibinfo {volume} {128}},\
  \bibinfo {pages} {227601} (\bibinfo {year} {2022})}\BibitemShut {NoStop}%
\bibitem [{\citenamefont {Wang}\ \emph {et~al.}(2020)\citenamefont {Wang},
  \citenamefont {Shih}, \citenamefont {Ghiotto}, \citenamefont {Xian},
  \citenamefont {Rhodes}, \citenamefont {Tan}, \citenamefont {Claassen},
  \citenamefont {Kennes}, \citenamefont {Bai}, \citenamefont {Kim},
  \citenamefont {Watanabe}, \citenamefont {Taniguchi}, \citenamefont {Zhu},
  \citenamefont {Hone}, \citenamefont {Rubio}, \citenamefont {Pasupathy},\ and\
  \citenamefont {Dean}}]{Wang2020}%
  \BibitemOpen
  \bibfield  {author} {\bibinfo {author} {\bibfnamefont {L.}~\bibnamefont
  {Wang}}, \bibinfo {author} {\bibfnamefont {E.-M.}\ \bibnamefont {Shih}},
  \bibinfo {author} {\bibfnamefont {A.}~\bibnamefont {Ghiotto}}, \bibinfo
  {author} {\bibfnamefont {L.}~\bibnamefont {Xian}}, \bibinfo {author}
  {\bibfnamefont {D.~A.}\ \bibnamefont {Rhodes}}, \bibinfo {author}
  {\bibfnamefont {C.}~\bibnamefont {Tan}}, \bibinfo {author} {\bibfnamefont
  {M.}~\bibnamefont {Claassen}}, \bibinfo {author} {\bibfnamefont {D.~M.}\
  \bibnamefont {Kennes}}, \bibinfo {author} {\bibfnamefont {Y.}~\bibnamefont
  {Bai}}, \bibinfo {author} {\bibfnamefont {B.}~\bibnamefont {Kim}}, \bibinfo
  {author} {\bibfnamefont {K.}~\bibnamefont {Watanabe}}, \bibinfo {author}
  {\bibfnamefont {T.}~\bibnamefont {Taniguchi}}, \bibinfo {author}
  {\bibfnamefont {X.}~\bibnamefont {Zhu}}, \bibinfo {author} {\bibfnamefont
  {J.}~\bibnamefont {Hone}}, \bibinfo {author} {\bibfnamefont {A.}~\bibnamefont
  {Rubio}}, \bibinfo {author} {\bibfnamefont {A.~N.}\ \bibnamefont
  {Pasupathy}},\ and\ \bibinfo {author} {\bibfnamefont {C.~R.}\ \bibnamefont
  {Dean}},\ }\bibfield  {title} {\bibinfo {title} {Correlated electronic phases
  in twisted bilayer transition metal dichalcogenides},\ }\href
  {https://doi.org/10.1038/s41563-020-0708-6} {\bibfield  {journal} {\bibinfo
  {journal} {Nature Materials}\ }\textbf {\bibinfo {volume} {19}},\ \bibinfo
  {pages} {861} (\bibinfo {year} {2020})}\BibitemShut {NoStop}%
\bibitem [{\citenamefont {Pan}\ \emph {et~al.}(2020)\citenamefont {Pan},
  \citenamefont {Wu},\ and\ \citenamefont {Das~Sarma}}]{Pan2020}%
  \BibitemOpen
  \bibfield  {author} {\bibinfo {author} {\bibfnamefont {H.}~\bibnamefont
  {Pan}}, \bibinfo {author} {\bibfnamefont {F.}~\bibnamefont {Wu}},\ and\
  \bibinfo {author} {\bibfnamefont {S.}~\bibnamefont {Das~Sarma}},\ }\bibfield
  {title} {\bibinfo {title} {Band topology, hubbard model, heisenberg model,
  and dzyaloshinskii-moriya interaction in twisted bilayer
  ${\mathrm{wse}}_{2}$},\ }\href
  {https://doi.org/10.1103/PhysRevResearch.2.033087} {\bibfield  {journal}
  {\bibinfo  {journal} {Phys. Rev. Research}\ }\textbf {\bibinfo {volume}
  {2}},\ \bibinfo {pages} {033087} (\bibinfo {year} {2020})}\BibitemShut
  {NoStop}%
\bibitem [{\citenamefont {Ghiotto}\ \emph {et~al.}(2021)\citenamefont
  {Ghiotto}, \citenamefont {Shih}, \citenamefont {Pereira}, \citenamefont
  {Rhodes}, \citenamefont {Kim}, \citenamefont {Zang}, \citenamefont {Millis},
  \citenamefont {Watanabe}, \citenamefont {Taniguchi}, \citenamefont {Hone},
  \citenamefont {Wang}, \citenamefont {Dean},\ and\ \citenamefont
  {Pasupathy}}]{Ghiotto2021}%
  \BibitemOpen
  \bibfield  {author} {\bibinfo {author} {\bibfnamefont {A.}~\bibnamefont
  {Ghiotto}}, \bibinfo {author} {\bibfnamefont {E.-M.}\ \bibnamefont {Shih}},
  \bibinfo {author} {\bibfnamefont {G.~S. S.~G.}\ \bibnamefont {Pereira}},
  \bibinfo {author} {\bibfnamefont {D.~A.}\ \bibnamefont {Rhodes}}, \bibinfo
  {author} {\bibfnamefont {B.}~\bibnamefont {Kim}}, \bibinfo {author}
  {\bibfnamefont {J.}~\bibnamefont {Zang}}, \bibinfo {author} {\bibfnamefont
  {A.~J.}\ \bibnamefont {Millis}}, \bibinfo {author} {\bibfnamefont
  {K.}~\bibnamefont {Watanabe}}, \bibinfo {author} {\bibfnamefont
  {T.}~\bibnamefont {Taniguchi}}, \bibinfo {author} {\bibfnamefont {J.~C.}\
  \bibnamefont {Hone}}, \bibinfo {author} {\bibfnamefont {L.}~\bibnamefont
  {Wang}}, \bibinfo {author} {\bibfnamefont {C.~R.}\ \bibnamefont {Dean}},\
  and\ \bibinfo {author} {\bibfnamefont {A.~N.}\ \bibnamefont {Pasupathy}},\
  }\bibfield  {title} {\bibinfo {title} {Quantum criticality in twisted
  transition metal dichalcogenides},\ }\href
  {https://doi.org/10.1038/s41586-021-03815-6} {\bibfield  {journal} {\bibinfo
  {journal} {Nature}\ }\textbf {\bibinfo {volume} {597}},\ \bibinfo {pages}
  {345} (\bibinfo {year} {2021})}\BibitemShut {NoStop}%
\bibitem [{\citenamefont {Zang}\ \emph
  {et~al.}(2021{\natexlab{a}})\citenamefont {Zang}, \citenamefont {Wang},
  \citenamefont {Cano},\ and\ \citenamefont {Millis}}]{Zhang2021}%
  \BibitemOpen
  \bibfield  {author} {\bibinfo {author} {\bibfnamefont {J.}~\bibnamefont
  {Zang}}, \bibinfo {author} {\bibfnamefont {J.}~\bibnamefont {Wang}}, \bibinfo
  {author} {\bibfnamefont {J.}~\bibnamefont {Cano}},\ and\ \bibinfo {author}
  {\bibfnamefont {A.~J.}\ \bibnamefont {Millis}},\ }\bibfield  {title}
  {\bibinfo {title} {Hartree-fock study of the moir\'e hubbard model for
  twisted bilayer transition metal dichalcogenides},\ }\href
  {https://doi.org/10.1103/PhysRevB.104.075150} {\bibfield  {journal} {\bibinfo
   {journal} {Phys. Rev. B}\ }\textbf {\bibinfo {volume} {104}},\ \bibinfo
  {pages} {075150} (\bibinfo {year} {2021}{\natexlab{a}})}\BibitemShut
  {NoStop}%
\bibitem [{\citenamefont {Haldane}(1988)}]{PhysRevLett.61.2015}%
  \BibitemOpen
  \bibfield  {author} {\bibinfo {author} {\bibfnamefont {F.~D.~M.}\
  \bibnamefont {Haldane}},\ }\bibfield  {title} {\bibinfo {title} {Model for a
  quantum hall effect without landau levels: Condensed-matter realization of
  the "parity anomaly"},\ }\href {https://doi.org/10.1103/PhysRevLett.61.2015}
  {\bibfield  {journal} {\bibinfo  {journal} {Phys. Rev. Lett.}\ }\textbf
  {\bibinfo {volume} {61}},\ \bibinfo {pages} {2015} (\bibinfo {year}
  {1988})}\BibitemShut {NoStop}%
\bibitem [{\citenamefont {Zang}\ \emph
  {et~al.}(2021{\natexlab{b}})\citenamefont {Zang}, \citenamefont {Wang},
  \citenamefont {Cano},\ and\ \citenamefont {Millis}}]{PhysRevB.104.075150}%
  \BibitemOpen
  \bibfield  {author} {\bibinfo {author} {\bibfnamefont {J.}~\bibnamefont
  {Zang}}, \bibinfo {author} {\bibfnamefont {J.}~\bibnamefont {Wang}}, \bibinfo
  {author} {\bibfnamefont {J.}~\bibnamefont {Cano}},\ and\ \bibinfo {author}
  {\bibfnamefont {A.~J.}\ \bibnamefont {Millis}},\ }\bibfield  {title}
  {\bibinfo {title} {Hartree-fock study of the moir\'e hubbard model for
  twisted bilayer transition metal dichalcogenides},\ }\href
  {https://doi.org/10.1103/PhysRevB.104.075150} {\bibfield  {journal} {\bibinfo
   {journal} {Phys. Rev. B}\ }\textbf {\bibinfo {volume} {104}},\ \bibinfo
  {pages} {075150} (\bibinfo {year} {2021}{\natexlab{b}})}\BibitemShut
  {NoStop}%
\bibitem [{\citenamefont {Ramond}(2010)}]{ramond2010group}%
  \BibitemOpen
  \bibfield  {author} {\bibinfo {author} {\bibfnamefont {P.}~\bibnamefont
  {Ramond}},\ }\href@noop {} {\emph {\bibinfo {title} {Group Theory: A
  physicist's survey}}}\ (\bibinfo  {publisher} {Cambridge University Press},\
  \bibinfo {year} {2010})\BibitemShut {NoStop}%
\bibitem [{\citenamefont {Xu}\ and\ \citenamefont
  {Balents}(2018)}]{PhysRevLett.121.087001}%
  \BibitemOpen
  \bibfield  {author} {\bibinfo {author} {\bibfnamefont {C.}~\bibnamefont
  {Xu}}\ and\ \bibinfo {author} {\bibfnamefont {L.}~\bibnamefont {Balents}},\
  }\bibfield  {title} {\bibinfo {title} {Topological superconductivity in
  twisted multilayer graphene},\ }\href
  {https://doi.org/10.1103/PhysRevLett.121.087001} {\bibfield  {journal}
  {\bibinfo  {journal} {Phys. Rev. Lett.}\ }\textbf {\bibinfo {volume} {121}},\
  \bibinfo {pages} {087001} (\bibinfo {year} {2018})}\BibitemShut {NoStop}%
\bibitem [{\citenamefont {Wu}\ \emph {et~al.}(2022)\citenamefont {Wu},
  \citenamefont {Wu},\ and\ \citenamefont
  {Yao}}]{https://doi.org/10.48550/arxiv.2203.05480}%
  \BibitemOpen
  \bibfield  {author} {\bibinfo {author} {\bibfnamefont {Y.-M.}\ \bibnamefont
  {Wu}}, \bibinfo {author} {\bibfnamefont {Z.}~\bibnamefont {Wu}},\ and\
  \bibinfo {author} {\bibfnamefont {H.}~\bibnamefont {Yao}},\ }\bibfield
  {title} {\bibinfo {title} {Pair-density-wave and chiral superconductivity in
  twisted bilayer transition-metal-dichalcogenides}\ }\href
  {https://doi.org/10.48550/ARXIV.2203.05480} {10.48550/ARXIV.2203.05480}
  (\bibinfo {year} {2022})\BibitemShut {NoStop}%
\bibitem [{\citenamefont {Classen}\ \emph {et~al.}(2019)\citenamefont
  {Classen}, \citenamefont {Honerkamp},\ and\ \citenamefont
  {Scherer}}]{PhysRevB.99.195120}%
  \BibitemOpen
  \bibfield  {author} {\bibinfo {author} {\bibfnamefont {L.}~\bibnamefont
  {Classen}}, \bibinfo {author} {\bibfnamefont {C.}~\bibnamefont {Honerkamp}},\
  and\ \bibinfo {author} {\bibfnamefont {M.~M.}\ \bibnamefont {Scherer}},\
  }\bibfield  {title} {\bibinfo {title} {Competing phases of interacting
  electrons on triangular lattices in moir\'e heterostructures},\ }\href
  {https://doi.org/10.1103/PhysRevB.99.195120} {\bibfield  {journal} {\bibinfo
  {journal} {Phys. Rev. B}\ }\textbf {\bibinfo {volume} {99}},\ \bibinfo
  {pages} {195120} (\bibinfo {year} {2019})}\BibitemShut {NoStop}%
\bibitem [{\citenamefont {Chubukov}\ \emph {et~al.}(2016)\citenamefont
  {Chubukov}, \citenamefont {Khodas},\ and\ \citenamefont
  {Fernandes}}]{PhysRevX.6.041045}%
  \BibitemOpen
  \bibfield  {author} {\bibinfo {author} {\bibfnamefont {A.~V.}\ \bibnamefont
  {Chubukov}}, \bibinfo {author} {\bibfnamefont {M.}~\bibnamefont {Khodas}},\
  and\ \bibinfo {author} {\bibfnamefont {R.~M.}\ \bibnamefont {Fernandes}},\
  }\bibfield  {title} {\bibinfo {title} {Magnetism, superconductivity, and
  spontaneous orbital order in iron-based superconductors: Which comes first
  and why?},\ }\href {https://doi.org/10.1103/PhysRevX.6.041045} {\bibfield
  {journal} {\bibinfo  {journal} {Phys. Rev. X}\ }\textbf {\bibinfo {volume}
  {6}},\ \bibinfo {pages} {041045} (\bibinfo {year} {2016})}\BibitemShut
  {NoStop}%
\bibitem [{\citenamefont {Cvetkovic}\ \emph {et~al.}(2012)\citenamefont
  {Cvetkovic}, \citenamefont {Throckmorton},\ and\ \citenamefont
  {Vafek}}]{Cvetkovic}%
  \BibitemOpen
  \bibfield  {author} {\bibinfo {author} {\bibfnamefont {V.}~\bibnamefont
  {Cvetkovic}}, \bibinfo {author} {\bibfnamefont {R.~E.}\ \bibnamefont
  {Throckmorton}},\ and\ \bibinfo {author} {\bibfnamefont {O.}~\bibnamefont
  {Vafek}},\ }\bibfield  {title} {\bibinfo {title} {Electronic multicriticality
  in bilayer graphene},\ }\href {https://doi.org/10.1103/PhysRevB.86.075467}
  {\bibfield  {journal} {\bibinfo  {journal} {Phys. Rev. B}\ }\textbf {\bibinfo
  {volume} {86}},\ \bibinfo {pages} {075467} (\bibinfo {year}
  {2012})}\BibitemShut {NoStop}%
\bibitem [{\citenamefont {Binz}\ \emph {et~al.}(2002)\citenamefont {Binz},
  \citenamefont {Baeriswyl},\ and\ \citenamefont {Dou{\c{c}}ot}}]{Binz2002}%
  \BibitemOpen
  \bibfield  {author} {\bibinfo {author} {\bibfnamefont {B.}~\bibnamefont
  {Binz}}, \bibinfo {author} {\bibfnamefont {D.}~\bibnamefont {Baeriswyl}},\
  and\ \bibinfo {author} {\bibfnamefont {B.}~\bibnamefont {Dou{\c{c}}ot}},\
  }\bibfield  {title} {\bibinfo {title} {Wilson's renormalization group applied
  to 2d lattice electrons in the presence of van hove singularities},\ }\href
  {https://doi.org/10.1140/e10051-002-0009-7} {\bibfield  {journal} {\bibinfo
  {journal} {The European Physical Journal B - Condensed Matter and Complex
  Systems}\ }\textbf {\bibinfo {volume} {25}},\ \bibinfo {pages} {69} (\bibinfo
  {year} {2002})}\BibitemShut {NoStop}%
\bibitem [{\citenamefont {Venderbos}(2016)}]{PhysRevB.93.115107}%
  \BibitemOpen
  \bibfield  {author} {\bibinfo {author} {\bibfnamefont {J.~W.~F.}\
  \bibnamefont {Venderbos}},\ }\bibfield  {title} {\bibinfo {title} {Symmetry
  analysis of translational symmetry broken density waves: Application to
  hexagonal lattices in two dimensions},\ }\href
  {https://doi.org/10.1103/PhysRevB.93.115107} {\bibfield  {journal} {\bibinfo
  {journal} {Phys. Rev. B}\ }\textbf {\bibinfo {volume} {93}},\ \bibinfo
  {pages} {115107} (\bibinfo {year} {2016})}\BibitemShut {NoStop}%
\bibitem [{\citenamefont {Yuan}\ and\ \citenamefont {Fu}(2022)}]{Noah2022}%
  \BibitemOpen
  \bibfield  {author} {\bibinfo {author} {\bibfnamefont {N.~F.~Q.}\
  \bibnamefont {Yuan}}\ and\ \bibinfo {author} {\bibfnamefont {L.}~\bibnamefont
  {Fu}},\ }\bibfield  {title} {\bibinfo {title} {Supercurrent diode effect and
  finite-momentum superconductors},\ }\href
  {https://doi.org/10.1073/pnas.2119548119} {\bibfield  {journal} {\bibinfo
  {journal} {Proceedings of the National Academy of Sciences}\ }\textbf
  {\bibinfo {volume} {119}},\ \bibinfo {pages} {e2119548119} (\bibinfo {year}
  {2022})}\BibitemShut {NoStop}%
\bibitem [{\citenamefont {Daido}\ \emph {et~al.}(2022)\citenamefont {Daido},
  \citenamefont {Ikeda},\ and\ \citenamefont
  {Yanase}}]{PhysRevLett.128.037001}%
  \BibitemOpen
  \bibfield  {author} {\bibinfo {author} {\bibfnamefont {A.}~\bibnamefont
  {Daido}}, \bibinfo {author} {\bibfnamefont {Y.}~\bibnamefont {Ikeda}},\ and\
  \bibinfo {author} {\bibfnamefont {Y.}~\bibnamefont {Yanase}},\ }\bibfield
  {title} {\bibinfo {title} {Intrinsic superconducting diode effect},\ }\href
  {https://doi.org/10.1103/PhysRevLett.128.037001} {\bibfield  {journal}
  {\bibinfo  {journal} {Phys. Rev. Lett.}\ }\textbf {\bibinfo {volume} {128}},\
  \bibinfo {pages} {037001} (\bibinfo {year} {2022})}\BibitemShut {NoStop}%
\bibitem [{\citenamefont {Ando}\ \emph {et~al.}(2020)\citenamefont {Ando},
  \citenamefont {Miyasaka}, \citenamefont {Li}, \citenamefont {Ishizuka},
  \citenamefont {Arakawa}, \citenamefont {Shiota}, \citenamefont {Moriyama},
  \citenamefont {Yanase},\ and\ \citenamefont {Ono}}]{Ando2020}%
  \BibitemOpen
  \bibfield  {author} {\bibinfo {author} {\bibfnamefont {F.}~\bibnamefont
  {Ando}}, \bibinfo {author} {\bibfnamefont {Y.}~\bibnamefont {Miyasaka}},
  \bibinfo {author} {\bibfnamefont {T.}~\bibnamefont {Li}}, \bibinfo {author}
  {\bibfnamefont {J.}~\bibnamefont {Ishizuka}}, \bibinfo {author}
  {\bibfnamefont {T.}~\bibnamefont {Arakawa}}, \bibinfo {author} {\bibfnamefont
  {Y.}~\bibnamefont {Shiota}}, \bibinfo {author} {\bibfnamefont
  {T.}~\bibnamefont {Moriyama}}, \bibinfo {author} {\bibfnamefont
  {Y.}~\bibnamefont {Yanase}},\ and\ \bibinfo {author} {\bibfnamefont
  {T.}~\bibnamefont {Ono}},\ }\bibfield  {title} {\bibinfo {title} {Observation
  of superconducting diode effect},\ }\href
  {https://doi.org/10.1038/s41586-020-2590-4} {\bibfield  {journal} {\bibinfo
  {journal} {Nature}\ }\textbf {\bibinfo {volume} {584}},\ \bibinfo {pages}
  {373} (\bibinfo {year} {2020})}\BibitemShut {NoStop}%
\bibitem [{\citenamefont {Davydova}\ \emph {et~al.}(2022)\citenamefont
  {Davydova}, \citenamefont {Prembabu},\ and\ \citenamefont {Fu}}]{Margarita}%
  \BibitemOpen
  \bibfield  {author} {\bibinfo {author} {\bibfnamefont {M.}~\bibnamefont
  {Davydova}}, \bibinfo {author} {\bibfnamefont {S.}~\bibnamefont {Prembabu}},\
  and\ \bibinfo {author} {\bibfnamefont {L.}~\bibnamefont {Fu}},\ }\bibfield
  {title} {\bibinfo {title} {Universal josephson diode effect},\ }\href
  {https://doi.org/10.1126/sciadv.abo0309} {\bibfield  {journal} {\bibinfo
  {journal} {Science Advances}\ }\textbf {\bibinfo {volume} {8}},\ \bibinfo
  {pages} {eabo0309} (\bibinfo {year} {2022})}\BibitemShut {NoStop}%
\bibitem [{\citenamefont {Gannot}\ \emph {et~al.}(2020)\citenamefont {Gannot},
  \citenamefont {Jiang},\ and\ \citenamefont {Kivelson}}]{PhysRevB.102.115136}%
  \BibitemOpen
  \bibfield  {author} {\bibinfo {author} {\bibfnamefont {Y.}~\bibnamefont
  {Gannot}}, \bibinfo {author} {\bibfnamefont {Y.-F.}\ \bibnamefont {Jiang}},\
  and\ \bibinfo {author} {\bibfnamefont {S.~A.}\ \bibnamefont {Kivelson}},\
  }\bibfield  {title} {\bibinfo {title} {Hubbard ladders at small $u$
  revisited},\ }\href {https://doi.org/10.1103/PhysRevB.102.115136} {\bibfield
  {journal} {\bibinfo  {journal} {Phys. Rev. B}\ }\textbf {\bibinfo {volume}
  {102}},\ \bibinfo {pages} {115136} (\bibinfo {year} {2020})}\BibitemShut
  {NoStop}%
\bibitem [{\citenamefont {Yao}\ and\ \citenamefont
  {Yang}(2015)}]{PhysRevB.92.035132}%
  \BibitemOpen
  \bibfield  {author} {\bibinfo {author} {\bibfnamefont {H.}~\bibnamefont
  {Yao}}\ and\ \bibinfo {author} {\bibfnamefont {F.}~\bibnamefont {Yang}},\
  }\bibfield  {title} {\bibinfo {title} {Topological odd-parity
  superconductivity at type-ii two-dimensional van hove singularities},\ }\href
  {https://doi.org/10.1103/PhysRevB.92.035132} {\bibfield  {journal} {\bibinfo
  {journal} {Phys. Rev. B}\ }\textbf {\bibinfo {volume} {92}},\ \bibinfo
  {pages} {035132} (\bibinfo {year} {2015})}\BibitemShut {NoStop}%
\bibitem [{\citenamefont {Classen}\ \emph {et~al.}(2020)\citenamefont
  {Classen}, \citenamefont {Chubukov}, \citenamefont {Honerkamp},\ and\
  \citenamefont {Scherer}}]{PhysRevB.102.125141}%
  \BibitemOpen
  \bibfield  {author} {\bibinfo {author} {\bibfnamefont {L.}~\bibnamefont
  {Classen}}, \bibinfo {author} {\bibfnamefont {A.~V.}\ \bibnamefont
  {Chubukov}}, \bibinfo {author} {\bibfnamefont {C.}~\bibnamefont
  {Honerkamp}},\ and\ \bibinfo {author} {\bibfnamefont {M.~M.}\ \bibnamefont
  {Scherer}},\ }\bibfield  {title} {\bibinfo {title} {Competing orders at
  higher-order van hove points},\ }\href
  {https://doi.org/10.1103/PhysRevB.102.125141} {\bibfield  {journal} {\bibinfo
   {journal} {Phys. Rev. B}\ }\textbf {\bibinfo {volume} {102}},\ \bibinfo
  {pages} {125141} (\bibinfo {year} {2020})}\BibitemShut {NoStop}%
\bibitem [{\citenamefont {Isobe}\ and\ \citenamefont {Fu}(2019)}]{Isobe2019}%
  \BibitemOpen
  \bibfield  {author} {\bibinfo {author} {\bibfnamefont {H.}~\bibnamefont
  {Isobe}}\ and\ \bibinfo {author} {\bibfnamefont {L.}~\bibnamefont {Fu}},\
  }\bibfield  {title} {\bibinfo {title} {Supermetal},\ }\href
  {https://doi.org/10.1103/PhysRevResearch.1.033206} {\bibfield  {journal}
  {\bibinfo  {journal} {Phys. Rev. Research}\ }\textbf {\bibinfo {volume}
  {1}},\ \bibinfo {pages} {033206} (\bibinfo {year} {2019})}\BibitemShut
  {NoStop}%
\bibitem [{\citenamefont {Vafek}\ and\ \citenamefont
  {Yang}(2010)}]{PhysRevB.81.041401}%
  \BibitemOpen
  \bibfield  {author} {\bibinfo {author} {\bibfnamefont {O.}~\bibnamefont
  {Vafek}}\ and\ \bibinfo {author} {\bibfnamefont {K.}~\bibnamefont {Yang}},\
  }\bibfield  {title} {\bibinfo {title} {Many-body instability of coulomb
  interacting bilayer graphene: Renormalization group approach},\ }\href
  {https://doi.org/10.1103/PhysRevB.81.041401} {\bibfield  {journal} {\bibinfo
  {journal} {Phys. Rev. B}\ }\textbf {\bibinfo {volume} {81}},\ \bibinfo
  {pages} {041401} (\bibinfo {year} {2010})}\BibitemShut {NoStop}%
\bibitem [{\citenamefont {Vafek}(2010)}]{PhysRevB.82.205106}%
  \BibitemOpen
  \bibfield  {author} {\bibinfo {author} {\bibfnamefont {O.}~\bibnamefont
  {Vafek}},\ }\bibfield  {title} {\bibinfo {title} {Interacting fermions on the
  honeycomb bilayer: From weak to strong coupling},\ }\href
  {https://doi.org/10.1103/PhysRevB.82.205106} {\bibfield  {journal} {\bibinfo
  {journal} {Phys. Rev. B}\ }\textbf {\bibinfo {volume} {82}},\ \bibinfo
  {pages} {205106} (\bibinfo {year} {2010})}\BibitemShut {NoStop}%
\end{thebibliography}%


%
 
\end{document}